\documentclass[10pt,reprint,prd,nofootinbib,superscriptaddress]{revtex4}

\usepackage{amsfonts}
\usepackage{amsmath}
\usepackage{amssymb}
\usepackage{mathrsfs}
\usepackage[english]{babel}
\usepackage{graphicx}
\usepackage{epsfig}
\usepackage{bm}
\usepackage{verbatim}
\usepackage[utf8]{inputenc}
\usepackage{booktabs}
\usepackage{multirow}
\usepackage{xcolor}
\usepackage[colorlinks=true,urlcolor=red,citecolor=red,hyperfootnotes=false]{hyperref}
\usepackage[font=small]{caption}
\usepackage{float}
\usepackage{blindtext}
\usepackage{subfigure}

\usepackage{cancel}
\usepackage{csquotes}
\usepackage{verbatim}
\usepackage[titletoc,title]{appendix}
\usepackage{dsfont}

\usepackage[T1]{fontenc}
\usepackage{textcomp}

\newbox{\ORCIDicon}
\sbox{\ORCIDicon}{\large\includegraphics[width=0.8em]{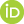}}

\newcommand{\nn}{\nonumber}

\newcommand{\bc}{\begin{center}}
\newcommand{\ec}{\end{center}}

\newcommand{\La}{\Lambda}

\newcommand{\mathsym}[1]{}

\definecolor{caribbeangreen}{rgb}{0.0, 0.8, 0.6}
\definecolor{parisgreen}{rgb}{0.31, 0.78, 0.47}
\definecolor{veronica}{rgb}{0.63, 0.36, 0.94}

\newcommand{\eq}[1]{Eq.(\ref{#1})}

\newcommand{\paren}[1]{\left(#1\right)}
\newcommand{\karen}[1]{\left[#1\right]}

\newcommand{\tx}[1]{\text{#1}}

\begin{document}

\title{Strongly coupled inert scalar sector with radiative neutrino masses}
\author{A. E. C\'arcamo Hern\'andez\,\href{https://orcid.org/0000-0002-2421-2732}{\usebox{\ORCIDicon}}\,}
\email{antonio.carcamo@usm.cl}
\affiliation{Universidad T\'{e}cnica Federico Santa Mar\'{\i}a, Casilla 110-V, Valpara\'{\i}so, Chile.}
\affiliation{Centro Científico-Tecnológico de Valparaíso, Casilla 110-V, Valparaíso, Chile.}
\affiliation{Millennium Institute for Subatomic physics at high energy frontier - SAPHIR. Fernandez Concha 700. Santiago. Chile.}
\author{Jeremy Echeverria Puentes\,\href{https://orcid.org/0000-0002-5832-103X}{\usebox{\ORCIDicon}}\,}
\email{jeremy.echeverria@usm.cl}
\affiliation{Universidad T\'{e}cnica Federico Santa Mar\'{\i}a, Casilla 110-V, Valpara\'{\i}so, Chile.}
\affiliation{Centro Científico-Tecnológico de Valparaíso, Casilla 110-V, Valparaíso, Chile.}
\author{R. Pasechnik\,\href{https://orcid.org/0000-0003-4231-0149}{\usebox{\ORCIDicon}}\,}   
\email{roman.pasechnik@cern.ch}
\affiliation{Department of Physics, Lund University, SE-223 62 Lund, Sweden.} 
\author{Daniel Salinas-Arizmendi\,\href{https://orcid.org/0000-002-0577-2005}{\usebox{\ORCIDicon}}\,}
\email{daniel.salinas@usm.cl}
\affiliation{Universidad T\'{e}cnica Federico Santa Mar\'{\i}a, Casilla 110-V, Valpara\'{\i}so, Chile.}
\affiliation{Centro Científico-Tecnológico de Valparaíso, Casilla 110-V, Valparaíso, Chile.}
\date{\today }

\begin{abstract}
We explore the phenomenological consequences of a model with an extended scalar sector, incorporating strongly coupled inert Higgs doublets. The model introduces three Higgs doublets: one that interacts with the $SU(2)$ symmetry of the Standard Model, and two inert doublets belonging to a strongly interacting sector, embedded in the  $SU(2)_2 \times SU(2)_1 \times U(1)_Y$ electroweak gauge symmetry. This symmetry structure is further supplemented by a spontaneously broken $\mathbb{Z}_2$ and a preserved $\mathbb{Z}_2'$ discrete symmetry. Our approach harnesses the exotic scalar fields emerging from this sector to implement mass-generation mechanisms. In particular, a one-loop seesaw mechanism accounts for the smallness of neutrino masses, while a universal seesaw mechanism naturally explains the hierarchy of charged fermion masses below the top quark mass. The model is shown to be consistent with current experimental constraints, including those from charged lepton flavor violation, electroweak precision observables, the Higgs diphoton decay rate, and the Higgs trilinear self-coupling. Notably, it also provides a viable interpretation of the 95~GeV diphoton excess, offering a distinctive signature of new physics beyond the Standard Model.

\end{abstract}
 
\maketitle

\section{\label{intro}Introduction}
The Standard Model (SM) is the theoretical framework that provides the most accurate description of the behavior of non-gravitational interactions. However, despite its outstanding agreement 
with the experimental data, challenges such as the origin of neutrino masses, the nature of dark matter and matter-antimatter asymmetry suggest the existence of new physics beyond the SM \cite{Planck:2018vyg,Fields:2019pfx,Super-Kamiokande:1998kpq,SNO:2001kpb}. In addition, some fundamental considerations, such as the hierarchy problem \cite{Weinberg:1975gm,Susskind:1978ms,Gildener:1976ai}, raise questions about the stability of the electroweak scale under 
quantum corrections, requiring a more robust formulation of the theoretical framework \cite{Yamada:2020bqe}.

One solution to the hierarchy problem without resorting to supersymmetry is to consider the Higgs boson as a composite state, product of a new strongly coupled dynamics. In this approach the mass of the Higgs boson is saturated at energy scales not much larger than the electroweak scale \cite{Contino:2009ez} and generates new physical states, such as massive vector resonances  \cite{Bando:1984ej,Bando:1984pw,Bando:1985rf,Bando:1987ym,Bando:1987br,Chivukula:2006cg,Barbieri:2009tx} and composite scalars \cite{Foadi:2007ue,Ryttov:2008xe,Sannino:2009za,CarcamoHernandez:2010qxf,CarcamoHernandez:2010wpm,CarcamoHernandez:2011pfx,Foadi:2012bb,Hapola:2011sd,Belyaev:2013ida,CarcamoHernandez:2013ydh,CarcamoHernandez:2017pei}, which can strongly interact with the scalar sector of the SM. These interactions not only provide observable corrections to the Higgs sector, but also produce phenomena that can be experimentally tested, endowing this class of models with additional appeal. In such models, the electroweak symmetry breaking (EWSB) is product of the emergence of an effective potential at low energies of this strongly coupled sector which exhibits a non-zero vacuum expectation value (VEV) \cite{Bardeen:1989ds}.

In this work, we propose a model based on an extended scalar sector including an active Higgs doublet and two inert Higgs doublets, accompanied by an extension of the electroweak gauge group to $SU(2)_2\times SU(2)_1\times U(1)$. In this framework, one of the $SU(2)_1$ groups is weakly coupled and contains the active Higgs doublet, while the other $SU(2)_2$ group is strongly coupled and contains the inert Higgs doublets. The model incorporates exotic scalar fields via a nonlinear sigma model in order to break the extended group symmetry, along with a symmetry breaking mechanism for a discrete symmetries $\mathbb{Z}_2\times\mathbb{Z}_2'$.
Models with two inert doublets have been extensively
worked on in the literature \cite{LopezHonorez:2006gr,Ivanov:2012fp,Keus:2013hya,Aranda:2012bv,Aranda:2014jua,Aranda:2019vda,Akeroyd:2016ssd,Hernandez:2021iss} and inert models with strongly coupled scalar sector in \cite{Zerwekh:2009yu,DiChiara:2016ybc,DeCurtis:2016scv,Rojas-Abatte:2017hqm,CarcamoHernandez:2017pei}.
The main achievement of this work is that the design of this model allows us to take advantage of the new fields to implement a radiative neutrino mass generation mechanism, supplemented by a seesaw scheme to explain the mass hierarchy in the fermionic sector. In addition, we explore phenomenological constraints derived from observables such as Higgs trilinear selfcoupling, the oblique parameters $S$, $T$, and $U$, Higgs diphoton decay rate and processes involving leptonic flavor and lepton number violation with the possibility of baryogenesis through leptogenesis.

This model represents a natural and minimal extension of the SM, addressing in a unified way the fundamental constraints such as the fermion mass hierarchy, the smallness of neutrino masses and the hierarchy problem, while remaining compatible with current experimental constraints. Its ability to provide viable and testable predictions with projected experimental data is also analyzed.

The structure of this paper is as follows: in Sec. \ref{sec::2} we describe the model, highlighting the construction of the strongly coupled scalar sector and its relation to the mass generation mechanisms. In Sec. \ref{sec::3} we present the scalar potential and analyze the emerging particle spectrum. Sec. \ref{sec::4} is devoted to the fermionic mass generation and hierarchy. In Section \ref{sec:trilinear}, the constraints imposed by the trilinear Higgs self-coupling are studied. In Sec. \ref{sec::5} we explore the parameter space compatible with the experimental constraints on the scalar sector, with emphasis in the Higgs two-photon decay. Sec. \ref{sec::6}, \ref{sec:leptogenesis} and \ref{sec::7} describe the results of the analysis of the phenomenology of the fermionic sector, including lepton flavor violating processes, the feasibility of leptogenesis within this theoretical framework and the oblique parameters. In Section \ref{sec:95gev}, we showed that the scalar particle spectrum can account for the excess of events observed around the $95$ GeV diphoton invariant mass, within a specific scenario of the model. Finally, in Sec. \ref{sec::8} we discuss the results and present our conclusions. The full model scalar potential and the analytical expression for the effective trilinear higgs coupling are shown in Appendices \ref{app:a} and \ref{app:b}, respectively.

\section{\label{sec::2}The model}

In this section we explain the theoretical framework underpinning the proposed model, detailing its symmetry structure, particle content, and mechanisms for fermion mass generation.

The extended symmetry group, $SU(3)_C\times SU(2)_2\times SU(2)_1\times U(1)_Y$, introduces a strongly coupled $SU(2)_2$ sector, whereas the $SU(2)_1$ symmetry retains the electroweak symmetry of the SM. The resulting gauge group introduces effective states, such as vector resonances, whose properties are crucial for the phenomenological analyses to be discussed in subsequent sections.

To recover the EWSB of the SM, the extended symmetry group is broken to the SM group at low energies via a hierarchical spontaneous symmetry breaking (SSB) mechanism. The first stage of this breaking is triggered by a nonlinear sigma model field, $\Sigma$, transforming as the fundamental representation of the $SU(2)_2\times SU(2)_1$ group. This field drives the transition to the SM gauge group by acquiring nonzero vacuum expectation values in its electrically neutral components \cite{Chivukula:2011ag,SekharChivukula:2009htk}. Additionally, the second stage of discrete SSB mechanism, related to the generation of fermion masses, is incorporated through the inclusion of a real singlet scalar field, $\sigma$. These SSB patterns are summarized schematically as follows:
\begin{gather}
SU(3)_{C}\times SU(2)_{2}\times SU(2)_{1}\times U(1)_{Y}\times \mathbb{Z}_2 \times \mathbb{Z}^{\prime}_2 \notag \\
\Downarrow v_{\Sigma},v_{\sigma}  \notag \\
\label{eq:symmetrygroup} SU(3)_{C}\times SU(2)_{L}\times U(1)_{Y}\times \mathbb{Z}_2^{\prime}   \\
\Downarrow v_{\phi}  \notag \\
SU(3)_{C}\times U(1)_{\text{em}}\times \mathbb{Z}_2^{\prime}  \notag
\end{gather}
where the discrete SSB mechanism is mediated by the singlet scalar field $\sigma$. In this hierarchy, the VEVs satisfy the condition $v_{\Sigma} \gg v_{\sigma} \gg v_\phi$, where $v_\phi$ denotes the electroweak VEV. The diagonal components of the \mbox{$SU(2)_2\times SU(2)_1$} scalar bidoblets, which are electrically neutral do acquire the VEV \mbox{$v_{\Sigma}$}. %

In addition to the previously mentioned scalar singlet $\sigma$ and scalar bidoblet $\Sigma$, the implementation of the seesaw mechanism for SM 
fermion mass generation, requires that the scalar sector also incorporates 
three Higgs doublets: $\phi$, $h_1$, and $h_2$. Here, $\phi$ represents the SM Higgs doublet, while $h_1$ and $h_2$ are inert doublets belonging the strongly coupled dark sector since they have non trivial charges under the preserved $\mathbb{Z}_2^{\prime}$ symmetry and transform as doublets under the $SU(2)_{2}$ group associated with a strongly interacting sector. The strongly coupled inert doublet $h_1$ and $h_2$ play a key role in the one-loop level radiative seesaw mechanism that yields the tiny masses of the active neutrinos. Furthermore, the gauge singlet scalar field $\sigma$ is crucial to generate mixings between heavy non SM charged fermions and the SM fermions then triggering a seesaw mechanism that generates the masses of SM charged fermions lighter than the top quark. Besides that, the scalar singlet $\sigma$ provides masses to the non SM heavy charged seesaw messengers.

\newpage

The particle content of the gauge and scalar sectors is summarized in Table \ref{tab:repre}. 
The $SU(2)_1$ and $SU(2)_2$ gauge bosons are denoted as $A_\mu^{(1)}$ and $A_\mu^{(2)}$, respectively. 
Throughout this work, the hypercharge ($Y$) operator is defined as follows
\begin{align}
    Q = T^{(3)} + Y = T_1^{(3)} + T_2^{(3)} + Y,
\end{align}
where $T^a$ denotes the generator of the diagonal subgroup $SU(2)_L \subset SU(2)_1 \times SU(2)_2$, namely $T^{(a)} \equiv T_1^{(a)} + T_2^{(a)}$, with $T_1^{(a)}$ and $T_2^{(a)}$ being the generators of the $SU(2)_1$ and $SU(2)_2$ groups, respectively. 
The orthogonal combination of $SU(2)_1$ and $SU(2)_2$ generators is broken, giving rise to the heavy vector states.

The most general Lagrangian for this model is expressed as:
\begin{equation}
\mathscr{L} = \mathscr{L}_\text{gauge} - \mathscr{L}_\text{fermion} - V(\phi, h_1, h_2, \sigma, \Sigma),
\end{equation}
where $\mathscr{L}_\text{gauge}$ includes the kinetic terms for the gauge bosons and scalar fields\footnote{The fermion kinetic terms have been omitted because they are the standard ones $i\bar\psi(\gamma^\mu\partial_\mu+m)\psi$.}, as well as the effective non-linear sigma model describing the low-energy dynamics of the strongly coupled dark sector. This term governs the behavior and self-interactions of the gauge fields. The fermionic Lagrangian, $\mathscr{L}_\text{fermion}$, contains the 
interactions responsible for generating the fermion mass spectrum. Lastly, $V(\phi, h_1, h_2, \sigma, \Sigma)$ represents the scalar potential, encoding the interactions among the scalar fields.

\begin{table}[]
\centering
\begin{tabular}{|c|c|c|c|c|c|c|c|c|c|c|c|c|c|c|c|c|c|c|c|c|}
\hline \hline
& $\phi$& $h_1$&$h_2$& $\sigma$ & $\Sigma$ & $B_\mu$ &$A_\mu^{(1)}$  &$A_\mu^{(2)}$  \\ \hline \hline
$SU(3)_C$ & $\textbf{1}$ & $\textbf{1}$  &$\textbf{1}$  &\textbf{1}& $\textbf{1}$  &$\textbf{1}$  & $\textbf{1}$ & $\textbf{1}$ \\ \hline 
$SU(2)_1$ & $\textbf{2}$ & $\textbf{1}$  &$\textbf{1}$  & \textbf{1}&$\textbf{2}$&$\textbf{1}$  & $\textbf{3}$ &$\textbf{1}$\\ \hline 
$SU(2)_2$ & $\textbf{1}$ & $\textbf{2}$  &$\textbf{2}$  &\textbf{1} & $\textbf{2}$&$\textbf{1}$  & $\textbf{1}$ & $\textbf{3}$ \\ \hline
$U(1)_Y$ & $1/2$ & $1/2$ &$1/2$ & $0$ & $0$ &$0$ & $0$ & $0$ \\ \hline
$\mathbb{Z}_2$& $0$& $0$&$0$& $1$& $1$&$0$& $0$& $1$\\ \hline 
$\mathbb{Z}^{\prime}_2$& $0$& $1$& $1$& $0$& $0$&$0$& $0$&$0$\\ \hline 
\end{tabular}
    \caption{Scalar and gauge assignments under the $SU(3)_{C}\times SU(2)_{2}\times SU(2)_{1}\times U(1)_{Y} \times \mathbb{Z}_2 \times \mathbb{Z}^{\prime}_2$ symmetry.}
    \label{tab:repre}
\end{table}

The gauge Lagrangian, $\mathscr{L}_\text{gauge}$, can be written explicitly as:
\begin{equation}
\begin{aligned}
\mathscr{L}_\text{gauge} = & -\frac{1}{4}\text{Tr}\left[F^{(1)}_{\mu\nu} F^{(1)\mu\nu}\right] - \frac{1}{4}\text{Tr}\left[F^{(2)}_{\mu\nu} F^{(2)\mu\nu}\right] - \frac{1}{4} B_{\mu\nu} B^{\mu\nu} + \frac{f_\Sigma^2}{2} \text{Tr}\left[\left(D^\mu \Sigma \right)^\dagger \left(D_\mu \Sigma \right)\right] \\
& + \left(D^\mu \phi\right)^\dagger \left(D_\mu \phi\right) + \left(D^\mu h_1\right)^\dagger \left(D_\mu h_1\right) + \frac{\beta^2}{2}\left(\phi^\dagger \phi\right) \text{Tr}\left[\left(D^\mu \Sigma\right)^\dagger \left(D_\mu \Sigma\right)\right] \\
& + \left(D^\mu h_2\right)^\dagger \left(D_\mu h_2\right) + \frac{1}{2} \partial_\mu \sigma \partial^\mu \sigma,
\label{eq:lagrangian}
\end{aligned}
\end{equation}
where $f_\Sigma$ represents the decay constant associated with the Goldstone bosons arising from the $\Sigma$ field, which plays a crucial 
role in the dynamics of the scalar sector. The parameter $\beta$, with mass dimension, has a direct influence on the masses of the gauge bosons, however, its physical constraints are related with the underlying strong dynamics so it can be treated as a free parameter \cite{CarcamoHernandez:2017pei}.

The covariant derivatives appearing in the gauge Lagrangian are defined as:
\begin{equation}
\begin{aligned}
D_\mu \Sigma &= \partial_\mu \Sigma - i g_1 A_{\mu}^{(1)} \Sigma + i g_2 \Sigma A_{\mu}^{(2)}, \\
D_\mu \phi &= \partial_\mu \phi - i g_1 A_{\mu}^{(1)} \phi - i \frac{g_y}{2} B_\mu, \\
D_\mu h_n &= \partial_\mu h_n - i g_2 A_{\mu}^{(2)} h_n - i \frac{g_y}{2} B_\mu,
\end{aligned}
\end{equation}
where $g_1$ and $g_2$ denote the gauge couplings for the $SU(2)_1$ and $SU(2)_2$ symmetries, respectively, while $g_y$ is the hypercharge coupling. The index $n = 1, 2$ is used to distinguish the scalar fields $h_1$ and $h_2$, and this convention will be maintained throughout this work unless otherwise specified.

In the fermionic sector, the left-handed quarks are denoted as $\overline{Q}_{iL}$, with $i = 1, 2, 3$ representing the family index, a convention adopted for the remainder of this study. The right-handed quarks are labeled as $u_{iR}$ and $d_{iR}$ for the \textit{up}-type and \textit{down}-type quarks, respectively.

We propose to motivate the mass generation mechanisms for the fermions of the model. First of all, the mass of the top quark will be generated at tree-level by a Yukawa interaction, as in the Standard Model:
\begin{align}
    \mathcal{O}_Y = \overline{Q}_{iL} \tilde{\phi} u_{3R}.
\end{align}

\begin{table}[]
\centering
\begin{tabular}{|c|c|c|c|c|c|c|c|c|c|c|c|c|c|}
\hline \hline
&  $Q_{iL}$   & $u_{nR}$&$u_{3R}$& $d_{iR}$& $T_{nL}$ & $T_{nR}$ & $B_{iL}$ & $B_{iR}$& $L_{iL}$ & $l_{iR}$ & $N_{R}$ &$E_{iL}$ & $E_{iR}$  \\ \hline \hline
$SU(3)_C$ & $\textbf{3}$ & $\textbf{3}$  &$\textbf{3}$  & $\textbf{3}$ & $\textbf{3}$ & $\textbf{3}$ & $\textbf{3}$ & $\textbf{3}$ &$\textbf{1}$ &$\textbf{1}$ &$\textbf{1}$ &$\textbf{1}$ &$\textbf{1}$ \\ \hline 
$SU(2)_1$ & $\textbf{2}$ &$\textbf{1}$  &$\textbf{1}$  & $\textbf{1}$ & $\textbf{1}$&$\textbf{1}$  & $\textbf{1}$  & $\textbf{1}$ &$\textbf{2}$ &$\textbf{1}$ &$\textbf{1}$ & $\textbf{1}$ & $\textbf{1}$\\ \hline 
$SU(2)_2$ & $\textbf{1}$ & $\textbf{1}$  &$\textbf{1}$  & $\textbf{1}$ & $\textbf{1}$& $\textbf{1}$ & $\textbf{1}$ &$\textbf{1}$ &$\textbf{1}$ & $\textbf{1}$& $\textbf{1}$& $\textbf{1}$& $\textbf{1}$ \\ \hline
$U(1)_Y$ & $1/6$ & $2/3$ &$2/3$ & $-1/3$ & $2/3$ & $2/3$ & $-1/3$ &$-1/3$ & $-1$ & $-1/2$ & $0$ & $-1/2$ & $-1$ \\ \hline
$\mathbb{Z}_2$& 0& $1$&0& 1& 0& 1& 0& 1&0&1&$0$&0&1\\ \hline 
$\mathbb{Z}^{\prime}_2$& $0$&$0$&$0$& $0$& $0$&$0$& $0$& $0$&$0$&$0$&$1$ & $0$& $0$\\ \hline 
\end{tabular}
    \caption{Fermionic assignments under the $SU(3)_{C}\times SU(2)_{2}\times SU(2)_{1}\times U(1)_{Y} \times \mathbb{Z}_2 \times \mathbb{Z}^{\prime}_2$ symmetry. In this case the index $\textit{i}=1,2,3;$ and the index $\textit{n}=1,2$.}
    \label{tab:charge_assign}
\end{table}

\begin{figure}
    \centering
\subfigure[]{\includegraphics[width=0.4\linewidth]{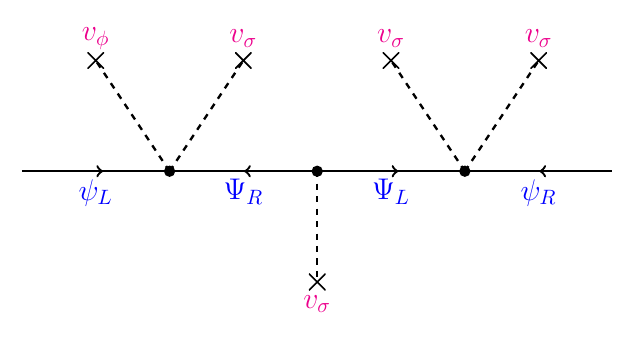}}
\subfigure[]{\includegraphics[width=0.4\linewidth]{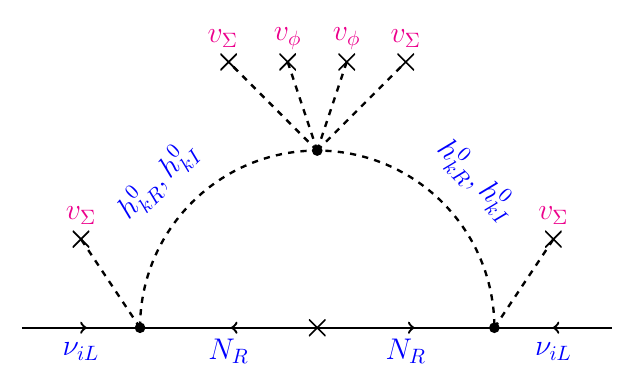}}
\caption{Fermion mass mechanism. In (a) Universal seesaw mechanism for quarks and charged lepton $\psi_L=\{ Q_{nL},e_{iL}\}$, $\psi_R=\{u_{nR},d_{nR}, e_{iR} \}$, $\Psi_L=\{T_{nL},B_{iL},E_{iL}\}$ and $\Psi_R=\{T_{nR},B_{iR},E_{iR}\}$, for $n=1,2$, $i=1,2,3$ (excluding top quark). In (b) One-loop radiative seesaw mechanism for neutrino sector.}
\label{fig:dia-mass}
\end{figure}

With this in mind, the particle content of the fermionic sector has been extended to include exotic fermions, singlets of $SU(2)_1\times SU(2)_2$. Two left-handed exotic \textit{up}-type quarks $T_{nL}$ and three \textit{down}-type $B_{iL}$ have been added in conjunction with $5$ right-handed $T_{nR},~B_{iR}$ singlets. For the leptonic sector, three exotic left-handed singlets $E_{iL}$ and three exotic right-handed singlets $E_{iR}$ have been added.

In order to generate the masses of the other charged and exotic fermions, we will use the universal seesaw mechanism, which requires forbidding the tree-level renormalizable operators that allow the generation of masses via Yukawa interaction:
\begin{align}
    \cancel{\bar{Q}_{iL} \tilde{\phi} u_{nR}},\quad \cancel{\bar{Q}_{iL} \phi d_{iR}},\quad \cancel{\bar{L}_L^i \phi l_{iR}},\quad \cancel{\bar{T}_{nL} \tilde{\phi} T_{nR}},\quad \cancel{\bar{B}_{iL}\phi B_{iR}},\quad \cancel{\bar{E}_{iL} \phi E_{iR}},
\end{align}
with the left-handed charged leptons given by $\bar{L}_L^i$ and the right-handed by $l_R^i$. To implement this prohibition, the model is equipped with a spontaneously broken global discrete symmetry $\mathbb{Z}_2$. The breaking of this discrete symmetry is due to the presence of a new real singlet scalar field, $\sigma$, charged under $\mathbb{Z}_2$, which acquires a VEV. On the other hand, all right-handed fermions, excluding the top quark, are charged under $\mathbb{Z}_2$.

Instead, the fermions can acquire their masses through non-renormalizable dimension-$5$ operators that are constructed using the SM Higgs doublet and the new singlet scalar field $\sigma$, whose form is:
\begin{align}
    \mathcal{O}_5\sim \left\{\frac{1}{\Lambda}\bar{Q}_{iL} \tilde{\phi} \sigma T_{mR},~\frac{1}{\Lambda}\bar{Q}_{iL} \phi \sigma B_{jR},~\frac{1}{\Lambda}\bar{L}_{iL} \phi \sigma E_{jR}\right\}.
\end{align}

From here, we can see the reason for the $\mathbb{Z}_2$ symmetry breaking: to give way to the usual Higgs mechanism at low energy. The presence of the $\sigma$ field allows new renormalizable Yukawa operators:
\begin{align}
\label{eq:opeYukawa}
   \mathcal{O}_Y\sim \left\{\overline{T}_{nL} \sigma u_{mR},~\overline{T}_{nL} \sigma T_{mR},~\overline{B}_{iL} \sigma d_{jR},~\overline{B}_{iL} \sigma B_{jR},~\overline{E}_{iL} \sigma l_{jR},~\overline{E}_{iL}\sigma E_{jR}\right\},
\end{align}
these together form a universal seesaw mechanism to provide mass to all fermions.

Regarding the 
neutrino sector, 
a one-loop level radiative seesaw mechanism will be used in order to generate the tiny masses for the active neutrinos. This implies that the smallness of active neutrino masses will be attributed to the loop suppression as well as to the small mass splitting between the physical dark scalars and dark pseudoscalars.  
The successfull implementation of such radiative seesaw mechanism requires the inclusion 
of a right-handed sterile heavy Majorana neutrino $N_R$ as well as a preserved %
discrete symmetry $\mathbb{Z}_2'$. This %
unbroken $\mathbb{Z}_2'$ symmetry is crucial %
to forbid tree-level masses for active neutrinos at all energy scales. %
To build %
the loop, we will use the inert Higgs doublets $h_1$ and $h_2$ in combination with the scalar bidoublet $\Sigma$ to construct the non-renormalizable operators:
\begin{align}
   \mathcal{O}_\nu \sim \left\{\frac{1}{\Lambda}\overline{L}_{iL} \tilde{h}_1 \Sigma N_R,~\frac{1}{\Lambda}\overline{L}_{iL} \tilde{h}_2 \Sigma N_R \right\},
\end{align}
and forbid others by making $h_1$, $h_2$ and $N_R$ charged under the preserved $\mathbb{Z}_2'$ symmetry. 
Such loop will be closed by the following non renormalizable scalar interactions:
\begin{align}
   \mathcal{O}_s \sim \left\{\frac{1}{\Lambda^2}\left(\phi^{\dagger} \Sigma h_1\right)^2,\frac{1}{\Lambda^2}\left(\phi^{\dagger} \Sigma h_2\right)^2\right\},
\end{align}
that will generate a small mass splitting between the dark CP even and dark CP odd scalars running in the internal lines of the neutrino loop diagram.

In conclusion, in this model, the masses of fermions are generated by a combination of a tree level universal seesaw mechanism for SM charged fermions lighter than the top quark and a radiative seesaw mechanism for active neutrinos, as we shown in Figure~\ref{fig:dia-mass}, which follows from the particle field assignments under the symmetries of the model %
displayed in Table~\ref{tab:charge_assign}.

Finally, with the particle content and symmetries specified in Table \ref{tab:charge_assign}, the following Yukawa interactions arise:  
\begin{equation}\label{eq:fermionlagrangian}
\begin{aligned}
\mathscr{L}_{\mathrm{fermion}}= & \sum_i Y_i^1 \overline{Q}_{iL} \tilde{\phi} u_{3R} +\sum_{i, m} \frac{ Y_{i m}^2}{\Lambda} \overline{Q}_{iL} \tilde{\phi} \sigma T_{mR}+\sum_{n, m} Y_{n m}^3 \overline{T}_{nL} \sigma u_{mR}+\sum_{n, m} Y_{n m}^4 \overline{T}_{nL} \sigma T_{mR} \\
& + \sum_{i, j} \frac{ Y_{i j}^5}{\Lambda} \overline{Q}_{iL} \phi \sigma B_{jR} 
+\sum_{i, j} Y_{i j}^6 \overline{B}_{iL} \sigma d_{jR}+\sum_{i, j} Y_{i j}^7 \overline{B}_{iL} \sigma B_{jR} +\sum_{i, j} y_{i j}^1 \overline{E}_{iL} \sigma l_{jR}  \\
& +\sum_{i, j} y_{i j}^2 \overline{E}_{iL}\sigma E_{jR}+\sum_{i, j} \frac{y_{i j}^3}{\Lambda} \overline{L}_{iL} \phi \sigma E_{jR} 
+\sum_{i} \frac{y_{i}^4}{\Lambda} \overline{L}_{iL} \tilde{h}_1 \Sigma N_R+\sum_{i} \frac{y_{i}^5}{\Lambda} \overline{L}_{iL} \tilde{h}_2 \Sigma N_R+m_N N_R \overline{N}_R^{c} + H.c,
\end{aligned}
\end{equation}
where $i,j=1,2,3$, $m,n=1,2$ and the model cutoff $\La  $ is the scale of the UV completion of the model.

\section{\label{sec::3}Scalar potential and scalar mass spectrum}

In this section, we will analyze the low-energy scalar potential of the model. The most general scalar potential invariant under the symmetry in Eq.~\eqref{eq:symmetrygroup} (see Table \ref{tab:repre}) reads:
\begin{align}\label{eq:potential}
V=&\ \mu_{\phi}^2\left(\phi \phi^{\dagger}\right)+\mu_{h_1}^2\left(h_1 h_{1}^{\dagger}\right)+\mu_{h_2}^2\left(h_2 h_{2 }^{\dagger}\right)+\mu_{h_{12}}^2\left(h_1 h_{2}^{\dagger}+\text {H.c.}\right)+\mu_{\Sigma}^2 \operatorname{Tr}\left(\Sigma \Sigma^{\dagger}\right)+\mu_{\Sigma'}^2 \operatorname{Tr}\left[\Sigma^2+(\Sigma^{*})^2\right]\\
\nn & +\mu_{\sigma}^2(\sigma \sigma) + \lambda_1\left(\phi \phi^{\dagger}\right)\left(\phi \phi^{\dagger}\right)+\lambda_2\left(\phi \phi^{\dagger}\right)\left(h_1 h_1^{\dagger}\right)+\lambda_3\left(\phi \phi^{\dagger}\right)\left(h_2 h_2^{\dagger}\right)+\lambda_4\left(\phi \phi^{\dagger}\right)\left(h_1 h_2^{\dagger}+\text {H.c.}\right) \\
\nn &+\lambda_5\left(\phi \phi^{\dagger}\right) \operatorname{Tr}\left(\Sigma \Sigma^{\dagger}\right)  +\lambda_6\left(\phi \phi^{\dagger}\right)(\sigma \sigma)+\lambda_7\left(h_1 h_1^{\dagger}\right)\left(h_1 h_1^{\dagger}\right)+\lambda_8\left(h_1 h_1^{\dagger}\right)\left(h_2 h_2^{\dagger}\right)+\frac{1}{2}\left[\lambda_9\left(h_1 h_2^{\dagger}\right)^2+\text {H.c.}\right] \\
\nn & +\lambda_{10}\left(h_1 h_1^{\dagger}\right)\left(h_1 h_2^{\dagger}+\text {H.c.}\right)+\lambda_{11}\left(h_1 h_1^{\dagger}\right) \operatorname{Tr}\left(\Sigma \Sigma^{\dagger}\right)+\lambda_{12}\left(h_1 h_1^{\dagger}\right)(\sigma \sigma)+\lambda_{13}\left(h_2 h_2^{\dagger}\right)\left(h_2 h_2^{\dagger}\right)\\
\nn & +\lambda_{14}\left(h_2 h_2^{\dagger}\right)\left(h_1 h_2^{\dagger}+\text {H.c.}\right)+\lambda_{15}\left(h_2 h_2^{\dagger}\right) \operatorname{Tr}\left(\Sigma \Sigma^{\dagger}\right)+\lambda_{16}\left(h_2 h_2^{\dagger}\right)(\sigma \sigma)+\lambda_{17}\left(h_1 h_2^{\dagger}+\text { H.c.}\right) \operatorname{Tr}\left(\Sigma \Sigma^{\dagger}\right) \\
\nn & +\lambda_{18}\left(h_1 h_2^{\dagger}+\text {H.c.}\right)(\sigma \sigma)+\lambda_{19} \operatorname{Tr}\left(\Sigma \Sigma^{\dagger}\right) \operatorname{Tr}\left(\Sigma \Sigma^{\dagger}\right)+\lambda_{20} \operatorname{Tr}\left[\left(\Sigma \Sigma^{\dagger}\right)^2\right]+\lambda_{21} \operatorname{Tr}\left(\Sigma \Sigma^{\dagger}\right)(\sigma \sigma)\\
\displaybreak
\nn &+\lambda_{22}(\sigma \sigma)(\sigma \sigma)+\lambda_{23}\operatorname{Tr}\left(\tilde\Sigma^\dagger\tilde\Sigma\right)\operatorname{Tr}\left(\tilde\Sigma^\dagger\tilde\Sigma\right)+\lambda_{24}\operatorname{Tr}\left[(\tilde\Sigma^\dagger\tilde\Sigma)^2\right]+\lambda_{25}\left(\phi \phi^{\dagger}\right) \operatorname{Tr}\left(\tilde\Sigma\tilde\Sigma^{\dagger}\right)+\lambda_{26}\left(h_1 h_1^{\dagger}\right) \operatorname{Tr}\left(\tilde\Sigma \tilde\Sigma^{\dagger}\right)\\
\nn &+\lambda_{27}\left(h_2 h_2^{\dagger}\right) \operatorname{Tr}\left(\tilde\Sigma \tilde\Sigma^{\dagger}\right)+\lambda_{28}\left(h_1 h_2^{\dagger}+\text { H.c. }\right) \operatorname{Tr}\left(\tilde\Sigma \tilde\Sigma^{\dagger}\right)+\lambda_{29} \operatorname{Tr}\left(\tilde\Sigma \tilde\Sigma^{\dagger}\right)(\sigma \sigma)+\lambda_{30}\operatorname{Tr}\left[\Sigma^2+(\Sigma^{*})^2\right]^2\\
\nn & +\lambda_{31}\left(\phi \phi^{\dagger}\right) \operatorname{Tr}\left[\Sigma^2+(\Sigma^{*})^2\right]+\lambda_{32}\left(h_1 h_1^{\dagger}\right) \operatorname{Tr}\left[\Sigma^2+(\Sigma^{*})^2\right]+\lambda_{33}\left(h_2 h_2^{\dagger}\right) \operatorname{Tr}\left[\Sigma^2+(\Sigma^{*})^2\right]\\
\nn & +\lambda_{34}\left(h_1 h_2^{\dagger}+\text { H.c.}\right) \operatorname{Tr}\left[\Sigma^2+(\Sigma^{*})^2\right]+\lambda_{35}\operatorname{Tr}\left[\Sigma^2+(\Sigma^{*})^2\right](\sigma \sigma)+\lambda_{36}\operatorname{Tr}\left(\Sigma \Sigma^{\dagger}\right)\operatorname{Tr}\left[\Sigma^2+(\Sigma^{*})^2\right]\\
\nn &+\lambda_{37}\operatorname{Tr}\left(\tilde\Sigma \tilde\Sigma^{\dagger}\right)\operatorname{Tr}\left[\Sigma^2+(\Sigma^{*})^2\right]+\lambda_{38}\operatorname{Tr}\left[\left(\Sigma^2+(\Sigma^{*})^2\right)^2\right]+\frac{\alpha_1}{\Lambda^2}\left|\phi^{\dagger} \Sigma h_1\right|^2+\frac{\alpha_2}{\Lambda^2}\left|\phi^{\dagger} \Sigma h_2\right|^2\\
\nn & +\frac{\alpha_3}{\Lambda^2}\left[\left(\phi^{\dagger} \Sigma h_1\right)^2+\left(h_1^{\dagger} \Sigma^{\dagger} \phi\right)^2\right] +\frac{\alpha_4}{\Lambda^2}\left[\left(\phi^{\dagger} \Sigma h_2\right)^2+\left(h_2^{\dagger} \Sigma^{\dagger} \phi\right)^2\right]+\frac{\alpha_5}{\Lambda^2}\left|\phi^{\dagger} \tilde\Sigma h_1\right|^2+\frac{\alpha_6}{\Lambda^2}\left|\phi^{\dagger} \tilde\Sigma h_2\right|^2\\
\nn &+\frac{\alpha_7}{\Lambda^2}\left[\left(\phi^{\dagger} \tilde\Sigma h_1\right)^2+\left(h_1^{\dagger} \tilde\Sigma^{\dagger} \phi\right)^2\right] +\frac{\alpha_8}{\Lambda^2}\left[\left(\phi^{\dagger} \tilde\Sigma h_2\right)^2+\left(h_2^{\dagger} \tilde\Sigma^{\dagger} \phi\right)^2\right].
\end{align}

Here, $\mu_{\phi,h_{1,2},\Sigma,\sigma}$ are the dimensionfull  
parameters of the SM Higgs doublet, the inert Higgs doublets, the scalar bi-doublet, and the scalar singlet, respectively; $\lambda_i$ are the dimensionless quartic couplings, and $\alpha_i$ are the dimensionless couplings of dimension-$6$ operators necessary for closing the loop in Figure \ref{fig:dia-mass}. The full scalar potential, including operators up to dimension-$6$, can be found in Appendix \ref{app:a}.

The scalar fields of the model can be expanded as:
\begin{equation}
\begin{aligned}
\phi = & \ \frac{1}{\sqrt{2}}\begin{pmatrix}
\sqrt{2} \ \phi^+\\ v_\phi+\phi^0_R + i\phi^0_I
\end{pmatrix},  \quad 
\Sigma=\frac{1}{\sqrt{2}}\left(\begin{matrix}
 v_{\Sigma}+ \Sigma^0_{\textbf{1}R}+ i \Sigma^0_{\textbf{1}I} & \sqrt{2}\  \Sigma_\textbf{2}^+\\
\sqrt{2} \ \Sigma_\textbf{1}^- & v_{\Sigma}+ \Sigma^0_{\textbf{2}R}+i \Sigma^0_{\textbf{2}I}
\end{matrix}\right),\\
h_k = & \ \frac{1}{\sqrt{2}} \begin{pmatrix}
\sqrt{2} \ h_k^+\\  h^0_{kR}+ i h^0_{kI}
\end{pmatrix}, \quad \sigma= \ v_\sigma+\tilde\sigma, \hspace{1cm} k=1,2.
\end{aligned}
\label{eq:parametrization}
\end{equation}
where $v_\phi$ is the VEV responsible for the EWSB.

\subsection{Gauge sector}
The masses of the electroweak gauge bosons arise from the gauge part of the Lagrangian in \eq{eq:lagrangian} after the high-energy symmetry breaking. Using the field parametrization of the $\Sigma$ field in \eq{eq:parametrization}, the Lagrangian takes the form:
\begin{equation}
\begin{aligned}
     \mathscr{L}_{\text{gauge}}=&-\frac{1}{4}\text{Tr}\left[F^{(1)}_{\mu\nu} F^{(1)\mu\nu}\right]-\frac{1}{4}\text{Tr}\left[F^{(2)}_{\mu\nu} F^{(2)\mu\nu}\right] - \dfrac{1}{4} B_{\mu\nu} B^{\mu\nu}+f_{\Sigma}^2v_{\Sigma}^2\left(\phi^{\dagger} \phi\right)\left(- i g_1 A_{\mu}^{(1)}+ i g_2A_{\mu}^{(2)}\right)^2\\
     &+\left(D^{\mu}\phi\right)^{\dagger}\left(D_{\mu}\phi\right)+ \left(D^{\mu}h_{1}\right)^{\dagger}\left(D_{\mu}h_{1}\right) +\left(D^{\mu}h_{2}\right)^{\dagger}\left(D_{\mu}h_{2}\right) +\beta^2v_{\Sigma}^2\left(\phi^{\dagger} \phi\right)\left(- i g_1 A_{\mu}^{(1)}+ i g_2A_{\mu}^{(2)}\right)^2,
\end{aligned}
\end{equation}
on the other hand, the EWSB will occur when $\phi$ acquires its VEV. Gauge bosons will be parametrized as usual:
\begin{align}
    A^{(1/2)}_\pm=\frac{1}{\sqrt{2}}\left(A^{(1/2)}_1\mp A^{(1/2)}_2\right).
\end{align}

After the SSB processes, the squared neutral mass matrix for the gauge fields in the $\left(B_\mu,A^{(1)}_3,A^{(2)}_3\right)$ basis is:
\begin{align}
M^2_N=\frac{ v_\phi^2}{4}\left(
\begin{array}{ccc}
 g_y^2 & -g_1 g_y & 0 \\
 -g_1 g_y & (1+c^2)g_1^2 & -c^2g_1g_2 \\
 0 & -c^2g_1g_2 & c^2g_2^2 \\
\end{array}
\right),
\end{align}
and the charged matrix in the $\left(A^{(1)}_\pm,A^{(2)}_\pm \right)$ basis:
\begin{align}
M_{CH}^2=\frac{ v_\phi^2}{4}\left(
\begin{array}{cc}
 \left(1+c^2\right) g_1^2 & -c^2 g_1 g_2 \\
 -c^2 g_1 g_2 & c^2 g_2^2 \\
\end{array}
\right).
\end{align}

The parameters $a$, $b$, and $c$ are defined as:
\begin{align}
    &a=f_\Sigma\frac{v_\Sigma}{v_\phi},\quad b=\beta\frac{v_\Sigma}{\sqrt{2}}\quad\text{and}\quad c^2=a^2+b^2.
\end{align}

The physical eigenstates can be defined by \cite{CarcamoHernandez:2017pei}:
\begin{align}
    \left(\begin{array}{c}
        A_\mu\\
        Z_\mu\\
        \rho^0_{\mu}
    \end{array}\right)=\left(\begin{array}{ccc}
        \cos \theta_W & \sin \theta_W & 0 \\
        -\cos \theta_N \sin \theta_W & \cos\theta_N \cos \theta_W & -\sin \theta_N \\
        -\sin \theta_N \sin \theta_W & \cos \theta_W \sin \theta_N & \cos \theta_N
    \end{array}\right)\left(\begin{array}{c}
        B_\mu\\
        A^{(1)}_{\mu,3}\\
        A^{(2)}_{\mu,3}
    \end{array}\right)
\end{align}
where $\theta_W$ is the electroweak mixing angle, $\tan\theta_W=g_y/g_1$, and the mixing angle from the strong sector $\theta_N$ is given by:
\begin{align}
    \tan2\theta_N=-\frac{2c^2g_1^2g_2}{(1+c^2)g_1^3+g_1g_y^2-c^2g_2^2\sqrt{g_1^2+g_y^2}},
\end{align}
and
\begin{align}
    \left(\begin{array}{c} W_\mu^{ \pm} \\ \rho_\mu^{ \pm}\end{array}\right)=\left(\begin{array}{cc}
    \cos\theta_{CH} & -\sin\theta_{CH}\\
    \sin\theta_{CH} & \cos\theta_{CH}
    \end{array}\right)\left(\begin{array}{c} A_{\mu,\pm}^{(1)} \\ A_{\mu,\pm}^{(2)}\end{array}\right).
\end{align}

The combination with extra $SU(2)_2$ gauge fields gives rise to new physical states in the form of two strong vector resonances $\rho^0_{\mu}$ and $\rho^\pm_{\mu}$. The mixing angle $\theta_{CH}$ is given by:
\begin{align}
    \tan{2\theta_{CH}}=-\frac{2c^2g_1g_2}{(1+c^2)g_1^2-c^2g_2^2}.
\end{align}

\begin{figure}[]
\centering
\subfigure[]{\includegraphics[width=0.49\linewidth]{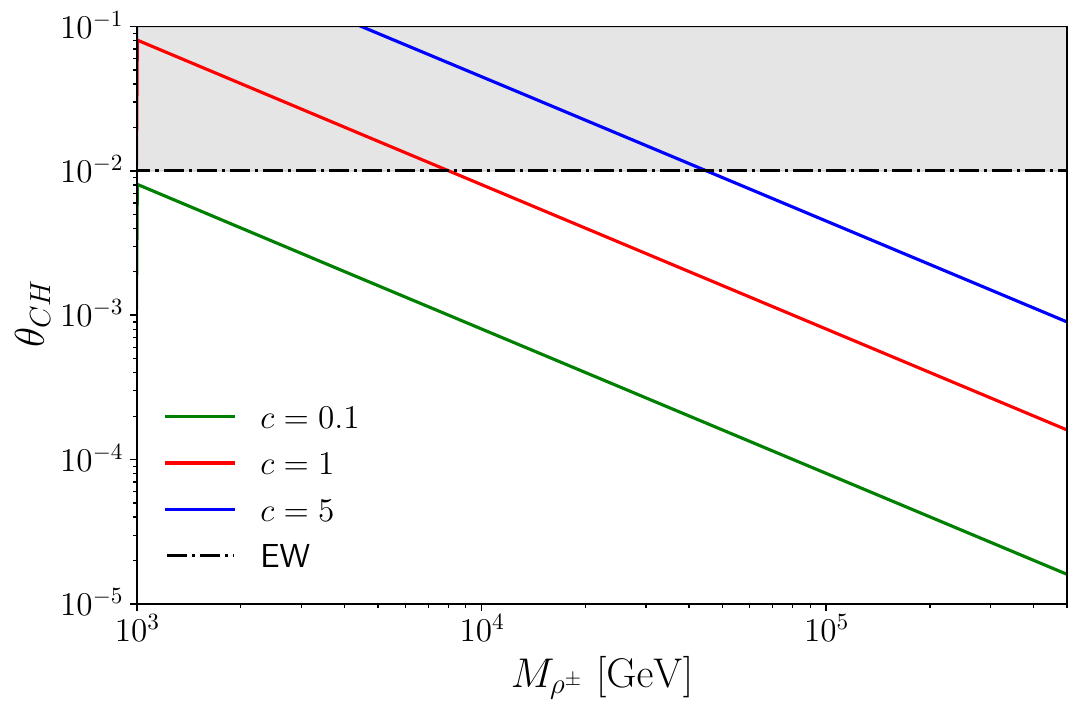}}\quad 
\subfigure[]{\includegraphics[width=0.49\linewidth]{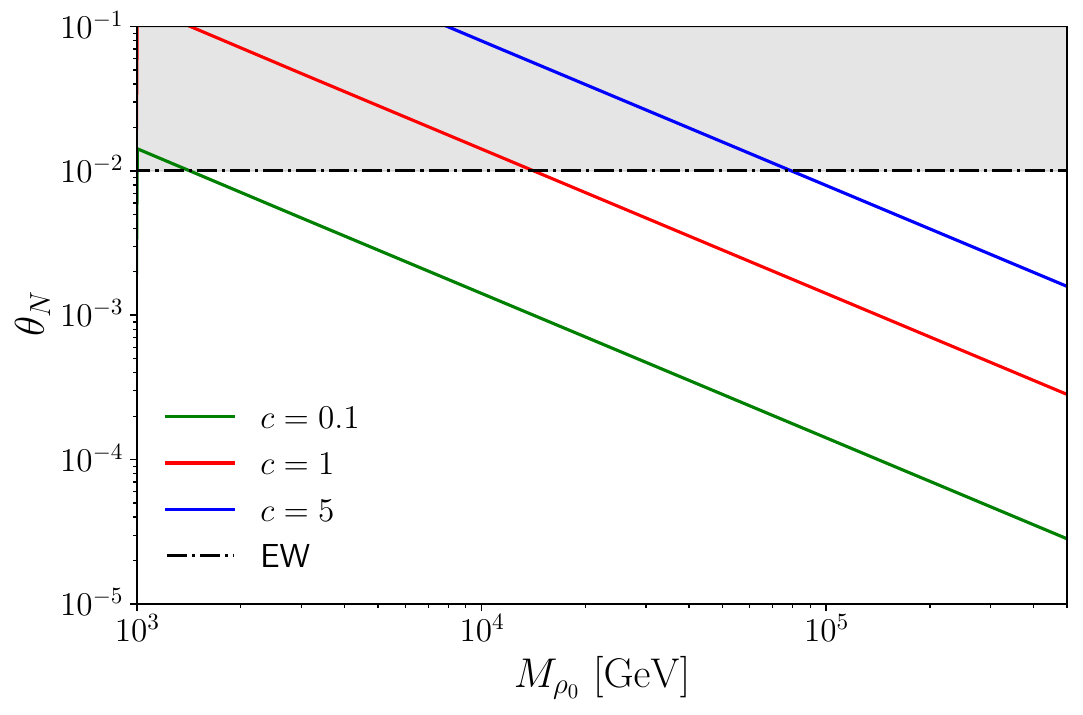}}\quad 
\caption{Panel (a): Behavior of the mixing angle $\theta_{CH}$ as a function of the charged vector resonance mass $M_{\rho^\pm}$ for different values of the $c$ parameter. The horizontal black dashed-dotted line indicates the exclusion region arising from the electroweak precision data and is denoted as ``EW'' \cite{Pankov:2021vzs}. Panel (b): Behavior of the mixing angle $\theta_{N}$ as a function of the neutral vector resonance mass $M_{\rho_0}$ for different values of the $c$ parameter. The horizontal black dashed-dotted line indicates the exclusion region resulting from the electroweak precision data (EW) \cite{Pankov:2021vzs}.}
\label{fig:gaugemixingangles}
\end{figure}

The mixing angles of exotic electroweak bosons have been strongly constrained in the literature, for instance in \cite{Osland:2020onj, Pankov:2021vzs}. Our numerical analysis shows that the mass of the vector resonances lies above the exclusion limits set by ATLAS and CMS Run 2 data \cite{Pankov:2021vzs}. In Fig.~\ref{fig:gaugemixingangles} the behavior of the mixing angles as a function of the masses of the exotic vector resonances is shown for different values of the $c$ parameter. Allowed values consistent with electroweak precision data are achieved for large vector resonance masses provided that the $c$  parameter is large. The predictions of this model for lighter mass resonances and their mixings with the Standard Model can be studied in detail in future works.

If $M_N^2$ and $M_{CH}^2$ are diagonalized in the limit $g_2 \gg g_1$ the following masses for the gauge bosons and strong vector resonances are obtained
\begin{equation}
\begin{aligned}
 M_A^2= & \ 0\quad \text{(exact)},\\
    M_Z^2= & \ \frac{v_\phi^2}{4}(g_1^2+g_y^2)+\mathcal{O}\left(
(g_1/g_2)^2
    \right),\\
    M_{\rho_0}^2= & \ \frac{v_\phi^2}{4}c^2g_2^2+\mathcal{O}\left((g_1/g_2)^2\right),\\
    M_{W^\pm}^2= & \ \frac{v_\phi^2}{4}g_1^2+\mathcal{O}\left((g_1/g_2)^2\right),\\
    M_{\rho^\pm}^2= & \ \frac{v_\phi^2}{4}c^2g_2^2+\mathcal{O}\left((g_1/g_2)^2\right).
    \label{eq:gbosonmasses}
\end{aligned}
\end{equation}

Full expressions for the masses are given in Appendix \ref{app:b2}.

\subsection{Scalar sector}

We are going to start applying the conditions that it has given rise to the SSB processes, we impose that the gradient of the scalar potential vanishes when the neutral components of $\phi$, $\Sigma$ and $\sigma$ acquire a $v_\phi$, $v_{\Sigma}$ and $v_\sigma$ VEV respectively. The dimension 6 operators are not considered because it is only relevant for high energies. The conditions impose the following restrictions on the $\mu-$ parameters:
\begin{align}
    \mu_\phi^2=&-\lambda_1v_\phi^2-(\lambda_5+\lambda_{25}+2\lambda_{35})v_\Sigma^2-\lambda_6v_\sigma^2,\\
    \nn\mu_{\Sigma}^2=&-\frac{1}{2}(\lambda_{5}+\lambda_{25}+2\lambda_{31})v_\phi^2-(2\lambda_{19}+\lambda_{20}+2\lambda_{23}+\lambda_{24}+4(2\lambda_{30}+\lambda_{36}+\lambda_{37}+\lambda_{38}))v_\Sigma^2\\
    &-(\lambda_{21}+\lambda_{29}+2\lambda_{35})v_\sigma^2-4\mu_{\Sigma'}^2,\\
    \mu_\sigma^2=&-\frac{1}{2}v_{\phi}^2\lambda_{6}-2\lambda_{22}v_{\sigma}^2-(\lambda_{21}+\lambda_{29}+2\lambda_{35})v_\Sigma^2.
    \label{eq:muconditions}
\end{align}
The next step is to ensure the tree-level stability of the potential, \textit{i.e.} to impose restrictions on the potential couplings as long as the potential is bounded from below. Using the following definitions for the bilinear combination of scalar fields:
\begin{equation}
\begin{aligned}
    &A=\phi^\dagger\phi,\quad B=h_1^\dagger h_1,\quad C=h_2^\dagger h_2,\quad D=\text{Tr}\left[\Sigma^\dagger\Sigma\right],\\
    E=\sigma^2,&\quad F=\text{Re}\left[h_1^\dagger h_2\right],\quad G=\text{Im}\left[h_1^\dagger h_2\right],\quad H^2=\text{Tr}\left[(\Sigma^\dagger\Sigma)^2\right].
\end{aligned}
\end{equation}
we can rewrite the quartic part of the potential as\footnote{The terms involving $\tilde\Sigma$ have been omitted since their behavior is identical to that of $\Sigma$.
}:
\begin{equation}
\begin{aligned}
V_4=& \ \left(\sqrt{\lambda_1}A-\sqrt{\lambda_7}B\right)^2+\left(\sqrt{\lambda_1}A-\sqrt{\lambda_{13}}C\right)^2+\left(\sqrt{\lambda_1}A-\sqrt{\lambda_{19}}D\right)^2+\left(\sqrt{\lambda_1}A-\sqrt{\lambda_{22}}E\right)^2\\
    &+\left(\sqrt{\lambda_7}B-\sqrt{\lambda_{13}}C\right)^2+\left(\sqrt{\lambda_7}B-\sqrt{\lambda_{19}}D\right)^2+\left(\sqrt{\lambda_7}B-\sqrt{\lambda_{22}}E\right)^2+\left(\sqrt{\lambda_{13}}C-\sqrt{\lambda_{19}}D\right)^2\\
    &+\left(\sqrt{\lambda_{13}}C-\sqrt{\lambda_{22}}E\right)^2+\left(\sqrt{\lambda_{19}}D-\sqrt{\lambda_{22}}E\right)^2+4(\lambda_8+2 \sqrt{\lambda_7} \sqrt{\lambda_{13}})F^2+\lambda_{22}H^2\\
& +2(\lambda_2+2 \sqrt{\lambda_1} \sqrt{\lambda_7})AB 
+2(\lambda_3+2 \sqrt{\lambda_1} \sqrt{\lambda_{13}})AC+2(\lambda_5+2 \sqrt{\lambda_1} \sqrt{\lambda_{19}})AD+2(\lambda_6+2 \sqrt{\lambda_1} \sqrt{\lambda_{22}})AE\\
& +2(\lambda_8+2(\lambda_{11}+2 \sqrt{\lambda_7} \sqrt{\lambda_{19}})BD +2 \sqrt{\lambda_7} \sqrt{\lambda_{13}})(BC-F^2-G^2)+2(\lambda_{12}+2 \sqrt{\lambda_7} \sqrt{\lambda_{22}})BE\\    
& +2(\lambda_{15}+2 \sqrt{\lambda_{13}} \sqrt{\lambda_{19}})CD+2(\lambda_{16}+2 \sqrt{\lambda_{13}} \sqrt{\lambda_{22}})CE
+2(\lambda_{21}+2 \sqrt{\lambda_{19}} \sqrt{\lambda_{22}})DE-2\text{Im}(\lambda_9)FG\\
 & +(\text{Re}(\lambda_9)-2\lambda_8-4 \sqrt{\lambda_7} \sqrt{\lambda_{13}})(F^2-G^2)+2\lambda_4AF+\lambda_{10}BF+\lambda_{14}CF+\lambda_{17}DF+\lambda_{18}EF,
\end{aligned}
\end{equation}

Using the method presented in \cite{Bhattacharyya:2015nca} we obtain the following restrictions for the couplings of the scalar potential:
\begin{equation}
\begin{aligned}
    &\qquad\qquad\lambda_{1},\lambda_{4},\lambda_{7},\lambda_{10},\lambda_{13},\lambda_{14},\lambda_{17},\lambda_{18},\lambda_{19}, \lambda_{20},\lambda_{22},\lambda_{23},\lambda_{24},\lambda_{28},\lambda_{30},\lambda_{34},\lambda_{38}\geq 0,\\
    &\lambda_2+2\sqrt{\lambda_1}\sqrt{\lambda_7}\geq 0,\quad \lambda_{3}+2\sqrt{\lambda_{1}}\sqrt{\lambda_{13}}\geq 0,\quad \lambda_{5}+2\sqrt{\lambda_{1}}\sqrt{\lambda_{19}}\geq0,\quad \lambda_{25}+2\sqrt{\lambda_{1}}\sqrt{\lambda_{23}}\geq0,\\
    &\lambda_{6}+2\sqrt{\lambda_{1}}\sqrt{\lambda_{22}}\geq 0,\quad\lambda_{8}+2\sqrt{\lambda_{13}}\sqrt{\lambda_{7}}\geq0,\quad  \lambda_{11}+2\sqrt{\lambda_{19}}\sqrt{\lambda_{7}}\geq0,\quad  \lambda_{26}+2\sqrt{\lambda_{23}}\sqrt{\lambda_{7}}\geq0,\\
    &\lambda_{12}+2\sqrt{\lambda_{22}}\sqrt{\lambda_{7}}\geq0,\quad \lambda_{15}+2\sqrt{\lambda_{13}}\sqrt{\lambda_{19}}\geq0,\quad \lambda_{27}+2\sqrt{\lambda_{13}}\sqrt{\lambda_{23}}\geq0,\quad \lambda_{25}+2\sqrt{\lambda_{1}}\sqrt{\lambda_{23}}\geq0,\\
    &\lambda_{31}+2\sqrt{\lambda_{1}}\sqrt{\lambda_{30}}\geq0,\quad \lambda_{32}+2\sqrt{\lambda_{7}}\sqrt{\lambda_{30}}\geq0,\quad \lambda_{33}+2\sqrt{\lambda_{8}}\sqrt{\lambda_{30}}\geq0\quad \lambda_{16}+2\sqrt{\lambda_{13}}\sqrt{\lambda_{22}}\geq0,\\
    &\lambda_{21}+2\sqrt{\lambda_{19}}\sqrt{\lambda_{22}}\geq0,\quad |\lambda_9|<2\lambda_8+4 \sqrt{\lambda_7} \sqrt{\lambda_{13}}.
    \label{eq:stability}
\end{aligned}
\end{equation}
Applying the conditions of \eq{eq:muconditions} we can obtain the mass matrices with respect to the components of the scalar fields. The sector formed by the components of the inert scalar fields, or dark sector, is separated from the sector formed by the active fields, or visible sector.

The squared mass matrices for the neutral CP-even visible scalar fields that transform trivially under the preserved $\mathbb{Z}_2\times \mathbb{Z}_2'$ symmetry in the basis $\left(\phi_R^0,\Sigma_{1R}^0,\Sigma_{2R}^0,\widetilde{\sigma}\right)$ are given by:
\begin{align}
M_{\text{CP-even}}^2=
\left(\begin{array}{cccc}
v_{\phi}^2 \lambda_1 &
\lambda_{B} &
\lambda_{B} &
v_{\sigma} v_{\phi} \lambda_6 \\
\lambda_{B} &
\lambda_{A} &
\lambda_{C} &
\lambda_{D} \\
\lambda_{B} &
\lambda_{C} &
\lambda_{A} &
\lambda_{D} \\
v_{\sigma} v_{\phi} \lambda_6 &
\lambda_{D} &
\lambda_{D} &
4 v_{\sigma}^2 \lambda_{22}
\end{array}\right),
\end{align}
where:
\begin{equation}
\begin{aligned}
    &\lambda_{A}=v_{\Sigma}^2 (\lambda_{19} + \lambda_{20} + \lambda_{23} + \lambda_{24} + 4 \lambda_{30} + 2 \lambda_{36} + 2 \lambda_{37} + 4 \lambda_{38})\\
    &\lambda_{B}=\frac{1}{2} v_{\Sigma} v_{\phi} (\lambda_{25} + 2 \lambda_{31} + \lambda_5)\\
    &\lambda_{C}=v_{\Sigma}^2 (\lambda_{19} + \lambda_{23} + 2 (2 \lambda_{30} + \lambda_{36} + \lambda_{37}))\\
    &\lambda_{D}=v_{\sigma} v_{\Sigma} (\lambda_{21} + \lambda_{29} + 2 \lambda_{35}),
\end{aligned}
\end{equation}
and which has four distinct eigenvalues. The lightest eigenstate is identified as the $126$ GeV SM-like Higgs boson, $h$, while the other three massive states correspond to physical fields which we name $\Sigma_1$, $\Sigma_2$, and $H_3^0$. Symbolic expression for the masses are very extensive to be included in this section, so we will simply work with their numerical values.

The squared mass matrix for the neutral CP-odd visible scalar fields in the basis $\left(\phi_I^0,\Sigma_{1I}^0,\Sigma_{2I}^0\right)$ is given by:
\begin{align}
M_{\text{CP-odd}}^2=
\left(\begin{array}{ccc}
0 & 0 & 0 \\
0 & 
\lambda_{E}
 & 0 \\
0 & 0 &
\lambda_{E}
\end{array}\right),
\end{align}
where 
\begin{align}
    \lambda_{E}=-v_{\phi}^2 \lambda_{31}- 2 v_{\Sigma}^2 (4 \lambda_{30} + \lambda_{36} + \lambda_{37}+2\lambda_{38})  - 2 (v_{\sigma}^2 \lambda_{35} + \mu_{\Sigma'}^2).
\end{align}
The massless state correspond to the Goldstone bosons associated with the breaking of gauge symmetry. One of these corresponds to the $G_0$ of the SM, which is the longitudinal component of the SM $Z$-boson. 
The non-zero eigenvalues corresponds to a CP-odd scalars $\xi_{1,2}$.

The mass matrix for the charged visible scalar fields in the basis $\left(\phi^\pm,\Sigma_{1}^\pm,\Sigma_{2}^\pm\right)$ is given by:
\begin{align}
M_{\text{charged}}^2 = \begin{pmatrix}
0 & 0 & 0 \\
0 & \lambda_{E} & -\lambda_{E} \\
0 & -\lambda_{E} & \lambda_{E}
\end{pmatrix},
\label{chmass}
\end{align}
which contains two Nambu–Goldstone bosons, $G_\pm$, that provide the longitudinal polarizations of the $W_\pm$ bosons, and one massive charged state, $\eta_2^\pm$.

\begin{figure}[]
    \centering
    \subfigure[]{
        \begin{minipage}{\textwidth}
            \centering
            \includegraphics[width=0.2\linewidth]{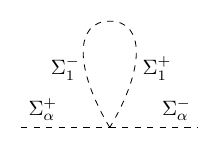}
            \includegraphics[width=0.2\linewidth]{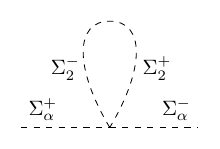}
            \includegraphics[width=0.2\linewidth]{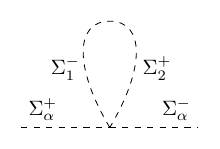}
            \includegraphics[width=0.2\linewidth]{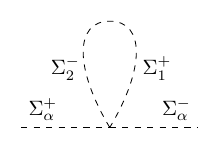}
        \end{minipage}
    }
    \subfigure[]{
        \begin{minipage}{\textwidth}
            \centering
            \includegraphics[width=0.2\linewidth]{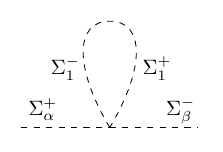}
            \includegraphics[width=0.2\linewidth]{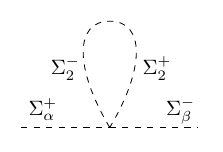}
            \includegraphics[width=0.2\linewidth]{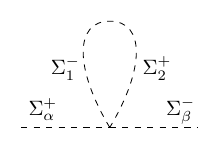}
            \includegraphics[width=0.2\linewidth]{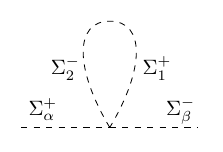}
        \end{minipage}
    }
    \caption{One-loop Feynman diagrams that contribute to the mass of the visible charged scalars. (a) Diagrams contributing to the diagonal entries of the mass matrix. Here $\alpha=1,2$. (b) Diagrams contributing to the off-diagonal entries of the mass matrix. Here $\alpha,\beta=1,2$ and $\alpha\neq\beta$.}
    \label{fig:chargedscalar}
\end{figure}

In addition, a massless charged scalar state arises due to the 
symmetry of the matrix of \eq{chmass}. This symmetry, however, is broken by quantum corrections, leading to a nonzero mass for $\eta_1^\pm$ at the one-loop level. The one-loop corrected mass matrix is given by:
\begin{align}
{M_{\text{charged}}^{(1-\text{loop})}}^2 =
\begin{pmatrix}
0 & 0 & 0 \\
0 & \lambda_E + \Delta(\mu) & -\lambda_E + \Delta'(\mu) \\
0 & -\lambda_E + \Delta'(\mu) & \lambda_E + \Delta(\mu)
\end{pmatrix},
\end{align}
where $\Delta(\mu)$ and $\Delta'(\mu)$ are the radiative corrections arising from the Feynman diagrams of Figure~\ref{fig:chargedscalar}, evaluated in the $\overline{\text{MS}}$ scheme:
\begin{equation}
\begin{aligned}
\Delta(\mu) &= \frac{\lambda_E}{16\pi^2}\left[(\gamma_1+\gamma_2)-(\gamma_1+\gamma_2-2\gamma_3)\ln{\frac{\lambda_E}{\mu^2}}\right], \\
\Delta'(\mu) &= \frac{\lambda_E}{16\pi^2}\left[2\gamma_3-(2\gamma_3-\gamma_2-\gamma_4)\ln{\frac{\lambda_E}{\mu^2}}\right],
\end{aligned}
\end{equation}
with the coefficients defined as
\begin{equation}
\begin{aligned}
\gamma_1 &= 4(\lambda_{19}+\lambda_{20}+\lambda_{23}+\lambda_{34}), \\
\gamma_2 &= 2(\lambda_{19}+\lambda_{23}+4\lambda_{30}+2\lambda_{38}), \\
\gamma_3 &= 4(\lambda_{36}+\lambda_{37}), \\
\gamma_4 &= 8(2\lambda_{30}+\lambda_{38}),
\end{aligned}
\end{equation}
where all quartic couplings originate from the Feynman diagrams shown in Figure~\ref{fig:chargedscalar}. These corrections visibly break the symmetry of the tree-level mass matrix, as the contributions to the off-diagonal elements differ from those ones associated to the diagonal ones.

\begin{figure}
    \centering
    \subfigure[]{\includegraphics[width=0.49\linewidth]{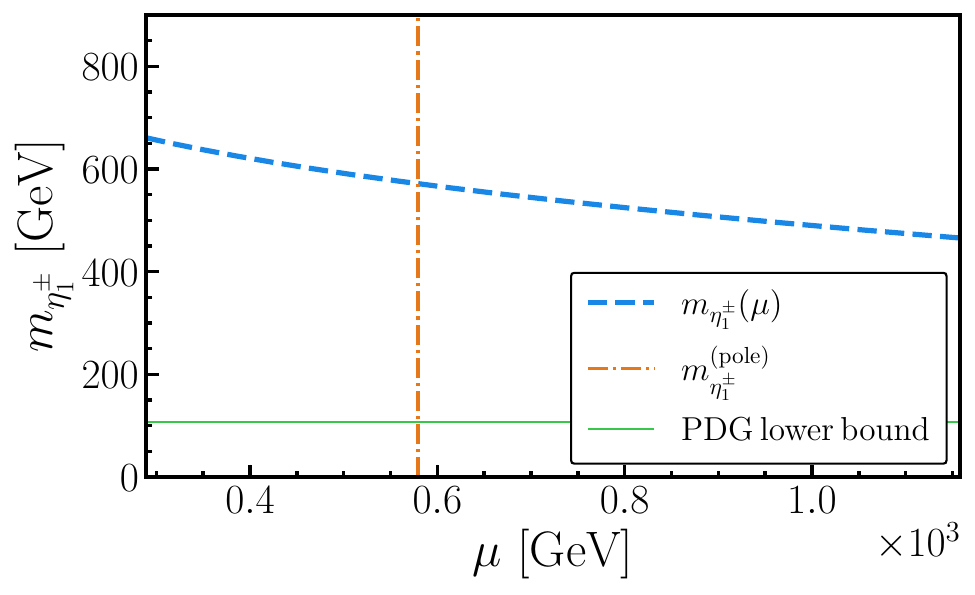}}
    \subfigure[]{\includegraphics[width=0.49\linewidth]{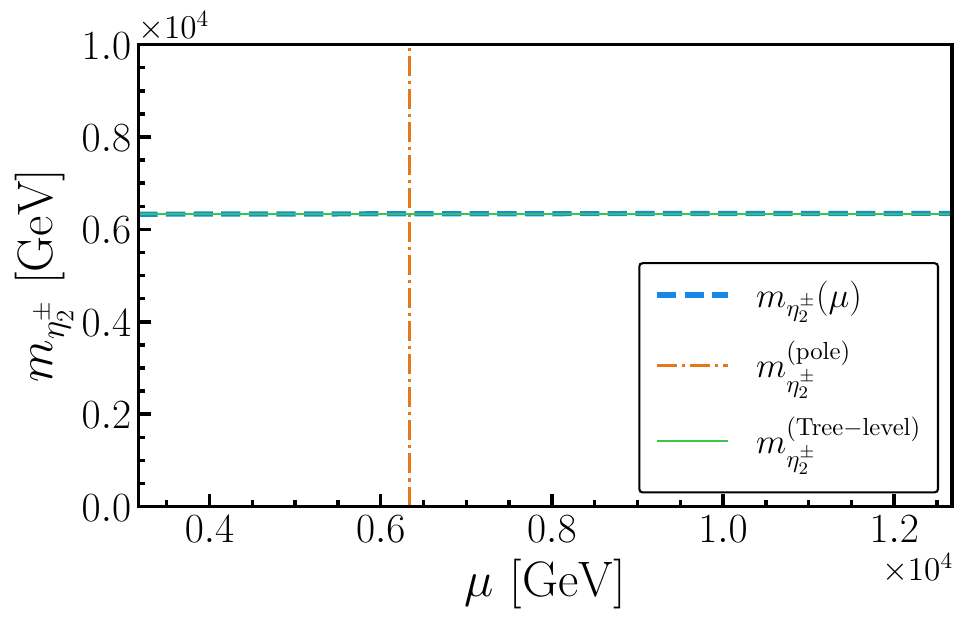}}
    \caption{The one-loop running of the charged scalar masses around the pole mass is shown. Panel (a) displays the running of $m_{\eta_1^\pm}$ in the range $\left[0.5m_{\eta_1^\pm}^{(\mathrm{pole})}, 2m_{\eta_1^\pm}^{(\mathrm{pole})}\right]$ and the lower bound for collider searches \cite{Workman:2022ynf}. Panel (b) presents the running of $m_{\eta_2^\pm}$ in the range $\left[0.5m_{\eta_2^\pm}^{(\mathrm{pole})}, 2m_{\eta_2^\pm}^{(\mathrm{pole})}\right]$, compared to its tree-level value.}
    \label{fig:running_masses}
\end{figure}

Using the coupling values from our benchmark point, we extract the numerical values of the charged scalar masses. As shown in Figure~\ref{fig:running_masses}, the one loop induced 
mass of $\eta_1^\pm$ is sufficiently large to lie beyond the current collider sensitivity for charged scalars~\cite{Workman:2022ynf}. In contrast, the quantum corrections to $\eta_2^\pm$ are small, resulting in a mass that remains nearly identical to its tree-level value.

For the dark sector, it is unnecessary to analyze particles individually because the mass matrices are degenerate and can be expressed as:
\begin{align}
\widetilde{M}^2=\frac{1}{4}\left(\begin{array}{cc}
\Delta_A & \Delta_B \\
\Delta_B & \Delta_C
\end{array}\right),
\end{align}
where the parameters are defined as:
\begin{align}
    &\Delta_A=v_{\sigma}^2 \lambda_{12} + \frac{1}{2} v_{\phi}^2 \lambda_2 + v_{\Sigma}^2 (\lambda_{11} + \lambda_{26} + 2 \lambda_{32}) + \mu_{h_1}^2,\\
    &\Delta_B=v_{\sigma}^2 \lambda_{18} + v_{\Sigma}^2 (\lambda_{17} + \lambda_{28} + 2 \lambda_{34}) + \frac{1}{2} v_{\phi}^2 \lambda_4 + \mu_{h_{12}}^2,\\
    &\Delta_C=v_{\sigma}^2 \lambda_{16} + \frac{1}{2} v_{\phi}^2 \lambda_3 + v_{\Sigma}^2 (\lambda_{15} + \lambda_{27} + 2 \lambda_{33}) + \mu_{h_2}^2.
\end{align}

This matrix has two non-zero eigenvalues, indicating the existence of two CP-even, CP-odd, and charged massive particles whose mass pairs are degenerate. Denoting the CP-even, CP-odd, and charged scalars by $H_{1,2}^0$, $\chi_{1,2}$, and $H_{1,2}^\pm$, their masses are:
\begin{align}
    m_{H^0_1}^2=m_{\chi_1}^2=m_{H_1^\pm}^2=&\frac{1}{8}\left(\Delta_A-\Delta_C+\sqrt{4\Delta_B^2+(\Delta_A-\Delta_C)^2}\right),\\
    m_{H^0_2}^2=m_{\chi_2}^2=m_{H_2^\pm}^2=&\frac{1}{8}\left(\Delta_A-\Delta_C-\sqrt{4\Delta_B^2+(\Delta_A-\Delta_C)^2}\right).
\end{align}

The degeneracy in the masses of the dark scalar sector is significant, as this is a scotogenic model \cite{Tao:1996vb,Ma:2006km}. The neutrino masses are related to the mass difference between the CP-even and CP-odd components of the inert doublets, as shown in \eq{eq::neutrinomassmatrix1} and \eq{eq::neutrinomassmatrix2}. While this initially suggests massless neutrinos, the contribution of dimension-6 operators in the scalar potential breaks this degeneracy, making the mass differences proportional to $\Lambda^{-1}$ and explaining the smallness of neutrino masses. For example:
\begin{align}
    m_{H_k^0}^2-m_{\chi_k}^2=\frac{v_{\Sigma}^2 v_\phi^2}{4 \Lambda^2}\left( \alpha_3+\alpha_4\right).
\end{align}

Hereafter, it is assumed that there exist values of the couplings capable of reproducing the observed neutrino mass spectrum.

\section{\label{sec::4}Standard model fermion mass hierarchy}

In this section, we demonstrate that the parameters of the 
fermionic sector, combined with those fitted in the scalar sector, successfully reproduce the mass hierarchy of the SM fermions.

\subsection{Quark sector}

From the interactions in Eq.~\eqref{eq:fermionlagrangian}, the mass matrices for the quark sector, $M_u$ and $M_d$, are obtained in the $(u_{L1},u_{L2},u_{L3},T_{L1},T_{L2})-(u_{R1},u_{R2},u_{R3},T_{R1},T_{R2})$ and $(d_{L1},d_{L2},d_{L3},B_{L1},B_{L2},B_{L3})-(d_{R1},d_{R2},d_{R3},B_{R1},B_{R2},B_{R3})$ bases, respectively, and are expressed as:
\begin{eqnarray}
    M_U &=&  \begin{pmatrix}
    T_{u} & A_{u} \\
    B_{u} & C_{u} \\
    \end{pmatrix}, \quad  M_D = \begin{pmatrix}
    0_{3\times 3} & A_{d}\\
    B_{d} & C_{d},\\
    \end{pmatrix}
\end{eqnarray}
with
\begin{eqnarray}
   T_{u}=\frac{v_\phi}{\sqrt{2}}\begin{pmatrix}
    0 & 0 & Y^1_{1}  \\
    0 & 0 & Y^1_{2} \\
    0 & 0 & Y^1_{3} \\
   \end{pmatrix},\quad && \karen{A_{u}}_{im}=\frac{v_\phi v_{\sigma}}{\sqrt{2}\Lambda} Y^2_{im},\quad \karen{A_{d}}_{ij}=\frac{v_\phi v_{\sigma}}{\sqrt{2}\Lambda} Y^5_{ij},\\
   && B_{u}=v_\sigma\begin{pmatrix}
    Y^3_{11} & Y^3_{12} & 0\\
    Y^3_{21} & Y^3_{22} & 0\\
\end{pmatrix},\quad \karen{B_{d}}_{ij}=v_\sigma Y^6_{ij},\\
    && \karen{C_{u}}_{nm}=v_\sigma  Y^4_{nm},\quad \karen{C_{d}}_{ij}=v_\sigma Y^7_{ij},
\end{eqnarray}
where $\Lambda$ is the energy scale associated with the masses of the new exotic fermions. While the up, charm, down, bottom, and strange quarks acquire their masses through the universal seesaw mechanism, the top quark obtains its mass via the SM tree-level mechanism. The effective mass matrices for the SM quarks can be simplified as:
\begin{equation}
\begin{aligned}
\widetilde{M}_{U} = & \ T_{u}-A_{u} C_{u}^{-1}B_{u},
\\
\widetilde{M}_{D} = & \ -A_{d} C_{d}^{-1}B_{d},
\end{aligned}
\end{equation}
The quark mass hierarchy in the SM is reproduced by carefully selecting the effective parameters and the energy scale. From the previous section, we infer that the VEVs have magnitudes $\mathcal{O}(v_\phi) \sim 10^{-1}$ TeV and $\mathcal{O}(v_\sigma)\sim 10$ TeV, yielding $v_\phi v_\sigma/\Lambda \sim 1$ TeV. Parameter scans indicate that for $\Lambda\sim 10^{2}$ TeV, the effective parameters remain within an order of magnitude of 
$\sim 0.1$ wich remains within the perturbative values. 

Using the same fitting strategy as in Eq.~\eqref{xisquare}, the effective parameters have been tuned to match the SM quark masses ($m_u,m_d,m_s,m_c,m_t,m_b$) \cite{Xing:2020ijf}, alongside the mixing angles $(\sin\theta_{12}^{(q)},\sin\theta_{13}^{(q)},\sin\theta_{23}^{(q)})$ and the Jarlskog invariant $J_q$ \cite{Workman:2022ynf}:
\begin{eqnarray}
    &m_u=(1.24 \pm 0.22)~\text{MeV},\quad m_d=(2.69\pm 0.19)~\text{MeV},\quad m_s =(53.5\pm 4.6)~\text{MeV},\nn\\
    &m_c=(0.63 \pm 0.02)~\text{GeV},\quad m_t=(172.9\pm 0.4)~\text{GeV},\quad m_b=(2.86\pm 0.03)~\text{GeV},\\
    &\sin\theta^{(q)}_{12}=0.2245\pm 0.00044,\quad \sin\theta^{(q)}_{23}=0.0421\pm 0.00076,\quad \sin\theta^{(q)}_{13}=0.00365\pm 0.00012,\nn\\
    &J_q=\left(3.18\pm 0.15\right)\times 10^{-5}.\nn
    \label{expquarks}
\end{eqnarray}

\begin{figure}[]
    \centering
    \subfigure[]{\includegraphics[scale=.5]{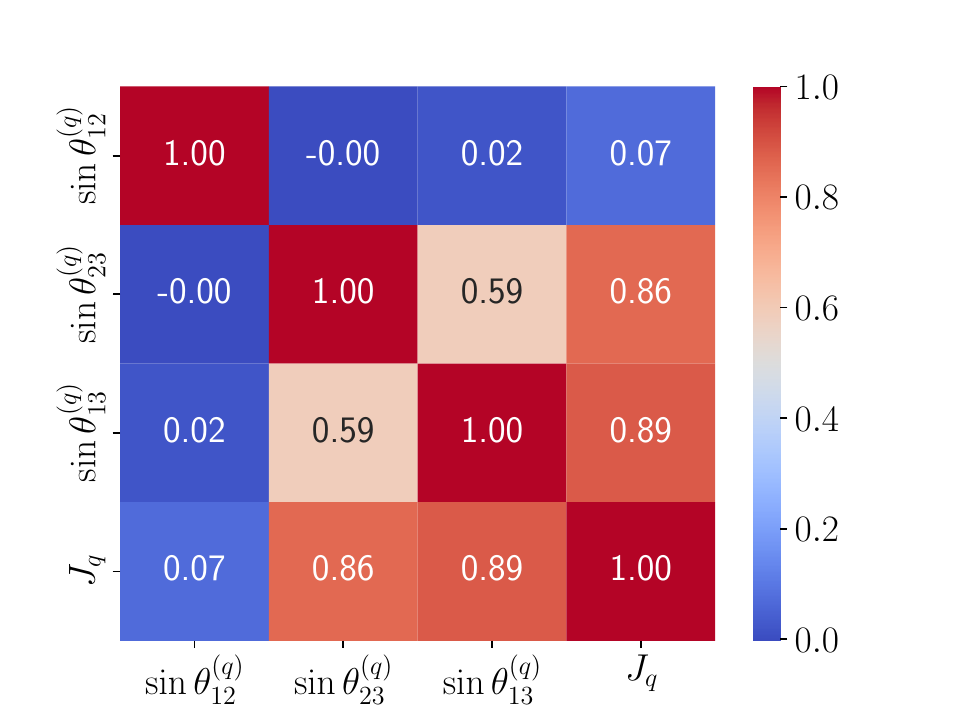}}
    \subfigure[]{\includegraphics[width=0.49\linewidth]{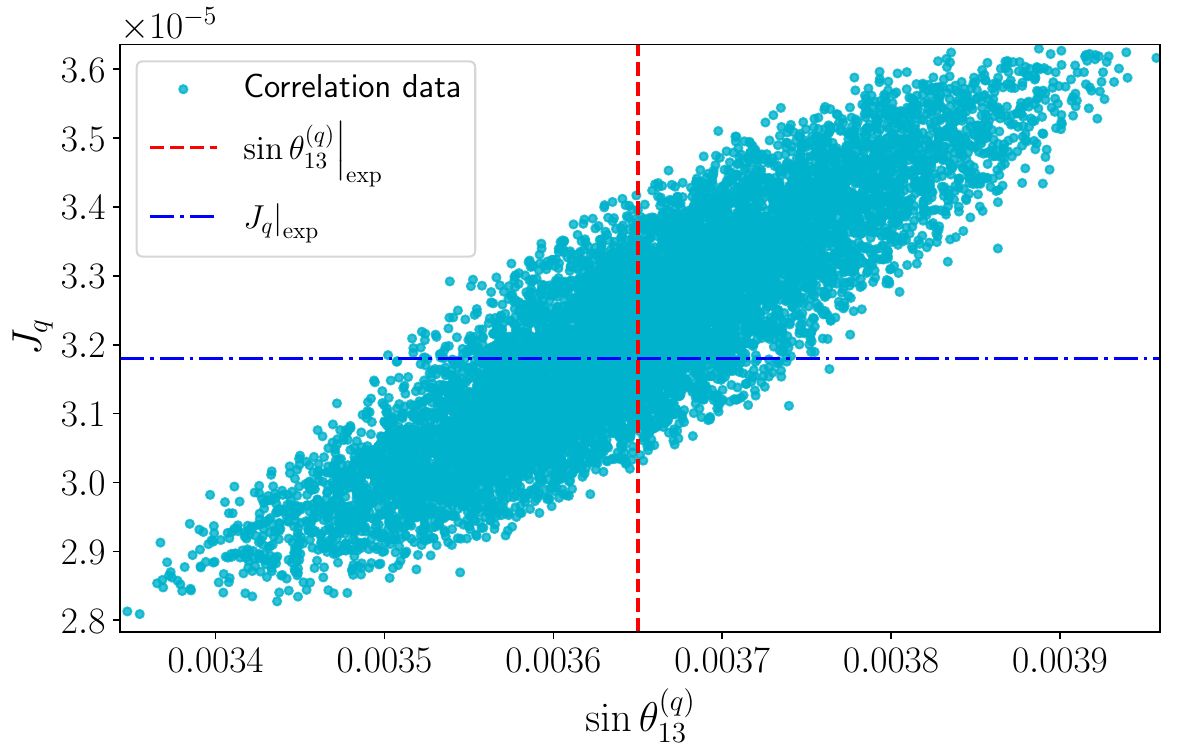}}
    \subfigure[]{\includegraphics[width=0.49\linewidth]{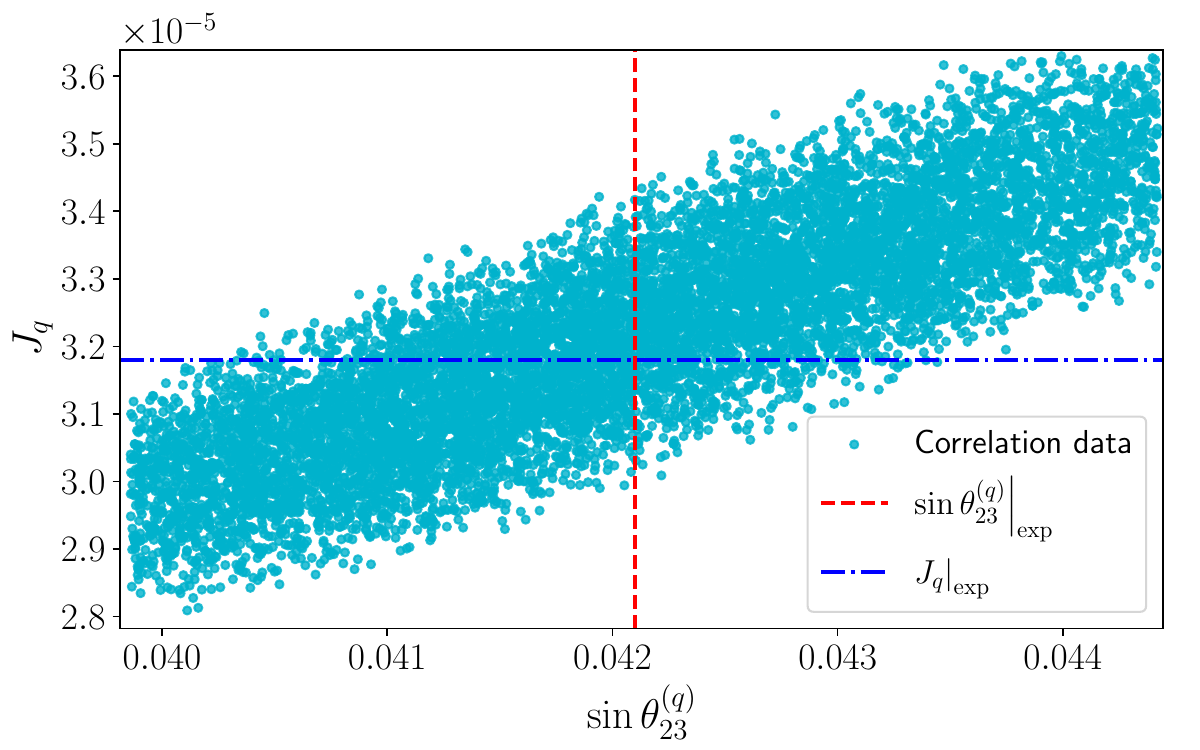}}
    \subfigure[]{\includegraphics[width=0.49\linewidth]{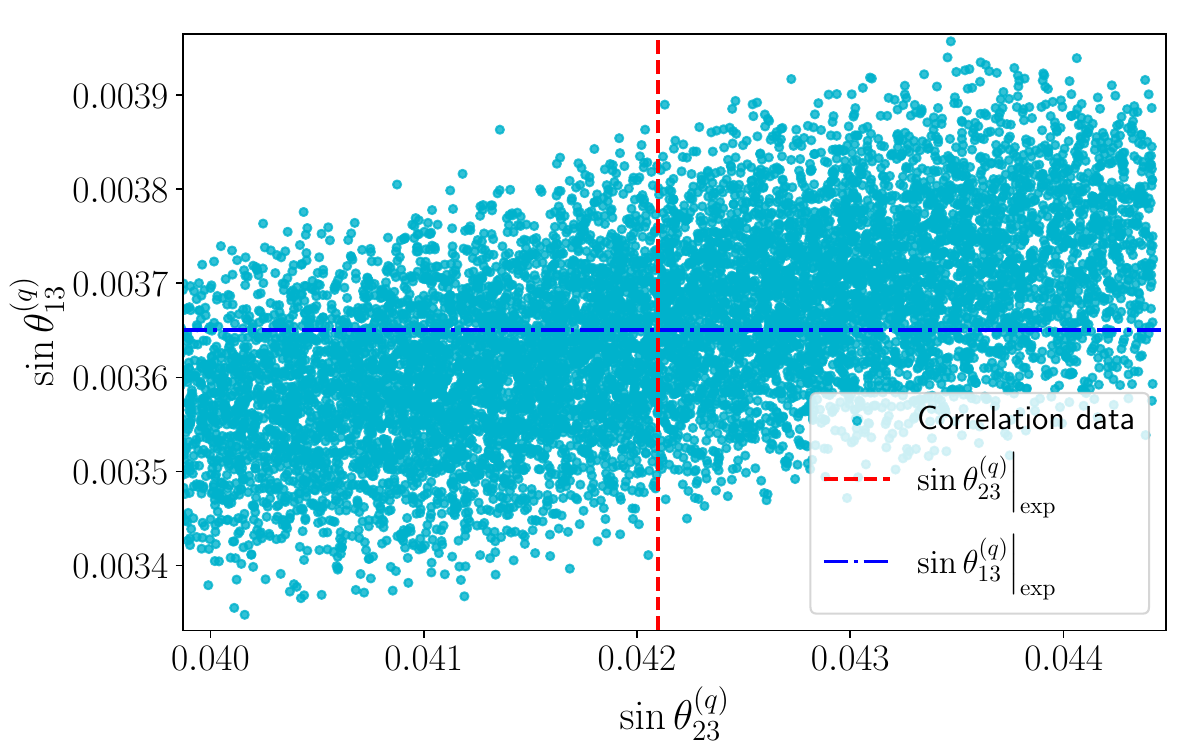}}
    \caption{(a) The correlation parameter matrix of the quark sector. The color scale corresponds to the degree of correlation between two parameters. (b, c, d) Correlation plot between the mixing angles of the quarks and the Jarlskog invariant obtained with our model. The dashed and dot-dashed lines correspond to the experimental values.}
    \label{fig:quarkcorrelation}
\end{figure}

The main correlations between these parameters are depicted in Figure \ref{fig:quarkcorrelation}. Panel (a) presents the correlation matrix for the quark sector parameters, where the heatmap quantifies the strength of correlation between parameter pairs. A strong correlation is observed between $\sin{\theta_{13}}$ and $\sin{\theta_{23}}$, as well as between these angles and the Jarlskog invariant $J_q$. In contrast, $\sin \theta_{12}$ shows a weak correlation with $J_q$, aligning with experimental data. Panels (b), (c), and (d) further illustrate these relationships through scatter plots, examining specific correlations between quark mixing angles and the Jarlskog invariant. The dashed and dot-dashed lines represent experimental values from Ref.~\cite{Workman:2022ynf}, enabling a visual comparison between the predictions of the model and experimental results. Within our model, values consistent with observations are achieved up to $3\sigma$.

\subsection{Charged lepton sector}

From the fermionic sector in \eq{eq:fermionlagrangian}, the mass matrix for the charged lepton sector is obtained. 

In the $(l_{L1},l_{L2},l_{L3},E_{L1},E_{L2},E_{L3})$ - $(l_{R1},l_{R2},l_{R3},E_{R1},E_{R2},E_{R3})$ basis, it is expressed as:
\begin{eqnarray}
M_\ell=\begin{pmatrix}
0_{3\times 3} & \frac{v_\phi}{\sqrt{2}}y^3_{ij}\\
v_\sigma y^1_{ij} & v_\sigma y^2_{ij}.
\end{pmatrix}
\end{eqnarray}

To simplify the analysis, the charged lepton mass matrix can be parametrized as:
\begin{equation}
    \widetilde{M}_{\ell} = A_{\ell} J^{-1}_{\ell} B_{\ell}^T,
\end{equation}
where:
\begin{equation}
    A_\ell =V_L^{(E)} M_e^{\frac{1}{2}} J_\ell^{\frac{1}{2}},\quad  B_\ell=V_R^{(E)} M_e^{\frac{1}{2}} J_\ell^{\frac{1}{2}}, 
\end{equation}
and the diagonal mass matrices are given by:
\begin{equation}
    M_e=\text{diag}(m_e,m_\mu,m_\tau),\quad  J_\ell=\text{diag}(m_{E_1},m_{E_2},m_{E_3}),
\end{equation}
with $V_L^{(E)}$ and $V_R^{(E)}$ representing the rotation matrices for the left- and right-handed charged leptons, respectively.

The mass hierarchy of the SM charged leptons is reproduced by exploring the parameter space, which also includes the masses of the exotic charged leptons. These masses have been set at an order of magnitude of approximately $100$~TeV to ensure compatibility with the experimental values for the charged leptons $(m_e,m_\mu,m_\tau)$ \cite{deSalas:2020pgw}:
\begin{eqnarray}
    &m_e=(0.4883266 \pm 0.0000017)~\mathrm{MeV},\quad m_\mu=(102.87267\pm 0.00021)~\mathrm{MeV},\quad m_\tau =(1747.43\pm 0.12)~\mathrm{MeV}.
    \label{expleptons}
\end{eqnarray}

The explicit values for the exotic charged lepton masses (in GeV) and the rotation matrices are as follows:
\begin{eqnarray}
    &m_{E_1}=20363.6,\quad m_{E_2}=61432.1,\quad m_{E_3}=14697.5,\nn\\
    &V_L^{(E)}=\begin{pmatrix}
    -0.359497 & 0.769444 & 0.527937 \\
    -0.528816 & -0.634131 & 0.564121 \\
    0.76884 & -0.0763817 & 0.634863
    \end{pmatrix},\quad 
    V_R^{(E)}=\begin{pmatrix}
    0.522082 & -0.752183 & 0.402058 \\
    -0.417024 & 0.186082 & 0.889643 \\
    0.74399 & 0.632134 & 0.216528
    \end{pmatrix}.
\end{eqnarray}

\subsection{Neutrino sector}

With respect to the neutrino sector, their interactions are designed to prohibit the generation of tree-level active neutrino masses, ensuring that the neutrino masses arise solely from one-loop quantum corrections. The form of the mass matrix is given by:
\begin{equation}
\left(M_\nu\right)_{ij}  = \sum_{k=2}^3 m_N \lambda^2 y_{i}^{(4,5)}y_{j}^{(4,5)} \mathcal{F}\left(
m_{H_k},m_{\chi_k}, m_N \right),
\label{eq::neutrinomassmatrix1}
\end{equation}
where the loop function is defined as:
\begin{equation}
\mathcal{F}\left(m_1, m_2, m_3\right)=  \frac{1}{16 \pi^2}\left[\frac{m_1^2}{m_1^2-m_3^2} \ln \left(\frac{m_1^2}{m_3^2}\right)-\frac{m_2^2}{m_2^2-m_3^2} \ln \left(\frac{m_2^2}{m_3^2}\right)\right].
\label{eq::neutrinomassmatrix2}
\end{equation}

The parameters of the SM neutrino sector, combined with the parameters of the PMNS matrix, are determined by exploring the values of 
the loop functions $\mathcal{F}\left(
m_{H_1},m_{\chi_1}, m_N \right)$ and $\mathcal{F}\left(
m_{H_2},m_{\chi_2}, m_N \right)$ 
and the Yukawa parameters. The reproduced observables include the experimental differences between neutrino masses $(\Delta m_{21},\Delta m_{31})$, the mixing angles $\left( \sin\theta_{12}^{(\ell)},\sin\theta_{13}^{(\ell)},\sin\theta_{23}^{(\ell)} \right)$, and the CP-violation phase $\delta_{\tx{CP}}^{(\ell)}$ \cite{deSalas:2020pgw}:
\begin{eqnarray}
    &\Delta m_{21}^2=7.50^{+0.22}_{-0.20}\times 10^{-5}~\tx{eV}^2,\quad \Delta m_{31}^2=2.55^{+0.02}_{-0.03}\times 10^{-3}~\tx{eV}^2,\nn\\
    &\sin^2\theta^{(\ell)}_{12}=0.318 \pm 0.016,\quad \sin^2\theta^{(\ell)}_{23}=0.574\pm 0.014,\quad \sin^2\theta^{(\ell)}_{13}=0.02200^{+0.00069}_{-0.00062},\nn\\
    &\delta^{(\ell)}_{\tx{CP}}=\left(194^{+24}_{-22}\right)^\circ.\nn
    \label{expleptons}
\end{eqnarray}

\begin{figure}[]
\centering
\subfigure[]{\includegraphics[scale=.5]{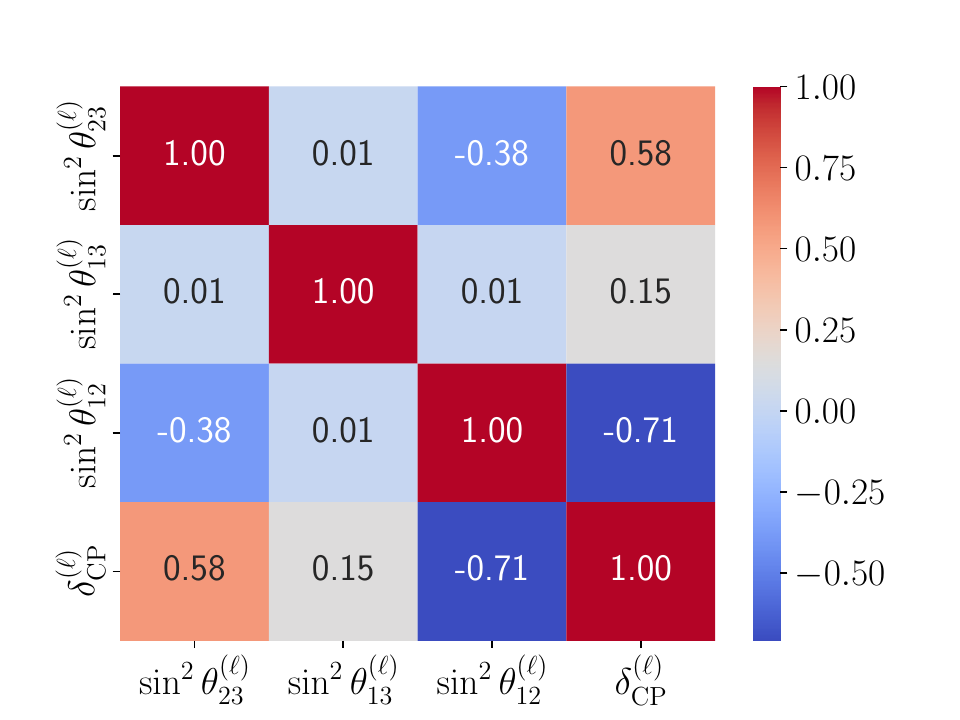}}
\subfigure[]{\includegraphics[scale=.38]{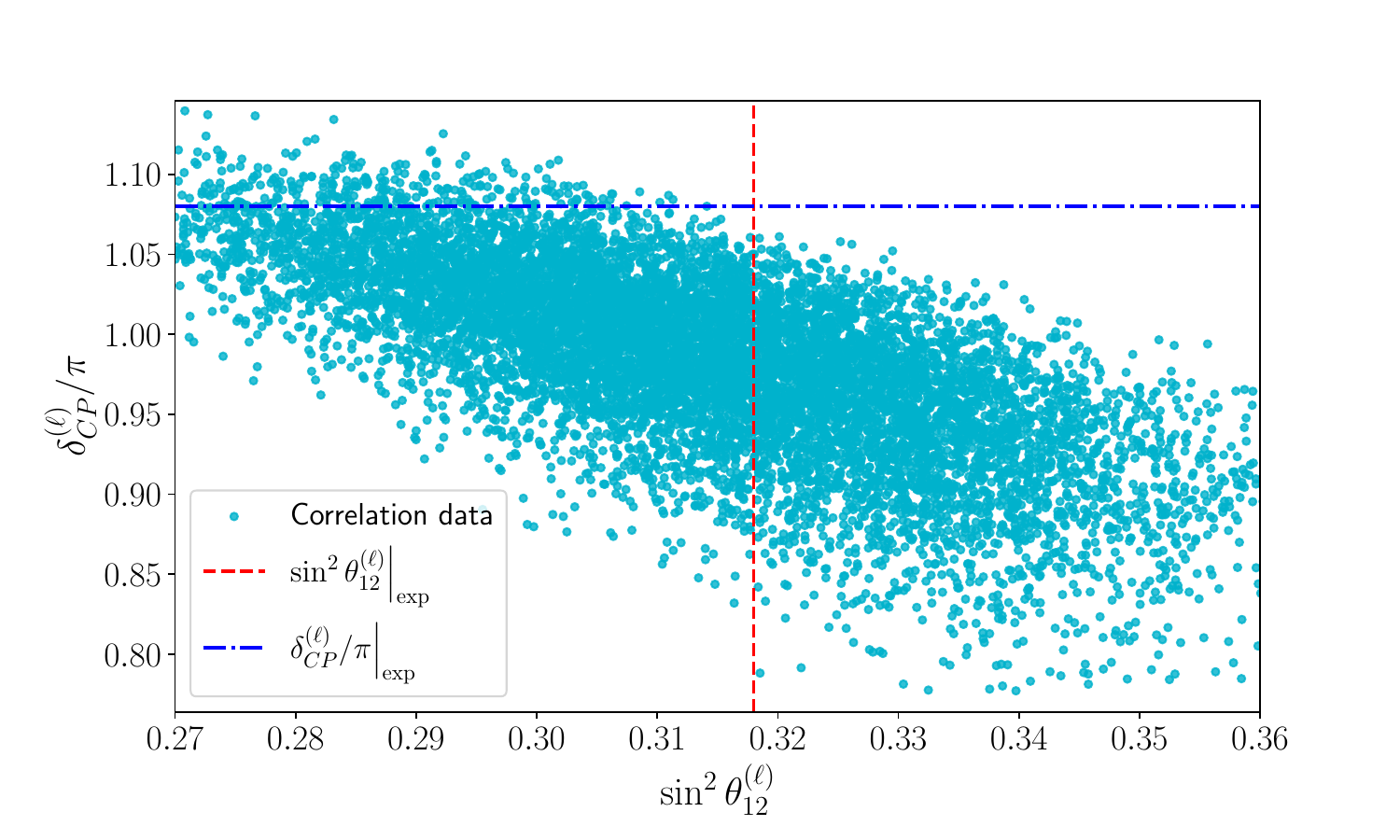}}
\subfigure[]{\includegraphics[width=0.5\linewidth]{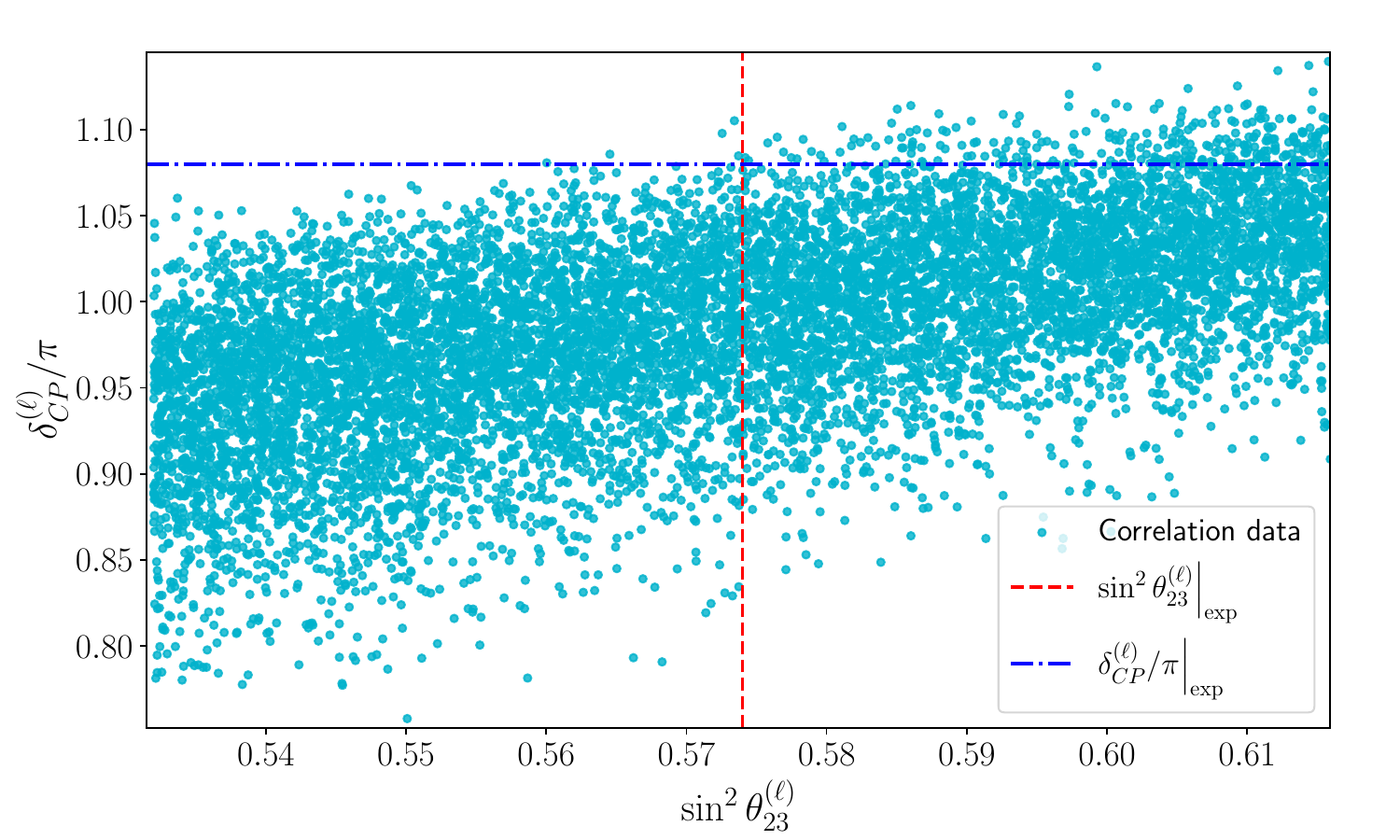}}
\caption{(a) The correlation parameter matrix of the lepton sector, where the color scale quantifies the degree of correlation between two parameters. 
(b, c) Scatter plots illustrating the correlations between the mixing angles in neutrino sector and $\delta_{\text{CP}}^{(\ell)}$
The dashed lines represent the experimental values, demonstrating the model's ability to replicate these observed correlations.}
 \label{fig:leptoncorrelation}
\end{figure}

Figure~\ref{fig:leptoncorrelation} presents the correlation analysis in the leptonic sector based on the predictions of the model. Panel (a) shows the correlation matrix of parameters, where the color scale indicates the strength of correlation between observable pairs. A strong negative correlation is observed between $\sin^2 \theta_{12}^{(\ell)}$ and the CP-violating phase $\delta_{\text{CP}}^{(\ell)}$, while $\sin^2 \theta_{13}^{(\ell)}$ and $\sin^2 \theta_{23}^{(\ell)}$ exhibit weaker correlations. Panels (b) and (c) illustrate scatter plots of these correlations, analyzing how $\delta_{\text{CP}}^{(\ell)}$ varies with $\sin^2 \theta_{12}^{(\ell)}$ and $\sin^2 \theta_{23}^{(\ell)}$, respectively. The dashed lines indicate experimental values, showing the model's ability to match observed data.

\section{Constraints from trilinear Higgs selfcoupling $H^3$}
\label{sec:trilinear}

The trilinear coupling of the Higgs boson, generally characterized by the parameter $\kappa_\lambda$, is a vital constraint for the scalar sector \cite{DiMicco:2019ngk,Abouabid:2024gms}:
\begin{align}
    \kappa_\lambda = \frac{\lambda_{h^3}^{\text{BSM}}}{\lambda_{h^3}^{\text{SM}}}.
\end{align}

This parameter determines the deviation of the Higgs coupling in a BSM model relative to that in the SM. The values of the Higgs self-couplings are crucial because they characterize the shape of the Higgs field potential, which in turn establishes the structure of the SSB mechanism of the SM. This mechanism governs the dynamics of mass generation for particles acquiring mass through this process.

The most precise experimental bounds on this parameter \cite{ATLAS:2022jtk} are:
\begin{align}
    &-0.4 \leq \kappa_\lambda \leq 6.3\label{eq:htri}.
\end{align}

Incorporating these bounds, a value for the parameter $\kappa_\lambda$ for the tree-level trilinear coupling 
\begin{align}
    \lambda_{h^3}^{\text{BSM}}=\left.\frac{\partial^3V_{\text{Phys}}}{\partial h^3}\right|_{\text{Fields}=0}   
\end{align}

has been found that lies within the experimental constraints described above as can be seen in the Table \ref{tab:numerical scalars}. This demonstrates that the proposed scalar sector, even with the strong dynamics associated with the new gauge symmetry, can reproduce a trilinear Higgs self-coupling consistent with current observational limits.

\begin{figure}[]
    \centering
        \begin{minipage}{\textwidth}
            \centering
            \includegraphics[width=0.25\linewidth]{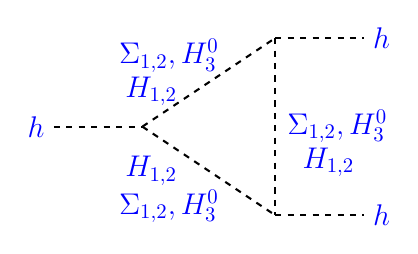}
            \hspace{-1.5em}
            \includegraphics[width=0.25\linewidth]{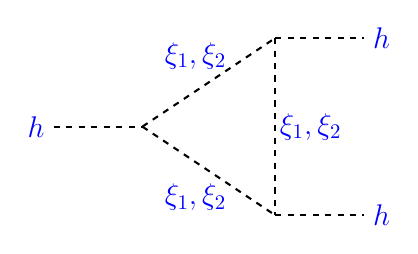}
            \hspace{-1.5em}
            \includegraphics[width=0.25\linewidth]{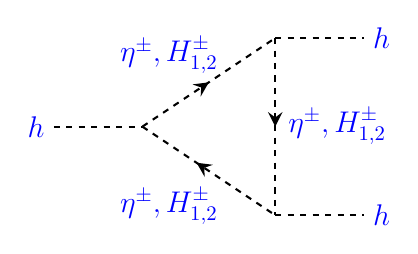}
            \includegraphics[width=0.25\linewidth]{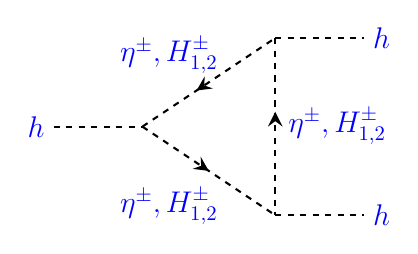}
            \hspace{-1.5em}
        \end{minipage}
        \begin{minipage}{\textwidth}
            \centering
            \includegraphics[width=0.22\linewidth]{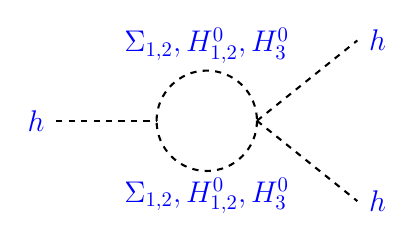}
            \includegraphics[width=0.22\linewidth]{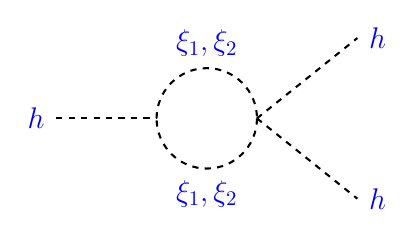}
            \hspace{-1.5em}
            \includegraphics[width=0.22\linewidth]{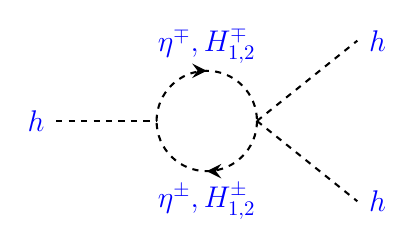}
        \end{minipage}
        \begin{minipage}{\textwidth}
            \centering
            \includegraphics[width=0.22\linewidth]{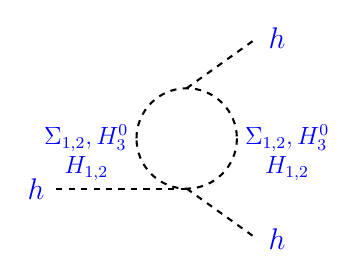}
            \hspace{-1.5em}
            \includegraphics[width=0.2\linewidth]{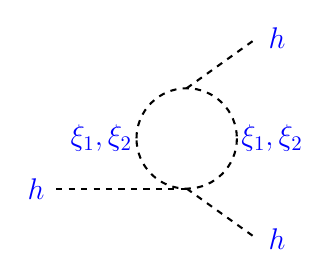}
            \hspace{-1.5em}
            \includegraphics[width=0.2\linewidth]{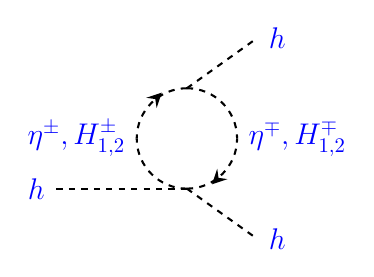}
        \end{minipage}
        \begin{minipage}{\textwidth}
            \centering
            \includegraphics[width=0.22\linewidth]{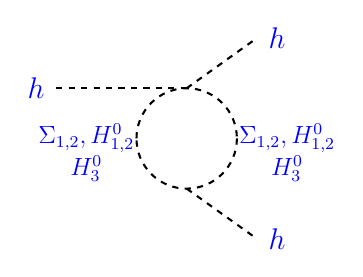}
            \hspace{-1.5em}
            \includegraphics[width=0.2\linewidth]{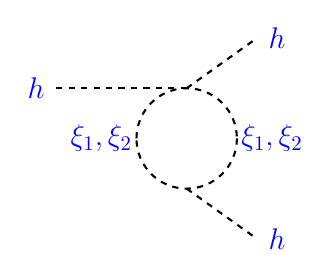}
            \hspace{-1.5em}
            \includegraphics[width=0.2\linewidth]{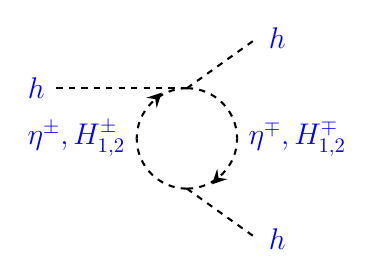}
        \end{minipage}
    \caption{Extra one-loop Feynman diagrams in the unitary gauge contributing to the Higgs trilinear selfcoupling.}
    \label{fig:trilinear}
\end{figure}

Beyond this, we evaluate the contribution of the scalar sector to the quantum one-loop corrections to the trilinear self-coupling. A detailed analysis of such corrections in extended Higgs sectors can be found in \cite{Arhrib:2015hoa, Kanemura:2016lkz, Moyotl:2016fdk, Kanemura:2017wtm}. The main Feynman diagrams contributing to this correction are shown in Figure \ref{fig:trilinear}.

For this study, we consider the off-shell decay process $h(q) \to h(q_1)h(q_2)$, where $q_1^2=q_2^2=m_h^2$ and $q^2 \neq m_h^2$. The treatment of ultraviolet divergences follows the renormalization procedure outlined in \cite{Arhrib:2015hoa}. The effective trilinear operator is given by:
\begin{align} 
\Gamma^R_{h^3}(q^2) = \lambda^{\text{BSM}}_{h^3}+ \Gamma^{\textrm{1-loop}}_{h^3}(q^2) + \delta\Gamma_{h^3}, 
\end{align}
where $\Gamma^R_{h^3}$ is the one-loop renormalized operator, $\Gamma^{\textrm{1-loop}}_{h^3}$ represents the contribution from the Feynman diagrams in Figure \ref{fig:trilinear}, and $\delta\Gamma_{h^3}$ accounts for the counterterm contributions. The explicit form of the one-loop terms can be found in Appendix \ref{app:b}.

The magnitude of the one-loop corrections is illustrated through the ratio:
\begin{align} 
\Delta \Gamma^{\textrm{1-loop}}_{h^3}(q^2) = \frac{\Gamma^R_{h^3}(q^2)}{\lambda_{h^3}^{\text{SM}}}. 
\end{align}

Numerical analysis reveals that, for our benchmark point, the renormalized effective coupling takes the value:
\begin{align} 
\Delta \Gamma^{\textrm{1-loop}}_{h^3}(q^2=M_Z^2) =0.34437. 
\end{align}

\begin{figure}[]
\centering
\subfigure[]{\includegraphics[width=0.45\linewidth]{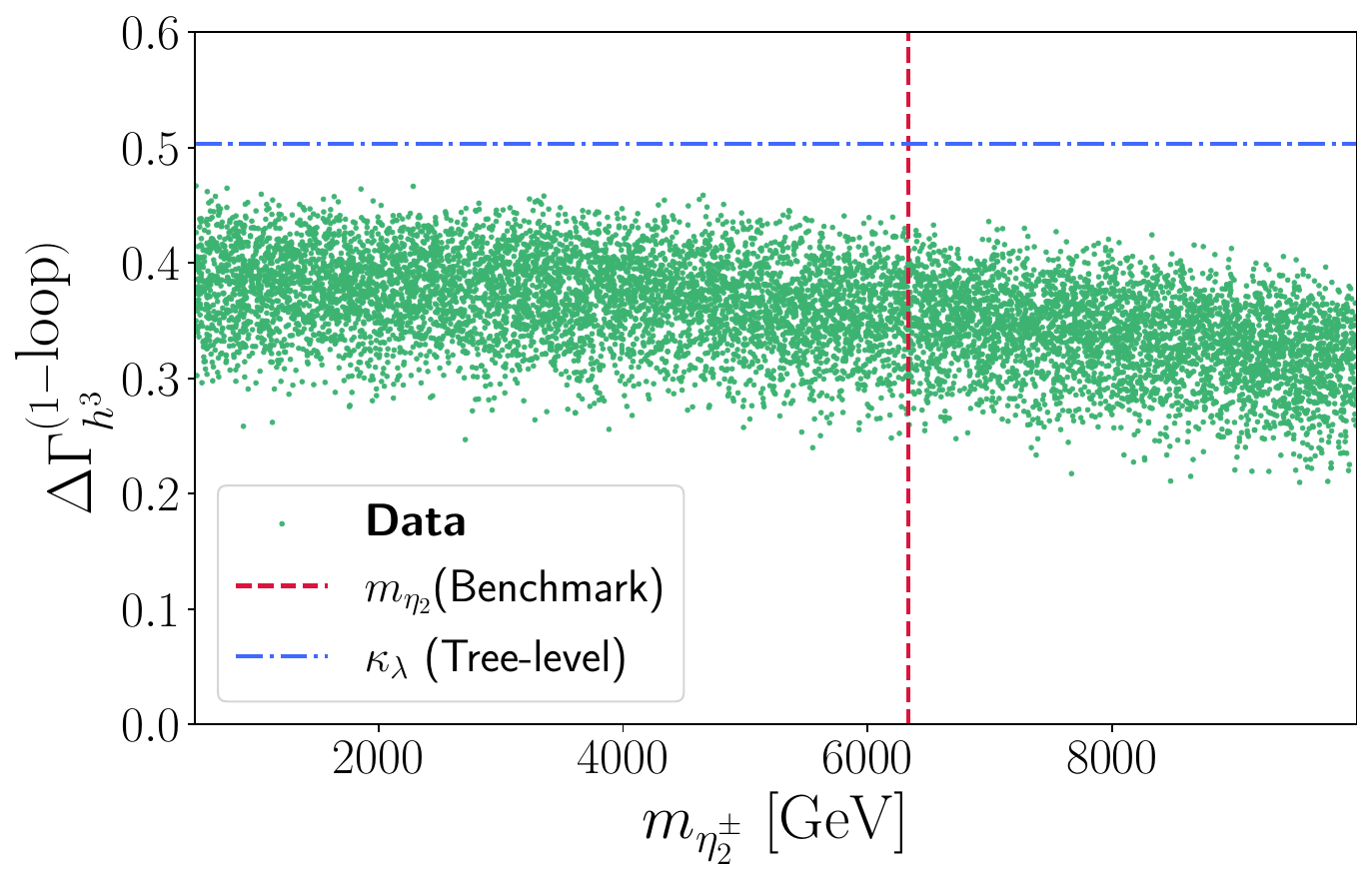}}
\subfigure[]{\includegraphics[width=0.46\linewidth]{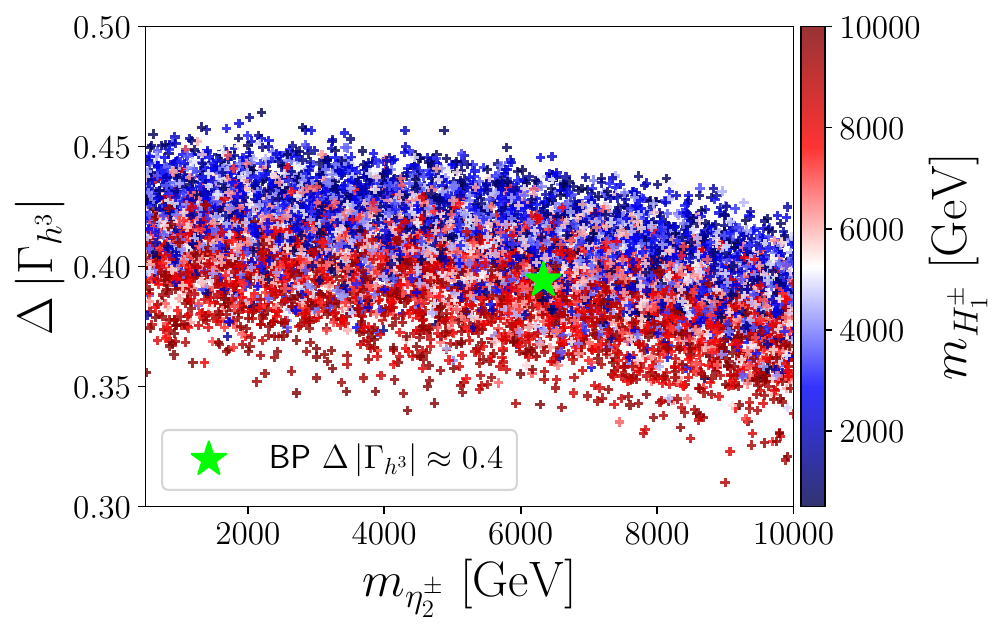}}
    \caption{One-loop radiative correction to the Higgs trilinear self-coupling. Panel (a) shows the variation of $\Delta \Gamma^{\textrm{1-loop}}_{h^3}$ for different masses of the exotic scalars, evaluated at the energy scale corresponding to the $Z$ boson mass. Panel (b) 
    presents the deviation of the radiative corrections from the new scalar sector of the model relative to the leading SM radiative corrections as a function of the exotic particle masses at the energy scale $q=M_Z$. The green star represents the deviation at our benchmark point.}
    \label{fig:trilineargraphs}
\end{figure}

This result indicates that radiative corrections shift the effective trilinear self-coupling away from the standard tree-level prediction; however, the deviation remains within the experimentally allowed range and does not lead to significant discrepancies. 
Figure \ref{fig:trilineargraphs}(a) illustrates that, by varying the masses of the exotic scalars between $500~\text{GeV}$ and $10~\text{TeV}$ while keeping the energy scale at the $Z$ boson mass, there are values for the one-loop corrections that remain consistent with experimental constraints.

Alternatively, we can analyze the deviation in the squared amplitude respect to the SM contribution by defining
\begin{align}
    \Delta \left|\Gamma_{h^3}(q^2)\right| =\sqrt{\frac{\left|\Gamma^R_{h^3}(q^2)+\lambda_{h^3}^{\text{BSM}}\Gamma^{SM}_{h^3}\right|^2}{\left|\lambda_{h^3}^{\text{SM}}(1+\Gamma^{SM}_{h^3})\right|^2}}, 
\end{align}
where $\Gamma^{SM}_{h^3}$ is the leading contribution of the SM to the effective coupling which will be considered only the correction due to the top quark \cite{Kanemura:2004mg,Kanemura:2002vm}
\begin{align}
    \Gamma^{SM}_{h^3}=-\frac{N_c}{3\pi^2}\frac{m_t^4}{v_\phi^2 m_h^2}\bigg\{1+\mathcal{O}\left(\mathrm{External}~ \mathrm{momenta}\right)\bigg\}.
\end{align}

In this case, the interference between both contributions leads to different outcomes. As shown in Figure \ref{fig:trilineargraphs}(b), varying the masses of the exotic scalars within the range of $500$ GeV to $10$ TeV can result in moderate percentage deviations of the radiative corrections from the exotic scalar sector relative to the dominant SM corrections. 
Notably, for our benchmark point (marked by the green star), the deviation is given by $\Delta \left|\Gamma_{h^3}(q^2)\right| = 0.394299$ wich still agree with the experimental bounds.

\section{\label{sec::5}Higgs diphoton rate}

The insertion of a strong sector in models with an extended number of higgs doublets can be a good approach in order to constrain the parametric space of that models, for example, this has been explored in a 2HDM models. In order to study the implications of the extra inert doublets in the decay of the $126$ GeV Higgs into a photon pair\footnote{The $\gamma \gamma$ channel is not the unique contribution to the total width of the Higgs boson decay, in a more complete study it is important to consider fermionic $c \bar{c}, b\bar{b}, \tau^+ \tau^-$ and vector $Z\gamma, W^+W^-, GG, \gamma \gamma$ decay channels, as well as additional invisible Higgs decay channels to scalar $H_k$ and pseudoscalar $\chi_k$ particles. For the regions where masses of scalar and pseudo-scalar $M_{H_k} (M_{\chi_k}) > m_h$, the invisible decay channels are kinematically closed.}, one introduces the Higgs diphoton signal strength $R_{\gamma \gamma}$, which is defined as:
\begin{equation}
R_{\gamma \gamma}=\frac{\sigma(p p \rightarrow h)_{\mathrm{BSM}} \Gamma(h \rightarrow \gamma \gamma)_{\mathrm{BSM}}}{\sigma(p p \rightarrow h)_{\mathrm{SM}} \Gamma(h \rightarrow \gamma \gamma)_{\mathrm{SM}}} ,
\end{equation}
where the expression is normalized by the $\gamma \gamma$ signal in the SM.
The dominant channel for Higgs boson production is gluon fusion, mediated by a top quark loop, therefore we can use the approximation \hbox{$\sigma\left(pp \rightarrow h\right)_{\text{BSM}} \simeq \mathcal{A}_{htt}^2 \sigma\left(pp \rightarrow h\right)_{\text{SM}} $}, and the ratio $R_{ \gamma \gamma}$ will involve only the Higgs decays. 

\begin{figure}[]
    \centering
    \subfigure[]{
        \begin{minipage}{\textwidth}
            \centering
            \includegraphics[width=0.3\linewidth]{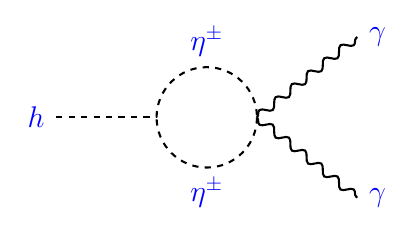}
            \hspace{0.02\linewidth} 
            \includegraphics[width=0.3\linewidth]{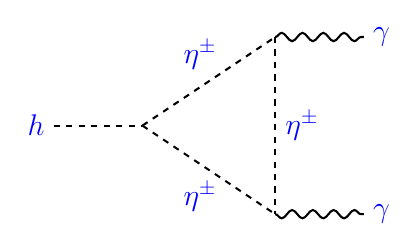}
        \end{minipage}
    }
    \subfigure[]{
        \begin{minipage}{\textwidth}
            \centering
            \includegraphics[width=0.3\linewidth]{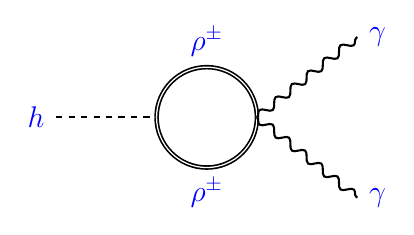}
            \hspace{0.02\linewidth} 
            \includegraphics[width=0.3\linewidth]{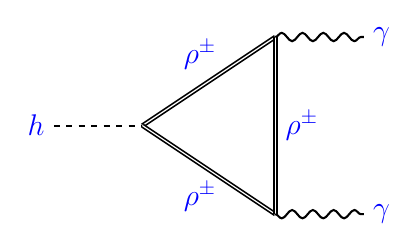}
        \end{minipage}
    }
    \caption{Extra one-loop Feynman diagrams in the unitary gauge contributing to the Higgs diphoton decay. In (a) the new charged scalars contribution and in (b) the vector resonances contributions}
    \label{fig:diagdiphoton}
\end{figure}

In the SM, the Higgs boson decay width into two photons is dominated by the interference between the $W-$boson and the top quark, meanwhile, in our model the presence of non-SM fields introduce loop corrections to the total amplitude of the decay. In Figure \ref{fig:diagdiphoton} the new one-loop Feynman diagrams can be seen.

The decay amplitude for the  $h \rightarrow \gamma \gamma$  process takes the form \cite{Bhattacharyya:2014oka,Logan:2014jla,CarcamoHernandez:2023dyz,CarcamoHernandez:2024ycd}:
\begin{equation}
\Gamma(h \rightarrow \gamma \gamma)=  \frac{\alpha_{\text{em}}^2 m_h^3}{256 \pi^3 v_\phi^2} \Bigg|
N_C \sum_f Q_f^2
\mathcal{A}_{hff}
F_{\frac{1}{2}}\left(\varrho_f\right)+\mathcal{A}_{hWW} F_1\left(\varrho_W\right)+
\mathcal{A}_{h \eta^{\pm} \eta^{\mp} } F_0\left(\varrho_{\eta^{\pm}}\right) 
 + \mathcal{A}_{h \rho^{\pm} \rho^{\mp}} F_1\left(\varrho_{\rho^{\pm}}\right) \Bigg|^2 ,
\end{equation}

where $\alpha_{\mathrm{em}}$ is the fine-structure constant, $N_C$ is the color factor $\left(N_C=3\right.$ for quarks and $N_C=1$ for leptons) and $Q_f$ is the electric charge of the fermion in the loop. The most significant contribution to the fermionic loop comes from the top quark, and it will be the only one considered.
Here, $\varrho$ represents the mass ratio $\varrho_i=4M_i^2/m_h^2$ with $M_i=m_f,M_W,M_{\rho^\pm},M_{\eta^\pm}$ and $k=1,2$. Furthermore $\mathcal{A}_{hff}$ and $\mathcal{A}_{hWW}$ are the deviation factors (normalized by the mass) from the SM Higgs-quark coupling and the SM Higgs-$W$ gauge boson coupling respectively, in the SM for the quark top the factor $\mathcal{A}_{htt} \simeq 1$ whereas the factor $\mathcal{A}_{hWW}$ can be written in the following form
\begin{equation}
    \mathcal{A}_{hWW}= \frac{1}{2} \frac{v_\phi}{M_{W^{\pm}}^2 }\left.\frac{\partial^3
    \mathscr{L}_{\text{gauge},\text{Phys}}}{\partial h\partial W^\pm\partial W^\mp}\right|_{\text{Fields}=0},
\label{eq:AWW}
\end{equation}
and the SM like Higgs  -charged Higgs boson and -vector resonance trilineal couplings are respectively
\begin{equation}
 \begin{aligned}
\mathcal{A}_{h \eta^{\pm} \eta^{\mp}} =  & \ \frac{1}{2} \frac{v_\phi}{M_{\eta^{\pm}}^2 }\left.\frac{\partial^3V_{\text{Phys}}}{\partial h\partial \eta^\pm\partial \eta^\mp}\right|_{\text{Fields}=0} ,\\
\mathcal{A}_{h\rho^\pm\rho^\mp}=& \frac{1}{2} \frac{v_\phi}{M_{\rho^{\pm}}^2 }\left.\frac{\partial^3\mathscr{L}_{\text{gauge},\text{Phys}}}{\partial h\partial \rho^\pm\partial \rho^\mp}\right|_{\text{Fields}=0}.
 \end{aligned}
\end{equation}
The form factors for the contributions of particles with spin-$0$, $1/2$ and $1$ are:
\begin{eqnarray}
F_0(\varrho	) & =& -\varrho(1-\varrho f(\varrho)),\\
F_{\frac{1}{2}}(\varrho) & =& 2\varrho(1+(1-\varrho) f(\varrho)), \\
F_1(\varrho) & =& -\left(2+3\varrho+3\varrho \left(2-\varrho\right) f(\varrho)\right),
\end{eqnarray}
with
\begin{equation}
f(\varrho)= \begin{cases}\arcsin ^2 \sqrt{\varrho^{-1}}, & \text { for } \varrho \geq 1 , \\ -\frac{1}{4}\left[\ln \left(\frac{1+\sqrt{1-\varrho}}{1-\sqrt{1-\varrho}}\right)-i\pi\right] ^2, & \text { for } \varrho<1.\end{cases}
\end{equation}

The decay rate in conjunction with the standard model scalar masses and couplings will be used to restrict the parametric space of the model. The couplings have been scanned over a range of numerical values compatible with the perturbative renormalization group, excepting for the $g_2$ coupling, on which the underlying strong sector is based. The parameters include the quartic couplings of the scalar potential, the VEVs of the $\Sigma$ and $\sigma$ fields, the coupling constants of the gauge group \hbox{$SU(2)_2\times SU(2)_1\times U(1)_Y$}, the decay parameter of the nonlinear Goldstone bosons $f_\Sigma$, the $\beta$ parameter, and the mixing angles.

Among the restrictions imposed on the parameter space are masses of the order of $1~\text{TeV}$ for new particles, in order to elude the LHC constraints on the detection of new particles, including strong vector resonances, and to place the exotic VEVs at a high scale of about $10~\text{TeV}$. Phenomenological constraints on the scan dictate that the parameter space reproduces the Higgs and $W$, $Z$ boson masses, the trilinear coupling between the Higgs and $W$ boson in \eq{eq:AWW}, the Higgs trilineal self-coupling in \eq{eq:htri} and the Higgs into two photons decay rate within a deviation of 3$\sigma$.

The adjustment strategy is based on the minimization of a $\chi^2$ function:
\begin{align}
    \chi^2=\sum_{o=1}^n\frac{\left(\mathcal{N}_{\text{o},\text{model}}-\mathcal{N}_{\text{o},\text{exp}}\right)^2}{(\Delta \mathcal{N}_{\text{o}}) ^2},
    \label{xisquare}
\end{align}
where $\mathcal{N}_{\text{o},\text{model}}$ and $\mathcal{N}_{\text{o},\text{exp}}$ refer to the numerical values of a physical observable predicted by the model with given parameter values and the experimental measurement of that observable and $\Delta\mathcal{N}_{\text{o}}$ is the reported error for the observable up to $3\sigma$ of deviation\footnote{The symbol $\sigma$ is used in the $\chi^2$ function. It is a symbol different from the singlet scalar field in this model.}. 

The best fit was found by preliminarily adjusting the boson masses, thereby finding a set of values over which the $\chi^2$ function is minimized through a random value scan equipped with a logic gate that selects smaller and smaller values of $\chi^2$, while discarding values that violate the stability conditions of the scalar potential in \eqref{eq:stability}.

The best-fit values for the main input parameters are:
\begin{align}
     & f_\Sigma=0.0553262,\quad v_{\Sigma}=17317.3~\mathrm{GeV},\quad v_\sigma=5044.37~\mathrm{GeV},\quad\beta=2.89831\times 10^{-4}~\mathrm{GeV^{-1}},
\end{align}
which yield the numerical values of the observables summarized in Table \ref{tab:numerical scalars}.

\begin{table}[]
\centering
\begin{tabular}{|ccc|} 
\hline\hline
\textbf{Observable} & \textbf{Model Value} & \textbf{Experimental Value}                        \\ 
\hline\hline
$m_h$        &         $125.245~\mathrm{GeV}$      & $\paren{125.25 \pm 0.17}~\mathrm{GeV}$ \cite{Workman:2022ynf}\\
$M_W$       &         $80.377~\mathrm{GeV}$      & $\paren{80.377 \pm 0.012}~\mathrm{GeV}$ \cite{Workman:2022ynf}\\
$M_Z$      &         $91.1876~\mathrm{GeV}$      & $\paren{91.1876 \pm 0.0021}~\mathrm{GeV}$ \cite{Workman:2022ynf}\\
$A_{hWW}$           &         $1.00464$      & $1.035 \pm 0.031$ \cite{ATLAS:2022vkf}\\
$R_{h\gamma\gamma}$ &         $0.994229$      & $1.04_{-0.09}^{+0.10}$\cite{ATLAS:2022tnm}
\\
$\kappa_\lambda$ & $0.503052$ & $[-0.4,6.3]$\cite{ATLAS:2022jtk} \\
\hline\hline
\end{tabular}
\caption{Comparison between predicted and experimental values of key observables in the model.}
\label{tab:numerical scalars}
\end{table}

\begin{figure}[]
\centering
\includegraphics[width=0.5\linewidth]{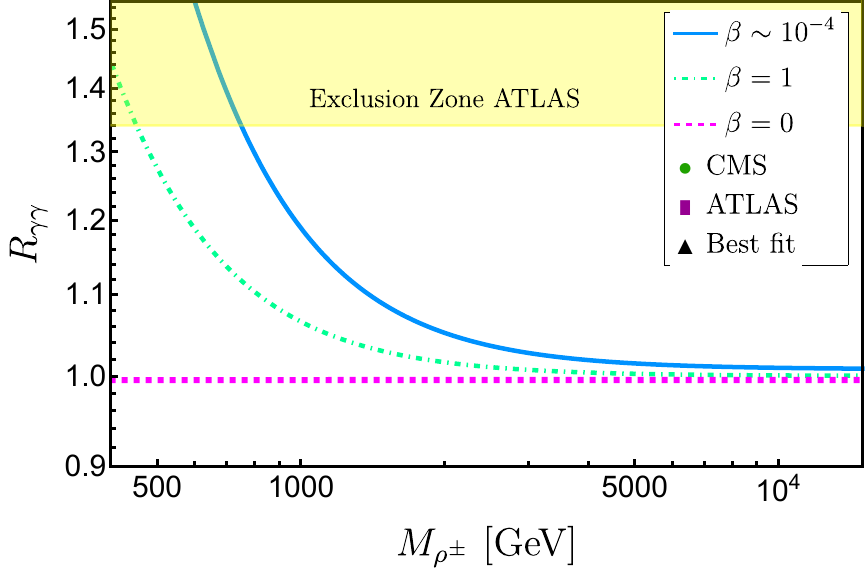}
\caption{The ratio $R_{\gamma \gamma}$ as a function of $M_{\rho^\pm}$ for several values of $\beta$ parameter. The circle and box correspond to the experimental predictions of the CMS $1.02_{-0.09}^{+0.11}$ \cite{Saha:2022cnz} and ATLAS $1.04_{-0.09}^{+0.10}$ \cite{ATLAS:2022tnm} collaborations, respectively, the triangle points correspond to best fit for our model. The shaded region corresponds to an exclusion zone superior to $3\sigma$ for the result of the CMS collaborations. 
}
\label{fig:diphoton}
\end{figure}

The Figure \ref{fig:diphoton} shows the dependence of the decay rate $R_{\gamma \gamma}$ with respect to the strong vector resonance mass $\rho^{\pm}$ for different values of the dimensionless parameter $\beta$. The blue solid line represents the case of the fit performed for $\beta$ in our model, in addition, the magenta dashed line corresponds to $\beta = 0$ and green dash-dotted line corresponds to $\beta = 1$.
In the region $M_{\rho^{\pm}}>500~\text{GeV}$, the value of the ratio $R_{\gamma \gamma}$ stabilizes, tending to the best fit represented by the triangular points in the figure. The experimental predictions of CMS and ATLAS are also included. 
This shows that although certain parameters, such as the trilinear coupling between the Higgs and the $W$ boson, are sensitive to the parameter $\beta$, the observables still show stability to changes in this parameter. This allows the theory to behave well both in the absence of this parameter and for values in the perturbative regime.

Finally, the significance of our result for $R_{\gamma \gamma}$ is contained in the present observations, suggesting that the theoretical framework revealed by our work could explain different deviations in the diphoton signal in future experiments. Theoretically, these deviations result from the incorporation of additional Feynman diagrams due to the new particles proposed by the model, modifying not only the decay signal but also the possible experimental signals. Any significant deviation could be a direct indication of the presence of new physics. In this sense, our results highlight the importance of considering these additional loop effects in the experimental analysis in order to discriminate between SM and BSM model predictions.

\section{\label{sec::6}The charged lepton flavor violating processes} \label{sec:CLFV}

The SM indicates that leptonic flavor is an invariant quantity, such that in any interaction process involving SM leptons, the total flavor is globally preserved. However, this does not preclude the possibility of flavor violations in other particles, such as quarks, within the SM. The reason for this distinction between leptons and quarks remains a significant challenge in particle physics. On the other hand, neutrino oscillations demonstrate that lepton flavor is not a true symmetry of nature, suggesting the existence of charged lepton flavor-violating (CLFV) processes.

To date, CLFV processes have not been directly observed. However, at higher energy scales, it is anticipated that such flavor violations could be detected in charged lepton interactions, similar to those observed in neutrinos. Within the SM, CLFV processes are suppressed by the Glashow–Iliopoulos–Maiani (GIM) mechanism to $\mathcal{O}(10^{-50})$ \cite{Ardu:2022sbt}, making them unobservable in practice. BSM frameworks provide promising scenarios where measurable CLFV processes could occur. The most stringent limits for CLFV are derived from measurements of decays involving a charged lepton into another charged lepton and a photon; specifically, the upper limit corresponds to the muon decay $\mu \rightarrow e \gamma$.

\begin{figure}[]
    \centering
    \subfigure[]{
        \begin{minipage}{\textwidth}
            \centering
            \includegraphics[width=0.3\linewidth]{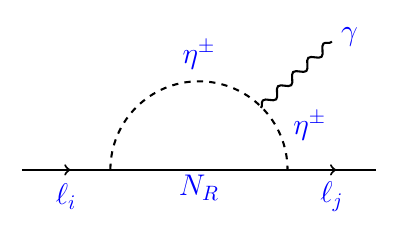}
            \hspace{-0.2em} 
            \includegraphics[width=0.35\linewidth]{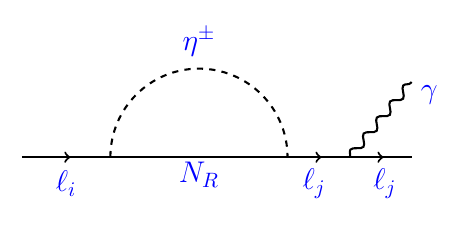}
            \hspace{-0.2em}
            \includegraphics[width=0.33\linewidth]{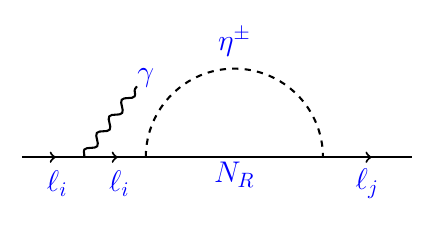}
        \end{minipage}
    }
    \subfigure[]{
        \begin{minipage}{\textwidth}
            \centering
            \includegraphics[width=0.35\linewidth]{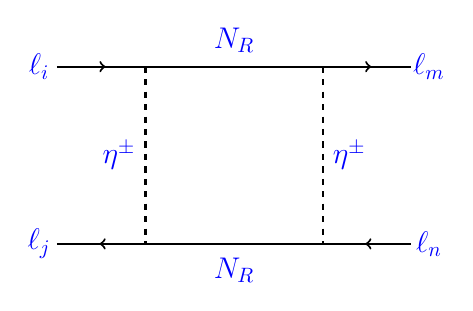}
            \hspace{0.02em} 
            \includegraphics[width=0.3\linewidth]{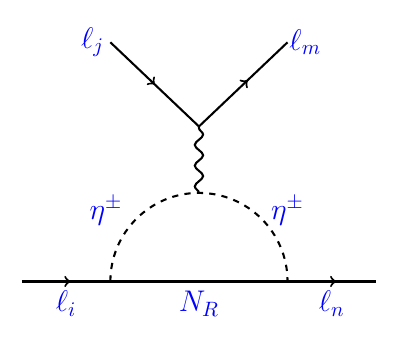}\\
            \includegraphics[width=0.35\linewidth]{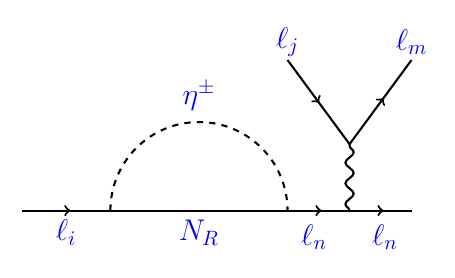}
            \hspace{0.02em} 
            \includegraphics[width=0.35\linewidth]{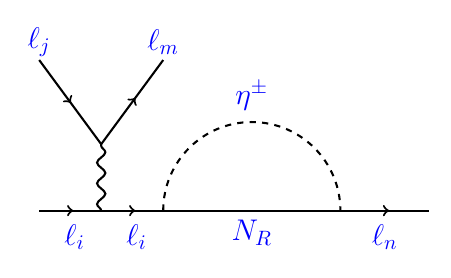}
        \end{minipage}
    }
    \caption{The one-loop Feynman diagrams leading to CLFV processes. In (a) the dipole-photon penguin diagrams contribution to $\ell_i \rightarrow \ell_j \gamma$ and in (b) the non-dipole photon penguin diagrams and box diagram contribution to $\ell_i\to \ell_j\ell_m\ell_n$}
    \label{fig:diagLFV}
\end{figure}

Due to the preserved $\mathbb{Z}^{\prime}_2$ symmetry, the extra scalars $h_k$ do not acquire VEVs, and therefore the neutrino masses do not arise at tree level. Consequently, the non-renormalizable operators, $\frac{1}{\Lambda}\overline{L}_L^i \Tilde{h}_k \Sigma N_R$, induce lepton flavor-violating decays, occurring at one-loop as depicted in Figure \ref{fig:diagLFV}(a). The branching ratio (Br) for $\ell_i \rightarrow \ell_j \gamma$ is given 
by \cite{Toma:2013zsa,Vicente:2014wga,Abada:2022dvm,Hernandez:2021iss}:
\begin{equation}
\text{Br}\left(\ell_i \rightarrow \ell_j \gamma\right)  =\frac{3(4 \pi)^3 \alpha_{\text{em}}}{4 G_F^2}\left| A_D^{(i,j)} \right|^2 \text{Br}\left(\ell_i \rightarrow \ell_j \nu_i \overline{\nu_j}\right),
\end{equation}
where $i\neq j$, and $G_F$ is the Fermi constant. The amplitude coming from the dipole-photon penguin diagram Figure \ref{fig:diagLFV} (a), usually known as the dipole form factor, is related to the transition magnetic moment between the lepton $\ell_j$ and the lepton $\ell_i$. In this case, $i,j$ are flavor indices, and it is defined as:
\begin{equation}
A_D^{(i,j)} = \frac{x_{i}^{(4,5)} x_{j}^{(4,5)}}{2(4 \pi)^2 m_{\eta^{ \pm}}^2} F_2\left(
\xi^2_{\eta^{\pm}} 
\right),
\end{equation}
where $x_{i}^{(4,5)}= \sum_{k=1}^3 y_{k}^{(4,5)}\left(V_{\ell L}^{\dagger}\right)_{ik}$, the mass ratio $\xi_{\eta^\pm_n} = m_N / m_{\eta^\pm_n}$, and $V_{\ell L}$ is the mixing matrix of the left-handed charged leptons. The loop function $F_2(x)$ is given by:
\begin{equation}
F_2(x)=\frac{1-6 x+3 x^2+2 x^3-6 x^2 \log x}{6(1-x)^4}.
\end{equation}

\begin{figure}[]
\centering
\includegraphics[scale=.45]{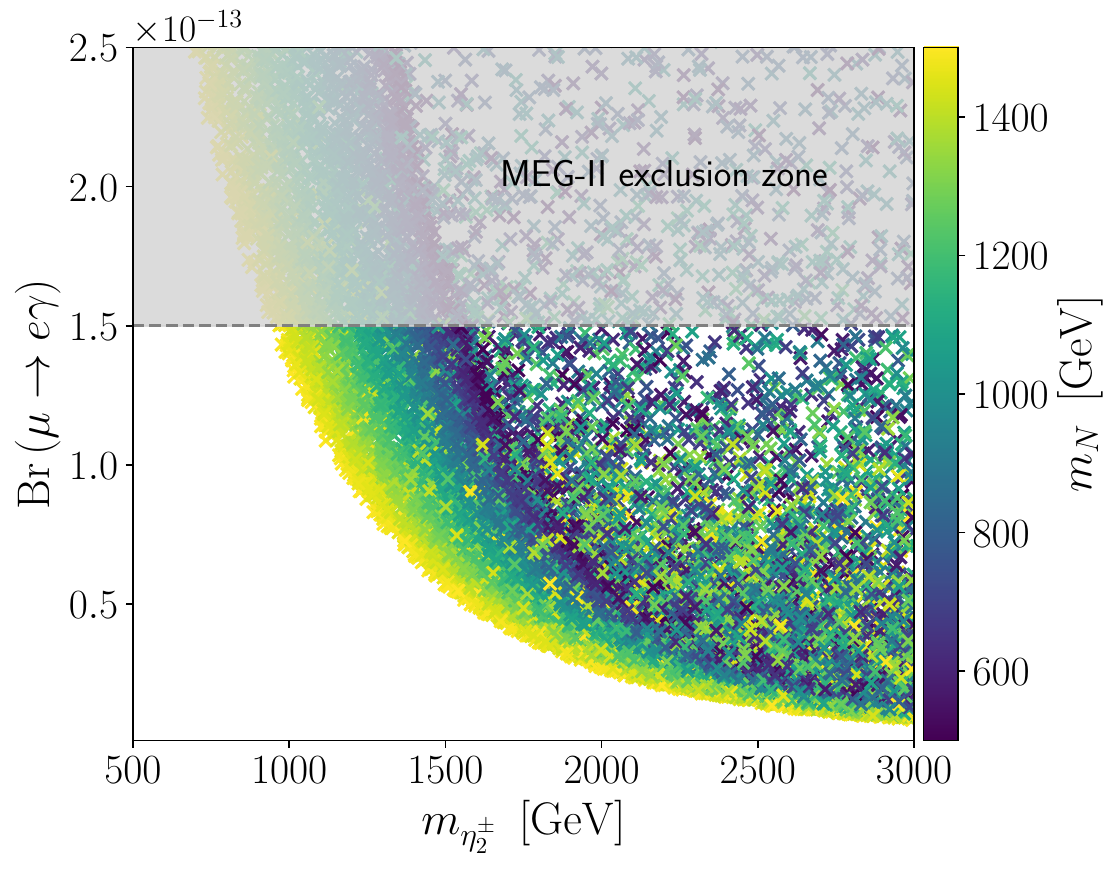}
\caption{Branching ratio $\text{Br}\left(\mu \rightarrow e \gamma \right)$ as a function of the mass of charged scalar masses $m_{\eta_{2}^{\pm}}$ and the color represents the scan in the plane $m_N$-$m_{\eta_{2}^{\pm}}$. The shadowed region is excluded by MEG II \cite{MEGII:2025gzr}.}
\label{fig:lfv1}
\end{figure}

Figure \ref{fig:lfv1} illustrates the relationship between the branching ratio for the CLFV process $\mu \rightarrow e \gamma$ and the charged scalar mass $m_{\eta_2^{\pm}}$, with a color gradient representing the range of heavy neutrino masses $m_N$. The scatter points depict the branching ratio as a function of $m_{\eta_2^{\pm}}$, showing that as the charged scalar mass increases, the branching ratio significantly decreases. The exclusion zone from the MEG-2016 experiment \cite{MEG:2016leq}, corresponding to an upper limit of $4.2 \times 10^{-13}$ for Br$(\mu \rightarrow e \gamma)$, is represented by the shaded region. Points within this region are excluded by current experimental data, while points below it are consistent with experimental constraints. The gradient colors, ranging from purple to yellow, indicate different values of $m_N$, with heavier neutrino masses favoring higher branching ratios.

Another important leptonic flavor violation observable is the branching ratio for $3$-body decays $\ell_i \rightarrow 3 \ell_j$, which is given by:
\begin{equation}
\begin{aligned}
\text{Br}\left(\ell_i \rightarrow \ell_j \overline{\ell_j} \ell_j \right)= & \ \frac{3(4 \pi)^2 \alpha_{\text{em}}^2}{8 G_F^2}\bigg[\left|A_{N D}^{(i,j)}\right|^2+\left|A_D^{(i,j)}\right|^2\left(\frac{16}{3} \log \left(\frac{m_i}{m_j}\right)-\frac{22}{3}\right)+\frac{1}{6}\left|B^{(i,j)}\right|^2 \\
& +\left(-2 A_{N D}^{(i,j)} A_D^{(i,j) *}+\frac{1}{3} A_{N D}^{(i,j)} B^{(i,j)*}-\frac{2}{3} A_D^{(i,j)} B^{(i,j)*}+H . c .\right)\bigg] \text{Br}\left(\ell_i \rightarrow \ell_j \nu_i \overline{\nu_j}\right),
\end{aligned}   
\end{equation}
where $A_{ND}^{(i,j)}$ is the non-dipole photon penguin diagram contribution, and $B^{(i,j)}$ is the contribution arising from the box diagram of Figure \ref{fig:diagLFV} (b). They can be expressed as:
\begin{eqnarray}
A_{N D}^{(i,j)} & =& \  \sum_{k=1}^2 \frac{x_{i,k} x_{j,k}^{*}}{6(4 \pi)^2} \frac{1}{m_{\eta_{k}^{ \pm}}^2} G_2\left(\xi^2_{\eta_{k}^{\pm}} \right),\\
 \label{eq:boxdiagram} e^2 B^{(i,j)} & = & \ \frac{1}{(4 \pi)^2 m_{\eta_{k}^{ \pm}}^2} \sum_{k=1}^2 \left[\frac{1}{2}
\left\vert x_{j,k}\right\vert^2 x_{j,k}^* x_{i,k}
D_1\left(\xi^2_{\eta_{k}^{\pm}}\right) 
+\xi^2_{\eta_{k}^{\pm}} \left\vert x_{j,k} \right\vert^2 x_{j,k}^* x_{i,k}
D_2\left(\xi^2_{\eta_{k}^{\pm}}\right)\right].
\end{eqnarray}

The different loop functions are given by \cite{CarcamoHernandez:2024ycd}:
\begin{equation}
\begin{aligned}
G_2(\varrho)  = & \frac{2-9 \varrho+18 \varrho^2-11 \varrho^3+6 \varrho^3 \log \varrho}{6(1-\varrho)^4}, \\
D_1(\varrho) = & \frac{-\varrho^2+2 \varrho \log (\varrho)+1}{(\varrho-1)^3}, \\
D_2(\varrho) = & \frac{-2 \varrho+(\varrho+1) \log (\varrho)+2}{(\varrho-1)^3}.
\end{aligned}
\end{equation}

The $\mu$-$e$ conversion in nuclei is another significant CLFV process, where a muon is converted into an electron in the presence of a nucleus. The conversion rate, denoted as $\text{CR}(\mu - e)$, quantifies the probability of this transition occurring within the atomic environment. This process is particularly sensitive to new physics contributions arising in BSM extensions like the one considered in this work, and provides complementary constraints to those obtained from other CLFV processes, such as $\mu \rightarrow e \gamma$. The branching ratio for $\mu^- - e^-$ conversion is defined as follows \cite{Lindner:2016bgg}:
\begin{equation}
\text{CR}(\mu-e) = \frac{\Gamma\left(\mu^{-}+\operatorname{Nucleus}(A, Z) \rightarrow e^{-}+\operatorname{Nucleus}(A, Z)\right)}{\Gamma\left(\mu^{-}+\operatorname{Nucleus}(A, Z) \rightarrow \nu_\mu+\operatorname{Nucleus}(A, Z-1)\right)}.
\end{equation}

For the radiative neutrino mass model considered in this work, the conversion rate, normalized to the charged lepton capture rate, takes the form \cite{Vicente:2014wga}:
\begin{equation}
\begin{aligned}
\text{CR}\left(\ell_i N \rightarrow \ell_j N\right) = & \ \frac{p_j E_j m_i^3 G_F^2 \alpha_{\text{em}}^3 Z_{\text {eff }}^4 F_p^2}{8 \pi^2 Z \Gamma_{\text {capt }}}\bigg[\left|(Z+N)\left(g_{L V}^{(0)}+g_{L S}^{(0)}\right)+(Z-N)\left(g_{L V}^{(1)}+g_{L S}^{(1)}\right)\right|^2 \\
& +\left|(Z+N)\left(g_{R V}^{(0)}+g_{R S}^{(0)}\right)+(Z-N)\left(g_{R V}^{(1)}+g_{R S}^{(1)}\right)\right|^2\bigg],
\end{aligned}
\end{equation}
where $Z$ and $N$ are the number of protons and neutrons in the nucleus, $Z_{\text{eff}}$ is the effective atomic charge \cite{Chiang:1993xz}, $F_p$ denotes the nuclear matrix element, and $\Gamma_{\text{capt}}$ represents the total muon capture rate. Numerical values extracted from \cite{Kitano:2002mt} are presented in Table \ref{tab:nucleo}. The coupling constants $g_{L / R ;S / V}^{(0,1)}$ are given by:
\begin{equation}
g_{L / R ;S / V}^{(0,1)} = \frac{1}{2} \sum_{q=u, d, s}\left(g_{L / R ;S / V (q)} G_{S,V}^{(q, p)}\pm g_{L / R ;S / V (q)} G_{S,V}^{(q, n)}\right),
\end{equation}
where the numerical values of the $G_{S,V}$ coefficients can be found in \cite{Kosmas:2001mv}.

\begin{table}[]
\centering
\begin{tabular}{|c|c|c|c|}
\hline
$Z^A \mathrm{Nucleus}$ & $Z_{\text {eff }}$ & $F_p$ & $\Gamma_{\text {capt }} (\mathrm{GeV})$ \\
\hline \hline
${ }_{22}^{48} \mathrm{Ti}$ & 17.6 & 0.54 & $1.70422 \times 10^{-18}$ \\
${ }_{51}^{197} \mathrm{Au}$ & 33.5 & 0.16 & $8.59868 \times 10^{-18}$ \\
${ }_{13}^{27} \mathrm{Al}$ & 11.5 & 0.64 & $4.64079 \times 10^{-19}$ \\
\hline
\end{tabular}
\caption{Values of $Z_{\text{eff}}$, $F_p$, and $\Gamma_{\text{capt}}$ for Titanium, Gold, and Aluminum nuclei \cite{Kitano:2002mt}.}
\label{tab:nucleo}
\end{table}

\begin{figure}[]
\centering
\includegraphics[scale=.45]{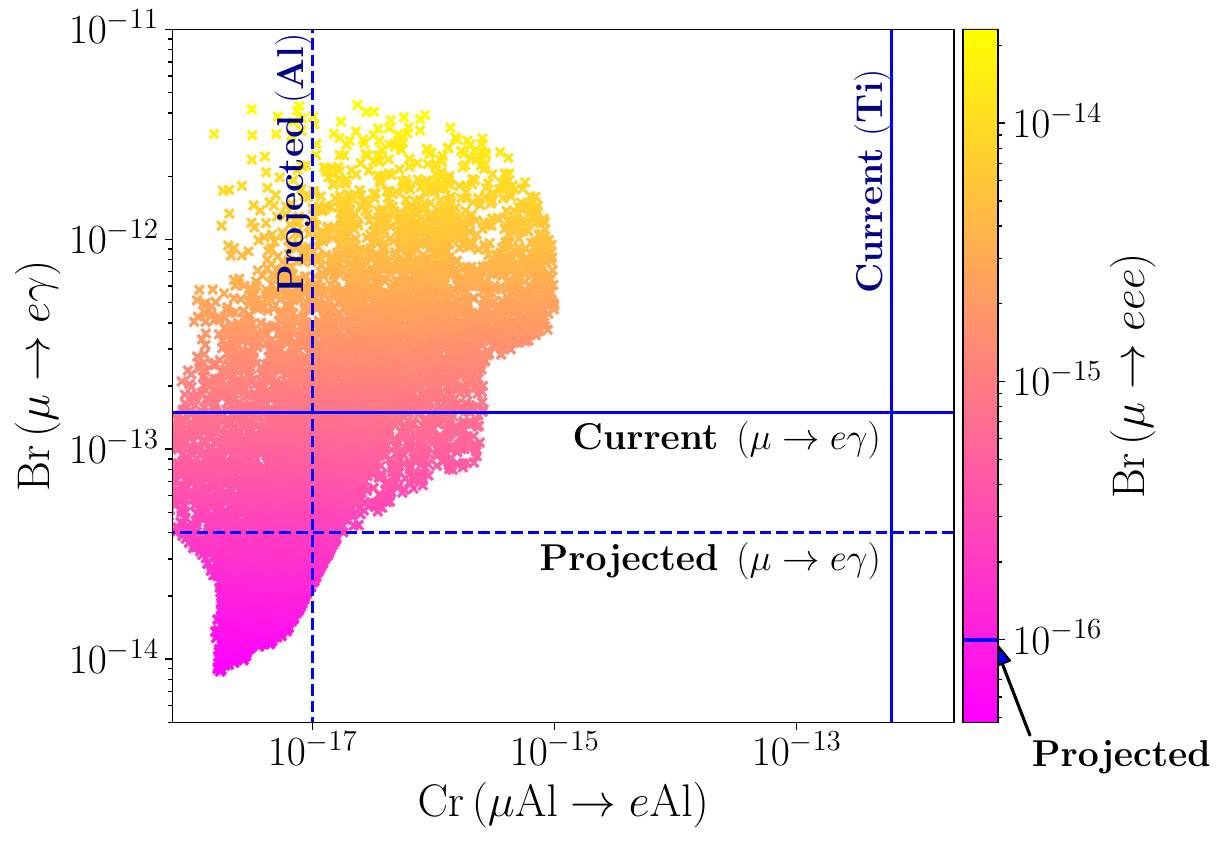}
\caption{Correlation between the LFV processes $\mu \rightarrow e$ conversion in aluminum nuclei and $\mu \rightarrow e \gamma$. The points are color-coded according to the branching ratio of the $\mu \rightarrow e e e$ process, as shown in the color bars. The current upper bounds are indicated by the solid blue lines, while the future sensitivities are marked by the dashed blue lines. The bars indicate the projected bounds for $\mu \rightarrow e e e$.}
\label{fig:lfv2}
\end{figure}

In Figure \ref{fig:lfv2}, the correlation between the CLFV processes $\mu \rightarrow e$ conversion in aluminum nuclei and $\mu \rightarrow e \gamma$ is illustrated. The points are color-coded based on the branching ratio of the $\mu \rightarrow eee$ process, as indicated by the color bar. The current experimental upper limits for these processes are shown as solid blue lines: Br$(\mu \rightarrow e \gamma)<1.5 \times 10^{-13}$ \cite{MEGII:2025gzr} and Cr$\left(\mu \mathrm{Ti} \rightarrow \mathrm{e}\mathrm{Ti}\right)<6.1 \times 10^{-13}$ \cite{SINDRUM:1987nra}. The projected future sensitivities are marked by dashed blue lines: Br$(\mu \rightarrow e \gamma)<4 \times 10^{-14}$ \cite{Mori:2016vwi} and Cr$\left(\mu \mathrm{Al} \rightarrow e \mathrm{Al}\right) \lesssim 10^{-17}$ \cite{Blondel:2013ia}. 

Additionally, the bars indicate the projected bounds for the $\mu \rightarrow e$ conversion in aluminum (Al) \cite{Wintz:1998rp} and titanium (Ti) \cite{Bernstein:2013hba} nuclei, allowing for a visual comparison between current constraints and those anticipated from future experimental upgrades. This correlation is crucial for evaluating the sensitivity of upcoming experiments in probing LFV processes and exploring a broader range of model parameters.

\section{Leptogenesis}
\label{sec:leptogenesis}

One of the central challenges in modern physics is understanding the baryon asymmetry of the Universe, which underpins the observed dominance of matter over antimatter. A dynamic explanation is often favored, as any preexisting asymmetry would likely have been erased by a potential cosmic inflationary period preceding the Big Bang. The current baryon asymmetry is quantified by the dimensionless parameter \cite{Planck:2018vyg}:
\begin{align} 
    Y_{\Delta B} = \frac{n_B - n_{\bar{B}}}{s}=(0.87 \pm 0.01)\times 10^{-10}, 
\end{align}
where $n_B$ and $n_{\bar{B}}$ represent the number densities of baryons and antibaryons, respectively, and $s$ denotes the entropy density.

For a physical model to generate a baryon asymmetry dynamically, it must satisfy the three Sakharov conditions \cite{Sakharov:1967dj}: (i) baryon number violation to enable a non-zero $Y_{\Delta B}$; (ii) violation of $C$ and $CP$ symmetries, allowing different dynamics for baryons and antibaryons; and (iii) a departure from thermal equilibrium to permit an asymmetry in number densities. While the SM accommodates these conditions in principle, the suppression of $CP$-violating effects and the insufficient strength of the baryon-number-violating processes render its contribution negligible, necessitating physics beyond the SM.

Leptogenesis offers an elegant mechanism by converting an initial lepton asymmetry into a baryon asymmetry through electroweak sphaleron processes. These processes violate the anomalous $B+L$ symmetry, while conserving $B-L$, enabling the transfer of a lepton asymmetry into the baryon sector. The topological nature of this anomaly plays a critical role, as sphaleron-induced transitions between distinct vacuum configurations of the gauge fields are non-perturbative phenomena that efficiently convert lepton number into baryon number in the early Universe.

\subsection{Lepton asymmetry parameter}

\begin{figure}[]
\begin{center}
    \includegraphics[scale=.7]{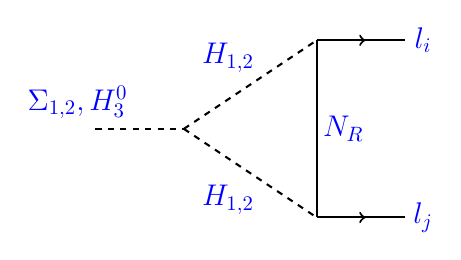}
\end{center}
\caption{One-loop contribution to lepton number violation processes.}
 \label{fig:lnvp}
\end{figure}

Our model predicts radiatively induced processes that violate lepton flavor, as illustrated in Figure \ref{fig:lnvp}. These processes emerge naturally from the extended scalar sector and the introduction of exotic doublets, which mediate flavor transitions through loop-level corrections. The associated asymmetry parameters, which quantify the generated lepton number violation, are defined as \cite{Datta:2021gyi,Hambye:2005tk,Lu:2016ucn}:
\begin{equation}
\begin{aligned}
    \epsilon_{S_\alpha}=&2\frac{\Gamma(S_\alpha\to l_il_j)-\Gamma(S_\alpha\to \bar l_i\bar l_j)}{\Gamma(S_\alpha\to l_il_j)+\Gamma(S_\alpha\to \bar l_i\bar l_j)}\\
    &=\frac{m_N}{2 \pi}\frac{v_{\Sigma}}{\sqrt{2}}\frac{\sum_i \mathrm{I m}\left[y^{4}_iy^{4}_i C_{S_\alpha h_{1}h_{1}}+y^{4}_iy^{5}_i C_{S_\alpha h_{1}h_{2}}+y^{5}_iy^{5}_i C_{S_\alpha h_{2}h_{2}}\right]}{C_{S_\alpha h_1h_1}^2+C_{S_\alpha h_1h_2}^2+C_{S_\alpha h_2h_2}^2} \ln \left(1+\frac{m_{S_\alpha}^2}{m_N^2}\right),
\end{aligned}
\end{equation}
where $S_\alpha=\Sigma_1,\Sigma_2,H^0_3$. On the other hand, the total decay width of the $S_\alpha$ particle is given by
\begin{align}
    \Gamma_{S_\alpha T}=\Gamma(S_\alpha\to l_il_j)+\Gamma(S_\alpha\to \bar l_i\bar l_j)=\frac{1}{8\pi}\left[\frac{C_{S_\alpha h_1h_1}^2+C_{S_\alpha h_1h_2}^2+C_{S_\alpha h_2h_2}^2}{m_{S_\alpha}}\right].
\end{align}

\begin{figure}[]
    \centering
    \subfigure[]{
        \begin{minipage}{\textwidth}
            \centering
            \includegraphics[width=0.43\linewidth]{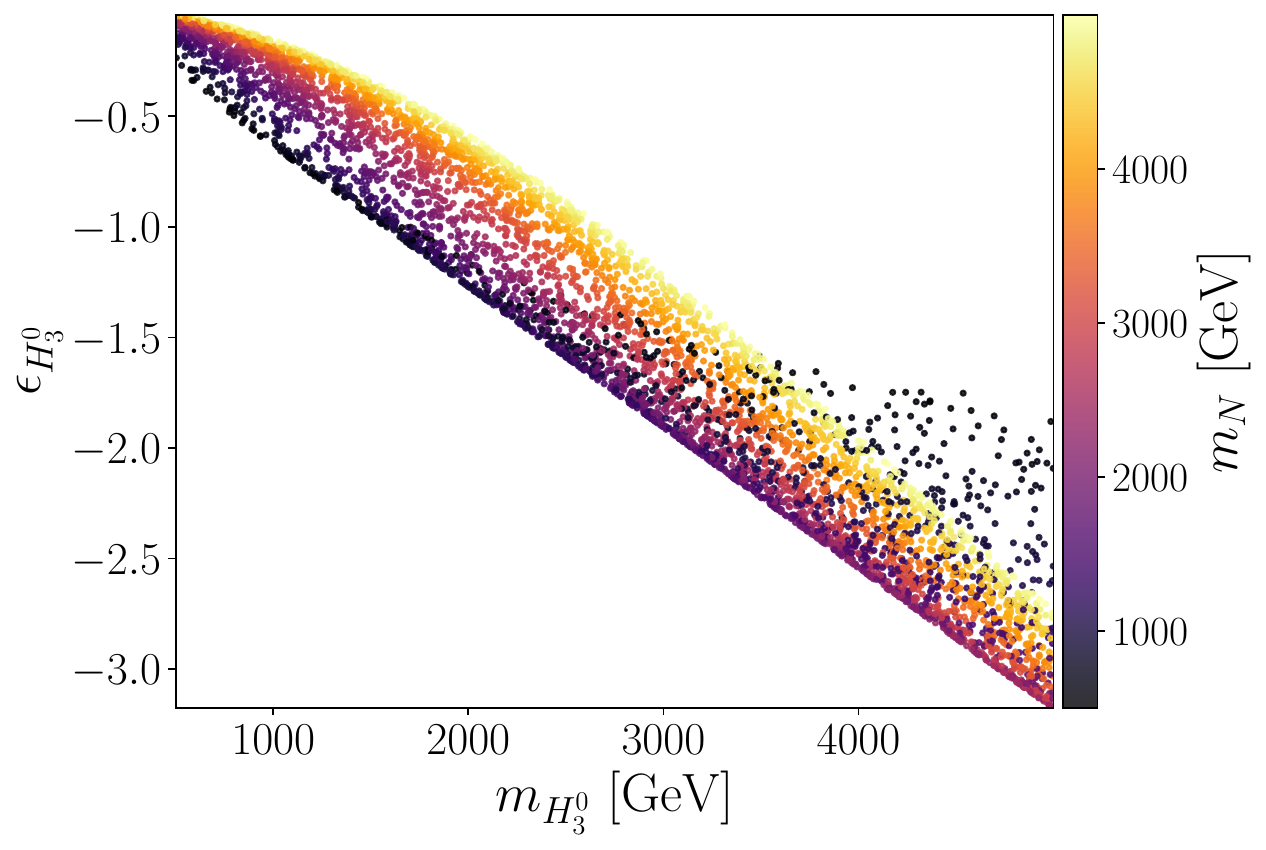}
            \hspace{-0.2em} 
            \includegraphics[width=0.45\linewidth]{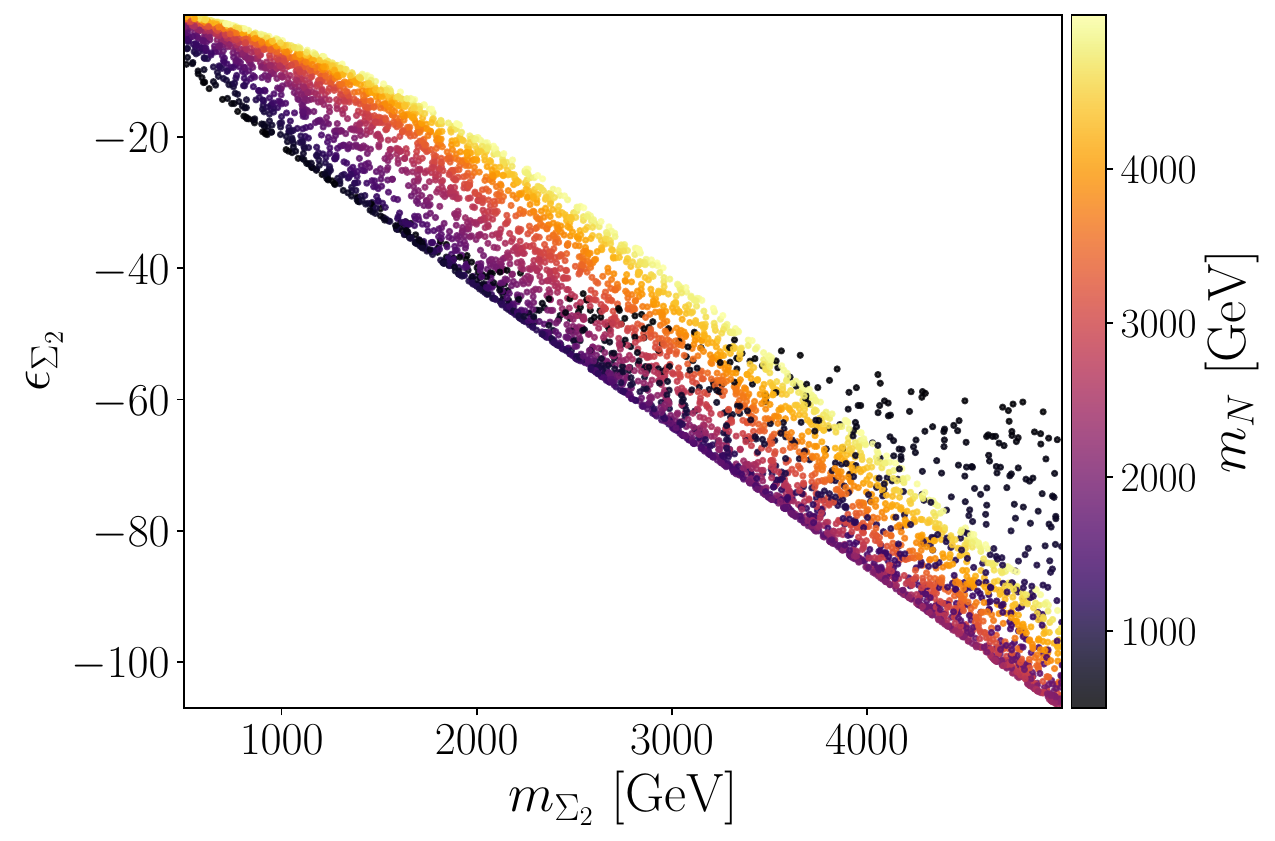}
        \end{minipage}
    }
    \subfigure[]{
        \begin{minipage}{\textwidth}
            \centering
            \includegraphics[width=0.45\linewidth]{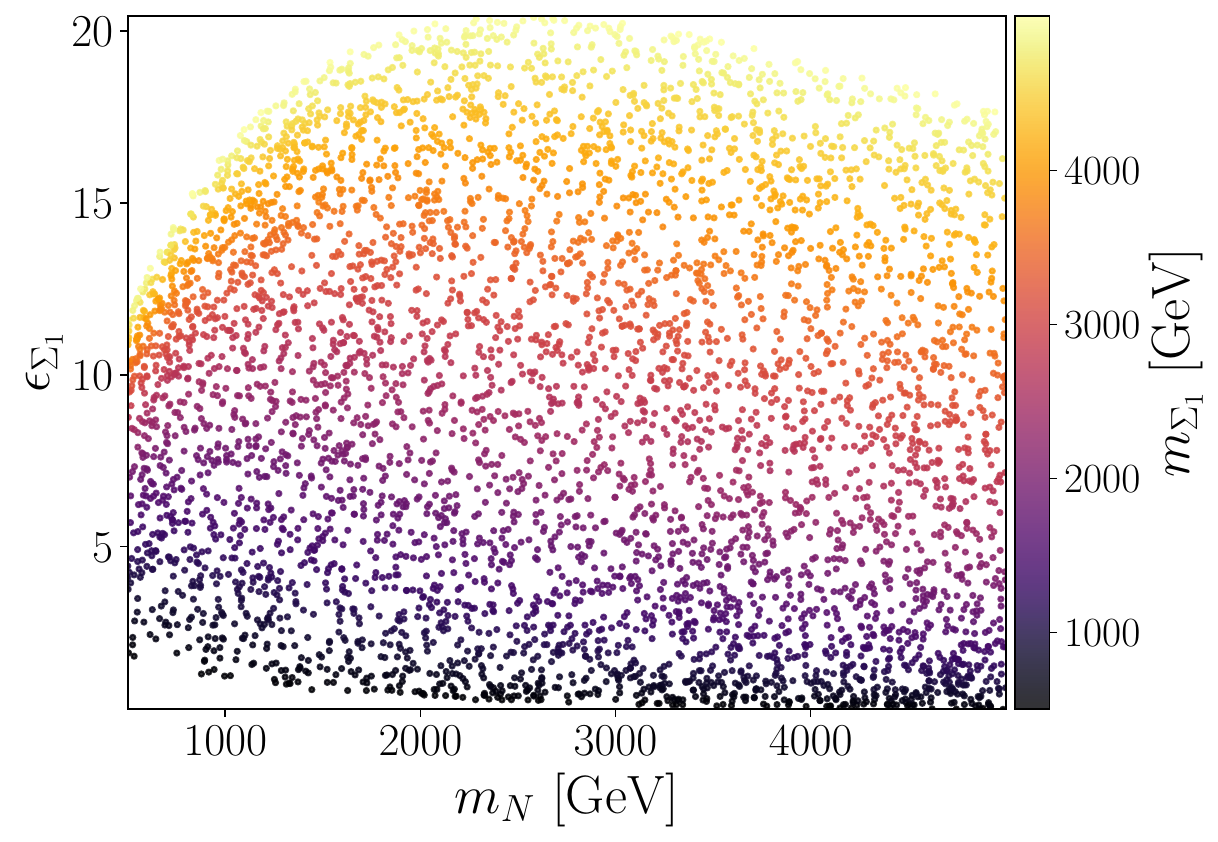}
        \end{minipage}
    }
    \caption{Colour plots of $CP$ asymmetries. In (a) we have the asymmetries of $H^0_3$ and $\Sigma_2$ as function of the masses of the scalars and the heavy neutrino while in (b) we have the asymmetry of $\Sigma_1$ as function of the mass of the heavy neutrino.}
    \label{fig:asympters}
\end{figure}

In Figure \ref{fig:asympters}, we present the dependence of the asymmetry parameters on the masses of the scalars and heavy neutrinos. The results indicate a partial dominance of the $CP$ asymmetry arising from the decay of the $\Sigma_2$ scalar. This is due to its slightly weaker coupling with the inert scalars, which contributes to a higher asymmetry parameter. Specifically, the asymmetry parameter associated with this decay is approximately one order of magnitude larger than those corresponding to other scalar decay processes.

\subsection{Baryon Asymmetry Parameter}

In order to test the capabilities of this model in leptogenesis processes we will estimate the baryonic asymmetry parameter, without resorting to explicitly solve the Boltzman equation, which is beyond the scope of this work.

We will start by calculating the washout parameters associated to each of the scalars as
\begin{align}
    K_\alpha=\left.\frac{\Gamma_{S_\alpha T}}{2 H(T)}\right|_{T=m_{S_\alpha}},
\end{align}
where $H(T)$ is the Hubble constant as function of the temperature
\begin{align}
    H(T)=\sqrt{\frac{8\pi^3g_*}{90}}\frac{T^2}{M_\text{Pl}},
\end{align}
$g_*=114.75$ is the number of relativistic degrees of freedom and $M_\text{Pl}=1.22\times 10^{19}~\text{GeV}$ is the Planck mass.

Since the washout parameters for this model using our benchmark scalar masses are very large we can use their relation to the baryon asymmetry parameter in the strong washout regime \cite{Pilaftsis:1997jf}
\begin{align}
    Y_{\Delta B}\equiv \frac{n_B-n_{\bar B}}{s}=-\frac{28}{79}\frac{0.3\sum_\alpha\epsilon_{S_\alpha}}{g_*K_\text{eff}(\ln K_\text{eff})^{0.6}}.
\end{align}

\begin{figure}
    \centering
    \includegraphics[width=0.5\linewidth]{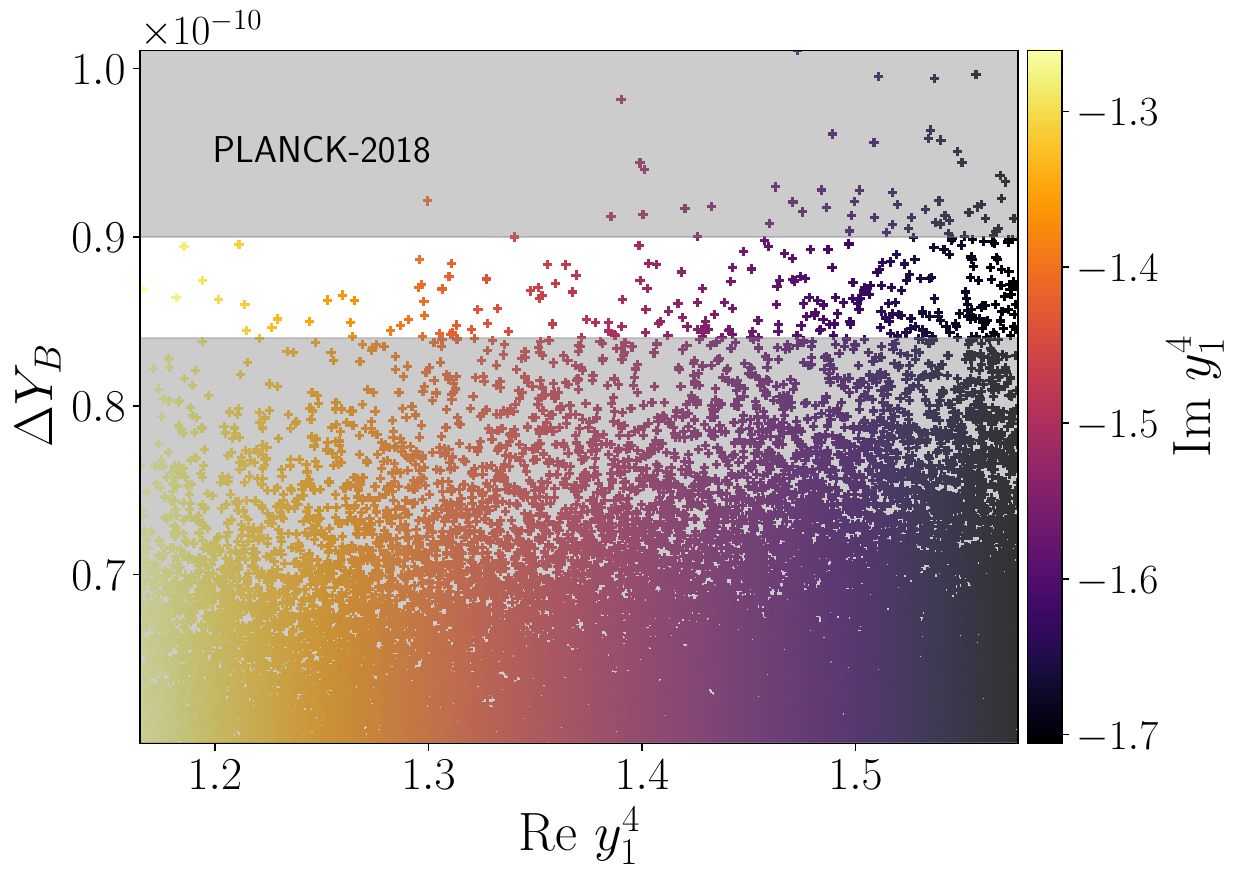}
    \caption{The baryon asymmetry parameter as a function of the real and imaginary parts of the Yukawa coupling $y^4_1$. The gray-shaded area denotes the parameter space excluded by the PLANCK collaboration \cite{Planck:2018vyg}.}
    \label{fig:DeltaYB}
\end{figure}

Figure \ref{fig:DeltaYB} shows a set of values for the baryon asymmetry parameter generated by our model, computed for different values of the Yukawa couplings and the Majorana neutrino mass, which spans in a range from $500$ GeV to $5$ TeV. The results from the parameter scan demonstrate that our model can successfully reproduce a scenario where the baryon asymmetry parameter falls within the experimental range provided by the PLANCK collaboration \cite{Planck:2018vyg}, thus supporting the feasibility of generating the observed baryon asymmetry via a leptogenesis mechanism.

\section{\label{sec::7}Oblique parameters}
The oblique parameters play a vital role in the so-called electroweak precision program of the SM due to their high sensitivity to new physics. Many of the observables used as precision tests of the SM are defined at the $Z$-boson pole, while the $W$-boson mass provides complementary information about radiative corrections independent of the $Z$-pole and asymmetries. Consequently, the renormalized propagators of the gauge bosons are highly sensitive to oblique corrections induced by new particles in models BSM.

In our model, the parameters $S$, $T$, and $U$ \cite{Peskin:1990zt,Altarelli:1990zd,Peskin:1991sw,Altarelli:1991fk} receive corrections due to the coupling of exotic scalar particles with the gauge bosons of the SM, modifying the standard corrections. The significance of oblique parameters in BSM models lies in their function as a window to high-energy physics, as heavy exotic particles can induce oblique corrections to the electroweak sector. Due to the stringent constraints on SM extensions, the study of the $S$, $T$, and $U$ parameters is a key criterion for assessing the validity of our model.

The oblique parameters $S$, $T$, and $U$ are defined in terms of the vacuum polarization tensors $\Pi_{ij}^{\mu \nu}(q^2)$ as:
\begin{equation}
\Pi_{ij}^{\mu \nu}\left(q^2\right) = g^{\mu \nu} \Pi_{ij}\left(q^2\right) - i q^\mu q^\nu \Delta_{ij}\left(q^2\right),
\end{equation}
where the quantities $\Pi_{ij}(q^2)$ represent the vacuum polarization amplitudes, with $i, j = 0, 1, 3$ corresponding to the $B$, $W_1$, and $W_3$ gauge bosons, respectively. The effects of new physics can be captured by expanding $\Pi_{ij}(q^2)$ in powers of $q^2$, truncated to linear order~\cite{Peskin:1991sw}:
\begin{equation}
\Pi_{ij}\left(q^2\right) = \left.\Pi_{ij}\left(q^2\right)\right|_{q^2=0} + q^2 \frac{d}{dq^2}\left.\Pi_{ij}\left(q^2\right)\right|_{q^2=0} + \mathcal{O}\left(q^4\right).
\end{equation}

The oblique corrections, parametrized by the well-known quantities $S$, $T$, and $U$, are defined as follows \cite{Peskin:1991sw,Altarelli:1990zd,Barbieri:2004qk}:
\begin{equation}
\begin{aligned}
S & = -\left.\frac{4 \cos\theta_W \sin\theta_W}{\alpha_{\mathrm{em}}} \frac{d}{dq^2} \Pi_{30}\left(q^2\right)\right|_{q^2=0}, \\
U & = \left.\frac{4 \sin\theta_W}{\alpha_{\mathrm{em}}} \frac{d}{dq^2}\left[\Pi_{11}\left(q^2\right) - \Pi_{33}\left(q^2\right)\right]\right|_{q^2=0}, \\
T & = \left.\frac{1}{\alpha_{\mathrm{em}} M_W^2}\left[\Pi_{11}\left(0\right) - \Pi_{33}\left(0\right)\right]\right|_{q^2=0}.
\end{aligned}
\end{equation}
were $\alpha_\text{em}$ is the electromagnetic fine structure constant and $\theta_W$ is the Weinberg angle. The calculations for the $2$HDM and $3$HDM models are extensively discussed in the literature \cite{CarcamoHernandez:2015mkh,CarcamoHernandez:2024ycd}, as well as for the generalized case of NHDM \cite{CarcamoHernandez:2015smi}, which considers a specific scenario where the rotation matrices for pseudoscalars and charged scalars are identical. In our model, as mentioned at the beginning of this section, the new fields associated with $\Sigma$, a bidoublet of $SU(2)_2\times SU(2)_1$, introduce additional couplings to the gauge bosons at the 1-loop level, thereby modifying the values of the oblique parameters. For these computations, we use the expressions derived in \cite{CarcamoHernandez:2023dyz,Grimus:2008nb}
\begin{align}
S \simeq & \ 
\frac{1}{12 \pi}
\sum_{i=1}^3 \sum_{k=1}^2 \left[\left(R_H\right)_{k i}\left(R_A\right)_{k 1}\right]^2 K_1\left(m_{H_i}^2, m_{\xi}^2, m_{\eta^{ \pm}}^2\right),\\
U \simeq & -S+\sum_{a=1}^{3}\sum_{k=1}^2  \left[ \left(R_{A}\right)_{a 1}\left(R_{C}\right)_{a k}\right]^2 G_3\left(m_{\xi}, m_{\eta^{\pm}}\right)  +\sum_{a=1}^3 \sum_{i=1}^3 \sum_{k=1}^2 \left[\left(R_H\right)_{a i}\left(R_{C}\right)_{a k}\right]^2 G_3\left(m_{H_i}, m_{\eta^{\pm}}\right),\\
\nn T \simeq & \ t_0 \left[ 
\sum_{k=1}^2 m_{\eta^{ \pm}}^2
+\sum_{a=1}^{3} \sum_{i=1}^3 \left[ \left(R_H\right)_{a i} \left(R_{A}\right)_{a 1}\right]^2 F_3\left(m_{H_i}, m_{\xi}\right)
\right. - \sum_{a=1}^{3}\sum_{i=1}^3 \sum_{k=1}^2  \left[\left(R_H\right)_{a i}\left(R_{C}\right)_{a k}\right]^2 F_3\left(m_{H_i}, m_{\eta^{\pm}}\right) \\
& \left. -  \sum_{a=1}^{3} \sum_{k=1}^2 \left[\left(R_{A}\right)_{a i}\left(R_{C}\right)_{a k}\right]^2 F_3\left(m_{\xi}, m_{\eta^{ \pm}}\right) \right]   ,
\end{align} 
where $t_0=\left( \pi^2 v_\phi^2 \alpha_{\text{em}}\left(M_Z\right)\right)^{-1}$, and $R_C$, $R_H$ and $R_A$ are the mixing matrices for the charged scalar fields, neutral scalar and pseudoscalars, respectively. 
Regarding to the masses of the scalar particles $H_i$ they correspond to $\Sigma_1$, $\Sigma_2$ and $H^0_3$ for $i=1,2,3$.
Furthermore, the following loop functions were introduced in \cite{CarcamoHernandez:2015smi,CarcamoHernandez:2015mkh}
\begin{eqnarray}
F_3\left(m_1,m_2\right)&= & \frac{m_1^2 m_2^2}{m_1^2-m_2^2} \ln \left(\frac{m_1^2}{m_2^2}\right),\\
G_3\left(m_1, m_2\right) &= & \frac{-5 m_1^6+27 m_1^4 m_2^2-27 m_1^2 m_2^4+6\left(m_1^6-3 m_1^4 m_2^2\right) \ln \left(\frac{m_1^2}{m_2^2}\right)+5 m_2^6}{6\left(m_1^2-m_2^2\right)^3},\\
K_1\left(m_1, m_2, m_3\right) & = & \frac{1}{\left(m_2^2-m_1^2\right)^3}\left\{m_1^4\left(3 m_2^2-m_1^2\right) \ln \left(\frac{m_1^2}{m_3^2}\right)-m_2^4\left(3 m_1^2-m_2^2\right) \ln \left(\frac{m_2^2}{m_3^2}\right)\right. \nonumber \\
&& \left.-\frac{1}{6}\left[27 m_1^2 m_2^2\left(m_1^2-m_2^2\right)+5\left(m_2^6-m_1^6\right)\right]\right\}.
\end{eqnarray}

\begin{figure}
    \centering
\subfigure[]{\includegraphics[width=0.49\linewidth]{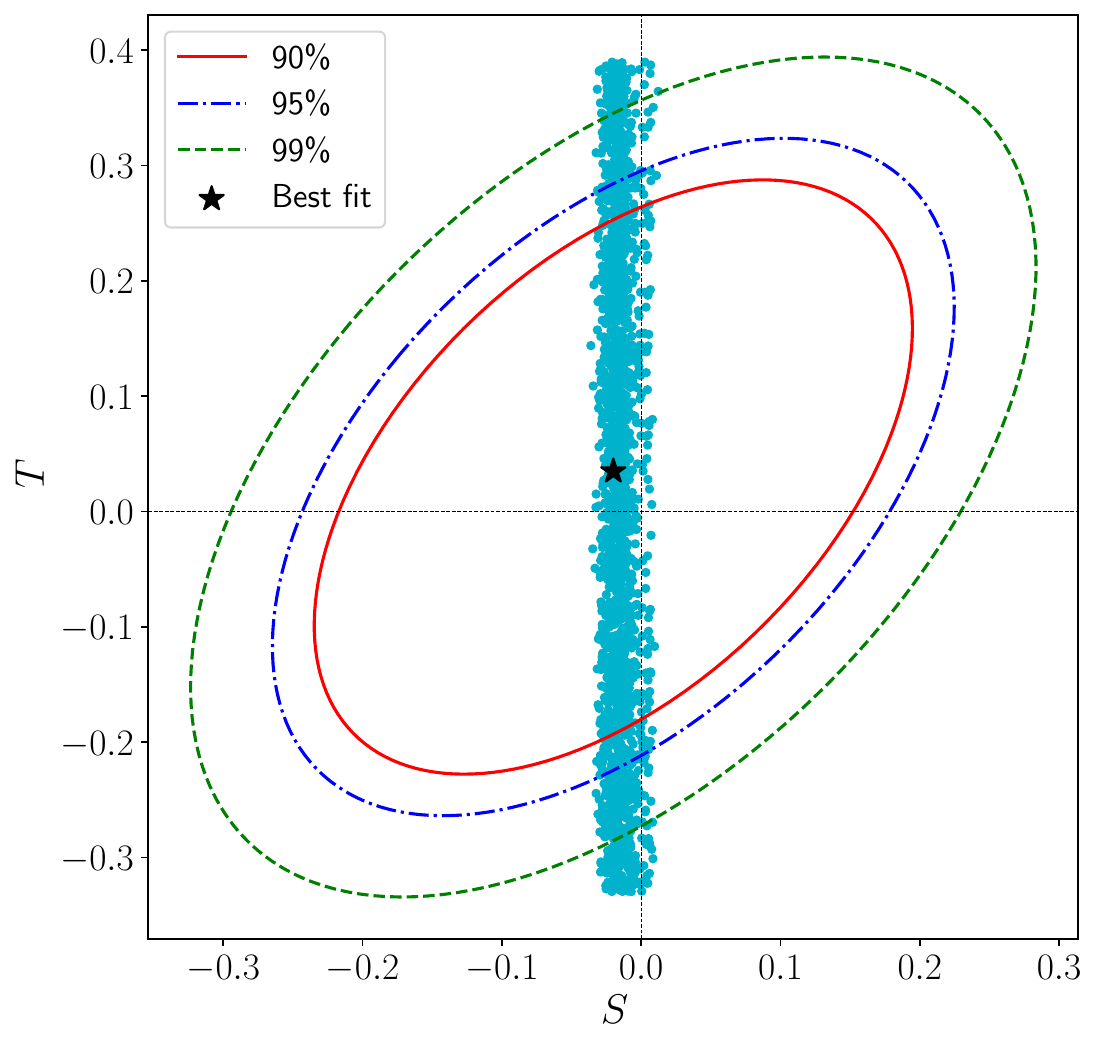}}\quad
\subfigure[]{\includegraphics[width=0.49\linewidth]{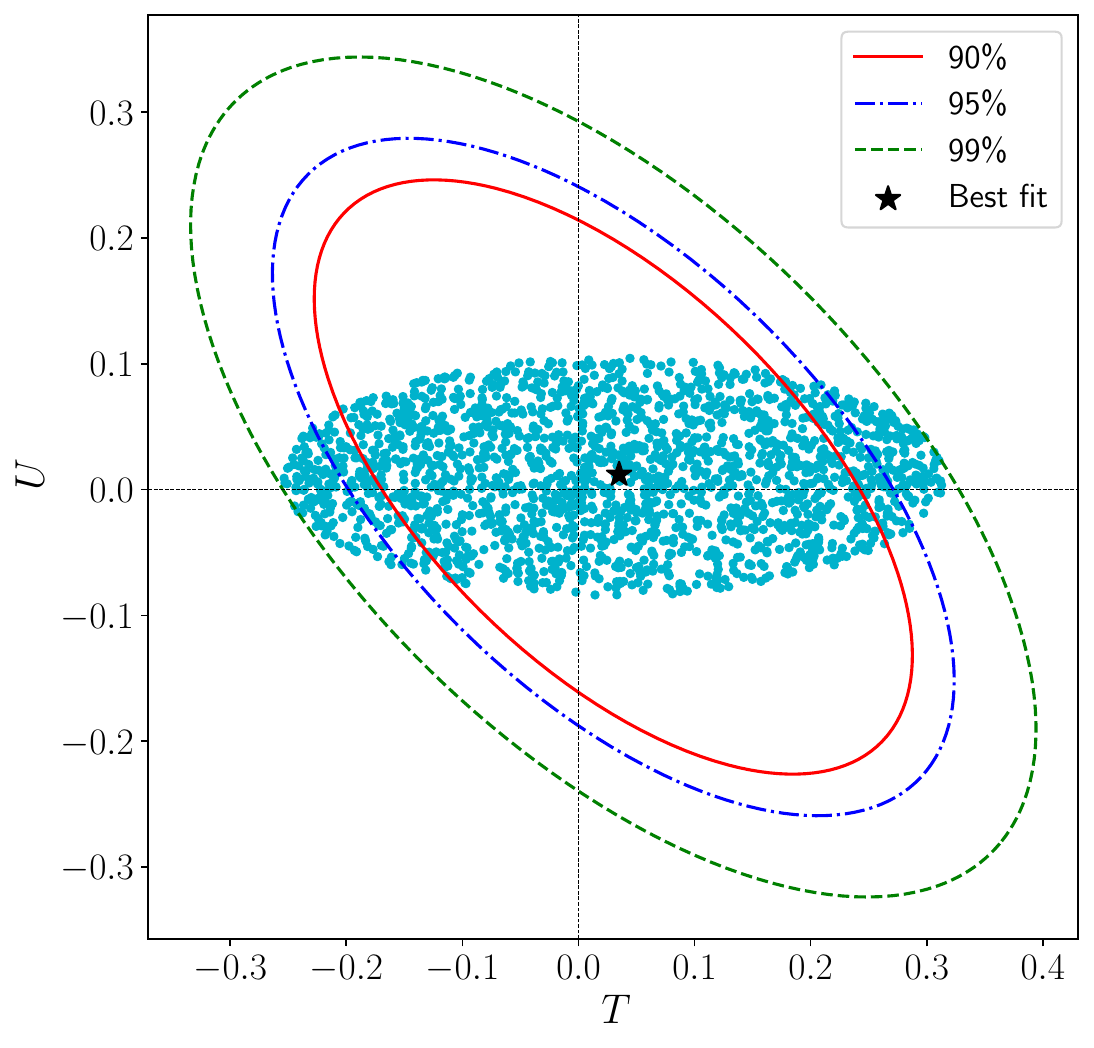}}
\caption{Allowed regions for the oblique parameters in the $S$-$T$ (left panel) and $U$-$T$ (right panel) planes. The contours represent confidence levels at $90\%$, $95\%$, and $99\%$, as indicated in the legend. The cyan points correspond to the model predictions, while the black star marks the best fit point, representing the values that minimize the $\chi^2$ function.}
\label{fig:oblicue}
\end{figure}

In the numerical analysis, Figure~\ref{fig:oblicue} displays the allowed regions in the oblique parameter spaces $S$-$T$ (left panel) and $U$-$T$ (right panel) at different confidence levels. The contours represent the $90\%$, $95\%$, and $99\%$ confidence levels, as indicated in the legend, and the cyan points correspond to the predictions of our model. These predictions include the contributions of the scalar bidoublet fields and their mixing effects, which play a critical role in modifying the vacuum polarization functions $\Pi_{ij}$ that define the oblique parameters $S$, $T$, and $U$.

The black star in both panels marks the best-fit point obtained from our model, corresponding to $S = -0.0200 \pm 0.0001$, $T = 0.035 \pm 0.007$, and $U = 0.012 \pm 0.003$, with a $\chi^2$ value of $6.9 \times 10^{-3}$. This best-fit point is consistent with the current experimental limits reported in Ref.~\cite{Workman:2022ynf}, which are $S_{\exp} = -0.02 \pm 0.1$, $T_{\exp} = 0.03 \pm 0.12$, and $U_{\exp} = 0.01 \pm 0.11$. Notably, the best-fit values lie within $3\sigma$ of the experimental bounds, demonstrating that the proposed model aligns well with existing experimental constraints.

In summary, the analysis of the oblique parameters $S$, $T$, and $U$ shows that the proposed model is consistent with current experimental limits, highlighting its viability as an extension of the SM. The results, particularly the best-fit point, emphasize the capability of the model's scalar sector to reproduce electroweak precision observables. Moreover, the contributions of the exotic scalar fields and their mixing effects provide a coherent and predictive explanation for deviations in electroweak precision measurements.

\section{A possible interpretation of the $95$ GeV diphoton excess}
\label{sec:95gev}

The CMS collaboration has recently reported an excess of events around $95$ GeV in the diphoton invariant mass spectrum, with a local significance approaching $3\sigma$~\cite{CMS:2018cyk,CMS:2023yay}. This intriguing anomaly, if confirmed, would suggest the existence of a new neutral scalar boson lighter than the $125$ GeV Higgs. Phenomenologically, such a signal would be highly relevant, representing the first direct evidence of an extended Higgs sector. 
Several theoretical frameworks—such as Left–Right symmetric models~\cite{Dev:2023kzu,Bonilla:2023wok} and Two-Higgs-Doublet Models~\cite{BrahimAit-Ouazghour:2024img,Sharma:2024vhv,Banik:2024ugs,Belyaev:2023xnv}—have been proposed to explain this signal. Consequently, it is essential to examine whether our strongly coupled iHDM can naturally reproduce the observed $95$ GeV diphoton excess.
In our 
model, there is an attractive candidate for this new state. Out of our benchmark scenario, we can identify the neutral scalar $H_3^0$ as the resonance responsible for the signal reported by CMS. It is worth mentioning that the scalar potential of our model has a good amount of parametric freedom that allows to successfully acommodate a physical CP even scalar with a mass of $95$ GeV. We denote the CP even scalar state responsible for the $95$ GeV diphoton excess, as $H_{95}$, reflecting its mass. In the following analysis, we will demonstrate that the $H_{95}$ state in the our model can indeed produce a diphoton signal of the observed magnitude without conflicting with current experimental constraints.

To quantify the diphoton signal of $H_{95}$, we define the signal strength in the $\gamma\gamma$ channel as the ratio between the production cross section times the decay branching fraction and that of a SM-like Higgs boson taken as reference. Specifically, focusing on gluon fusion production and the diphoton decay channel, we have
\begin{equation}\label{eq:95GeV}
\mu\left(H_{95}\right)_{\gamma \gamma}=\frac{\sigma (gg \to H_{95})}{\sigma_{\tx{SM}}(gg \to h)} \times  
\frac{\tx{Br}\left( H_{95} \to \gamma \gamma\right)}{\tx{Br} \left(h \to \gamma \gamma \right) },
\end{equation}
where $\sigma(gg\to H_{95})$ is the production cross section of the $95$ GeV scalar via gluon fusion, $\text{Br}(H_{95}\to\gamma\gamma)$ is its diphoton branching fraction, and the denominator corresponds to SM values for the Higgs boson.

The experimental measurement for the signal strength of the observed excess at $95$ GeV reported by CMS is~\cite{CMS:2023yay,Biekotter:2023jld}:
\begin{equation}\label{eq:95GeVEXP}
\mu_{\gamma \gamma}^{\text{exp}} = 0.35 \pm 0.12.
\end{equation}

Unlike the Standard Model, where the process $h\to \gamma\gamma$ receives loop contributions from $W^{\pm}$ gauge bosons and the top quark, in our model the scalar $H_{95}$ does not possess these tree-level couplings. Nevertheless, it can decay into two photons via loop contributions involving new particles. One of the dominant contributions arises from the vector-like sector; specifically, the model includes heavy quarks $T_n$ ($n=1,2$), $B_i$, and heavy leptons $E_i$ ($i=1,2,3$), which couple to the scalar field associated with $\sigma$ through Yukawa operators shown in Eq.~\eqref{eq:opeYukawa}. The corresponding loop contribution is analogous to the one arising from the top-quark loop in the SM Higgs decay into two photons, i.e., it is proportional to the squared electric charge of the fermion $Q_f^2$ and the Yukawa couplings $y_{T_n}$, $y_{B_i}$ and $y_{E_i}$ ($\sim M_{f_i}/v_{\sigma}$). In the strongly coupled regime or for small values of $v_\sigma$, these loop contributions become positive and significant. Another significant contribution originates from loop effects arising from the extended scalar sector, involving the virtual exchange of the electrically charged scalar fields $\eta^{\pm}$. The spontaneous symmetry breaking yields the trilinear scalar coupling $C_{\sigma \eta^\pm \eta^\mp}$, then giving rise to an interaction of the form $H_{95} \eta^{+}\eta^{-}$, whose corresponding coupling can be expressed as:
\begin{equation}
C_{\sigma \eta^{\pm} \eta^{\mp}} =  \left.\frac{\partial^3V_{\text{Phys}}}{\partial \sigma \partial \eta^\pm\partial \eta^\mp}\right|_{\text{Fields}=0}, \quad k=1,2.
\end{equation}

The magnitude of this contribution to the $95$ GeV diphoton signal is inversely proportional to the square of the masses of the charged scalar fields and directly proportional to the aforementioned trilinear scalar coupling. It is important to note that the scalar loops carry an opposite sign compared to the fermionic loops, provided that the trilinear scalar coupling is positive then implying that in this case 
their net effect strongly depends on the model parameters. In Refs~\cite{Bonilla:2023wok,Huong:2025uwx}, detailed expressions are provided for the cross section of the diphoton scalar resonance production, as well as for the corresponding decay widths of the resonance into photon and gluon pairs, within BSM theories containing contributions from an extended scalar and exotic charged vector-like fermion sectors.

\begin{figure}
\centering
\subfigure[]{\includegraphics[width=0.49\linewidth]{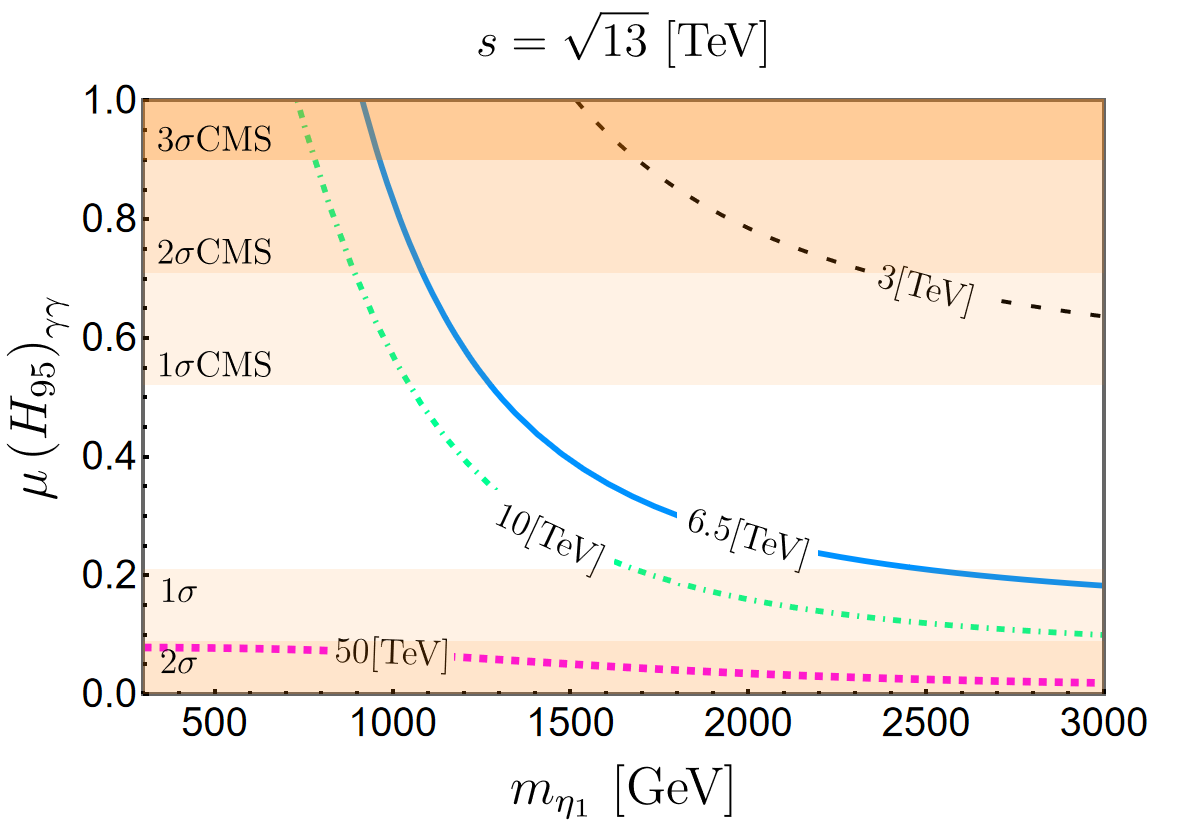}}
\subfigure[]{\includegraphics[width=0.49\linewidth]{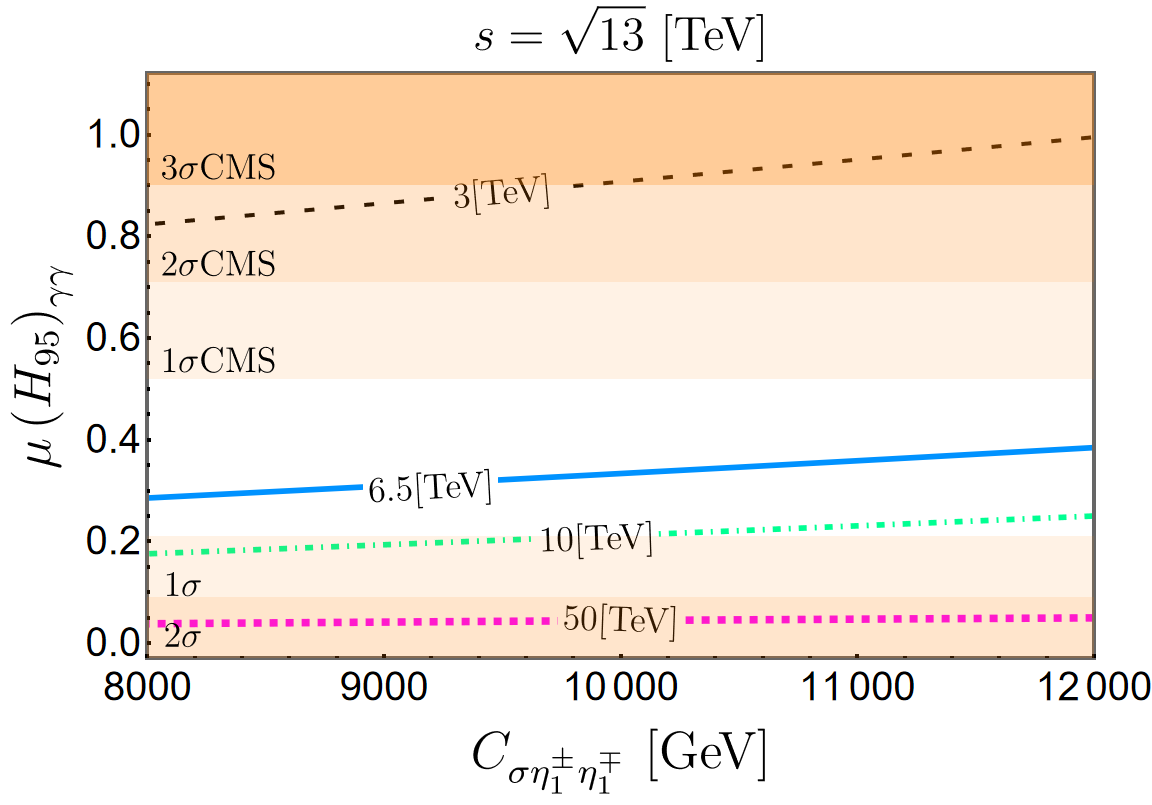}}
\caption{Predictions for the $H_{95}$ boson signal in the $\gamma\gamma$ channel. 
The signal strength $\mu(H_{95})_{\gamma\gamma}$ is shown as a function of (a) the mass $m_{\eta_1}$ and (b) the coupling coefficient $C_{\sigma \eta_1^\pm \eta_1^\mp}$. The shaded orange bands represent the exclusion regions at $1\sigma$, $2\sigma$, and $3\sigma$ from the CMS analysis. The values indicated on each curve correspond to signal predictions for different values of $v_\sigma$.}
\label{fig:95GeV1}
\end{figure}

Figure~\ref{fig:95GeV1} summarizes the behavior of the signal strength $\mu(H_{95})_{\gamma\gamma}$ as a function of the charged scalar mass $m_{\eta_1}$ and the effective coupling 
$H_{95}\eta_1^+\eta_1^-$. The shaded orange bands indicate the experimental $1\sigma$, $2\sigma$, and $3\sigma$ regions reported by the CMS collaboration for the mass window around $95$ GeV.
Figure~\ref{fig:95GeV1}(a) illustrates that, for most of the viable parameter space, the charged scalar loop interferes constructively with fermionic loops, enhancing the branching ratio $\text{Br}(H_{95}\to \gamma\gamma)$. However, the scalar loop effect rapidly decreases as the scalar mass increases. Additionally, the impact of $v_\sigma$ shows that for smaller values around $3$–$4$ TeV, significantly larger values of $\mu_{\gamma\gamma}$ are obtained compared to the range of $6$–$10$ TeV for the same scalar mass. Within this latter range, there exists a favorable parameter space consistent with the CMS $1\sigma$ region. Meanwhile, the lower curve (corresponding to larger $v_\sigma$ values) remains below $\mu_{\gamma\gamma}\sim 0.2$, within the $2\sigma$ regime, indicating that the model would yield a low signal strength in the weak coupling regime. Overall, this figure highlights a substantial parameter region for moderately low masses of the scalar $\eta_1^{\pm}$ and $v_{\sigma}$ values consistent with the $95$ GeV excess.
The curves in Figure~\ref{fig:95GeV1}(b) corroborate the parameter-space scenario described in Figure~\ref{fig:95GeV1}(a). We observe that the signal strength increases monotonically with the trilinear scalar coupling $C_{\sigma \eta^\pm \eta^\mp}$. A larger coupling implies a greater scalar-loop contribution to the amplitude of $H_{95}\to \gamma\gamma$, thus enhancing the decay branching fraction. If the trilinear interaction is weak, the signal contribution remains small, with the lower baseline arising from fermionic loops. In our model, it is quite natural to have a moderately large trilinear coupling; hence, achieving the required $\mu_{\gamma\gamma}$ does not constitute a stringent condition, but rather represents a generic possibility within the strongly coupling regime. Consequently, the viable parameter space for obtaining a significant diphoton signal is reinforced.

For the numerical calculations shown in Figure~\ref{fig:95GeV1}, we use a center-of-mass energy of $\sqrt{s}=13\,\mathrm{TeV}$, corresponding to the LHC conditions. In the factorization process, we adopt the scale $\mu=m_{H_{95}}$, and for the total cross section we employ the MSTW next-to-leading-order (NLO) gluon distribution functions~\cite{Martin:2009iq}, together with the running strong coupling constant $\alpha_s(M_Z) = 0.1179 \pm 0.0010$.

\begin{figure}
\centering
\subfigure[]{\includegraphics[width=0.49\linewidth]{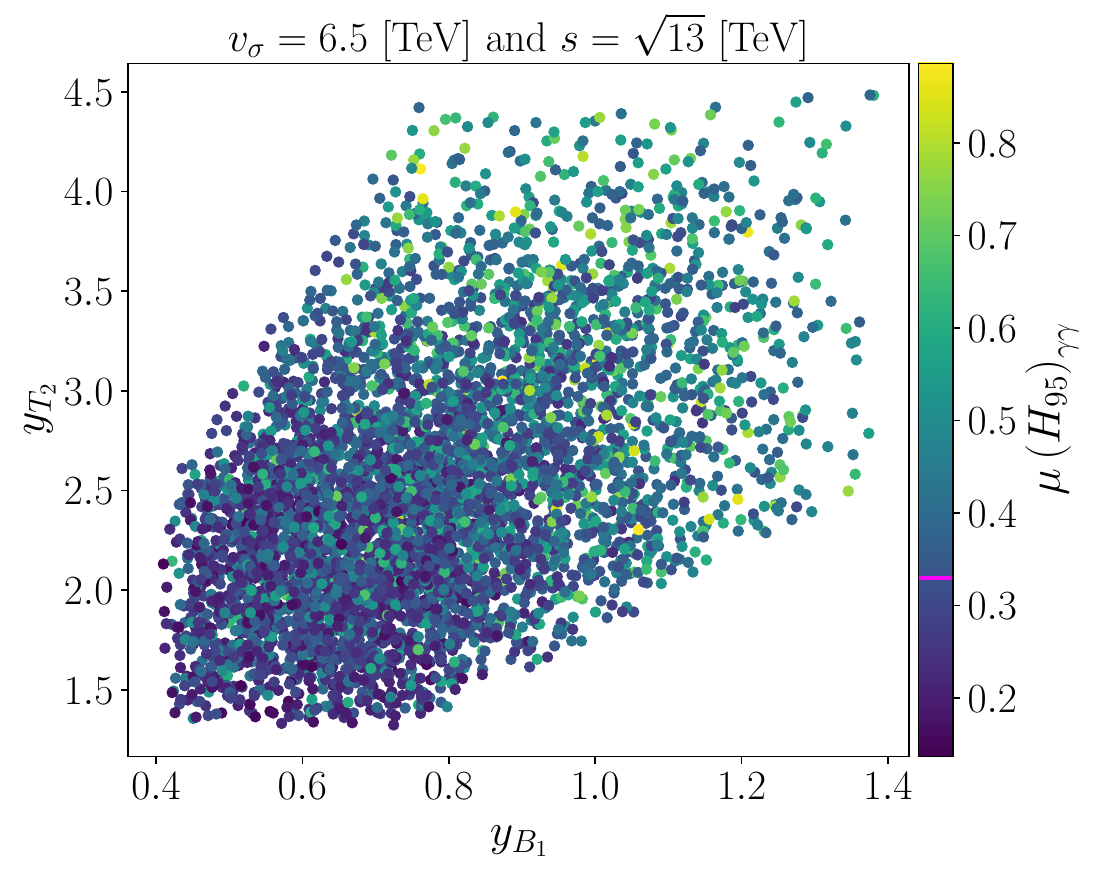}}
\subfigure[]{\includegraphics[width=0.49\linewidth]{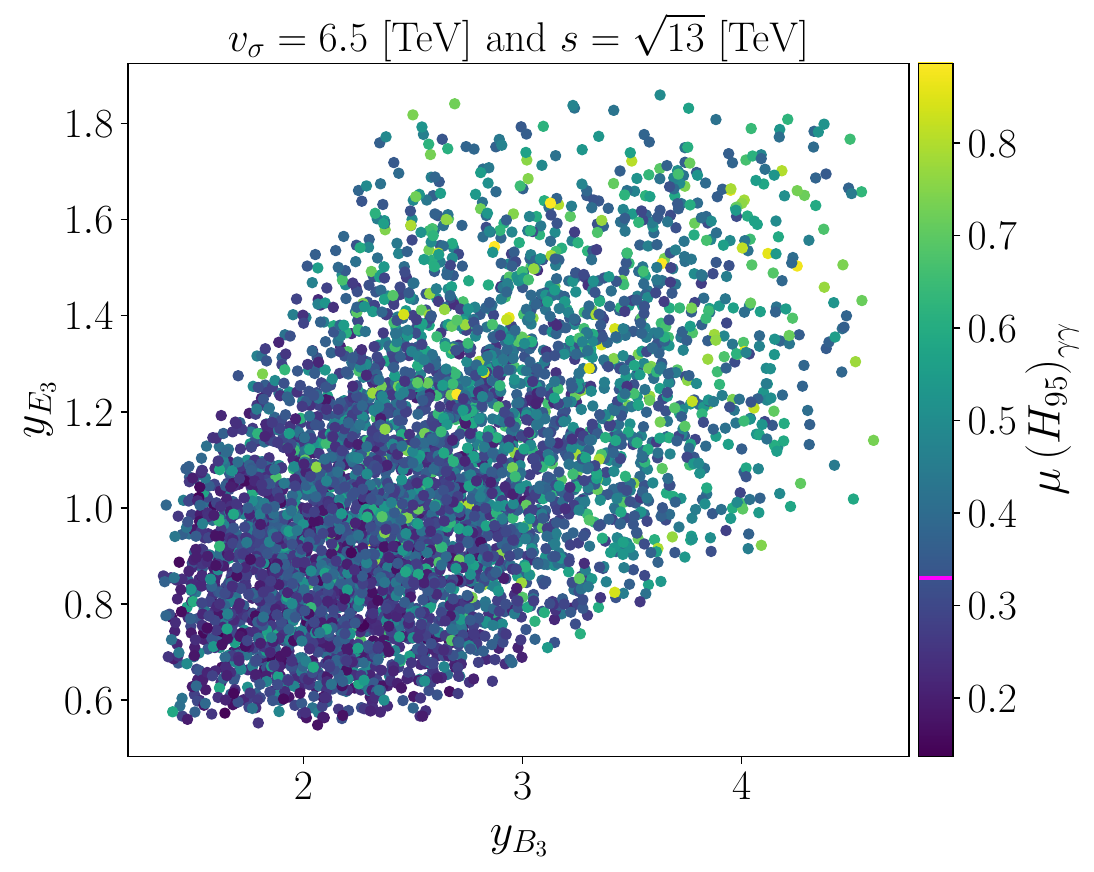}}
\caption{Signal strength $\mu(H_{95})_{\gamma\gamma}$ in the parameter space of the vector-like Yukawa couplings. Panel (a) shows the plane of Yukawa couplings $y_{B_1}$ versus $y_{T_2}$, while panel (b) displays the plane $y_{B_3}$ versus $y_{E_3}$. The color bars indicate the magnitude of $\mu(H_{95})_{\gamma\gamma}$, and the horizontal magenta line represents the central value of the CMS measurement at $95$ GeV.}
    \label{fig:95GeV2}
\end{figure}

Figure~\ref{fig:95GeV2} provides a broader perspective by depicting regions of the vector-like Yukawa coupling space that yield a given $\mu(H_{95})_{\gamma\gamma}$. Figure~\ref{fig:95GeV2}(a) shows the $y_{B_1}$ versus $y_{T_2}$ plane consistent with the observed $\mu_{\gamma \gamma}$, where the horizontal magenta line on the color bar corresponds to the central CMS measurement value given by Eq.~\eqref{eq:95GeVEXP}. Similarly, Figure~\ref{fig:95GeV2}(b) presents the $y_{B_3}$ versus $y_{E_3}$ plane, with the color bar indicating $\mu(H_{95})_{\gamma\gamma}$ as before. Viable regions for $\mu_{\gamma\gamma}$ in both panels correspond to Yukawa coupling values in the range of $1$–$3$, comfortably within the perturbative regime. Thus, our model can accommodate the large loop contributions required without entering a non-perturbative dynamic regime.

In summary, our model provides a viable interpretation for the $95$ GeV diphoton excess without resorting to excessively large couplings or conveniently chosen bound states. The interaction naturally arises through the interplay between the VEV of $\sigma$ and the masses of the vector-like fermions. This explanation is consistent with the current experimental data and simultaneously leads to a significant set of predictions for future studies.

\section{\label{sec::8}Summary and Conclusions}

We have constructed an
extension of the SM based on the gauge structure $SU(2)_2 \times SU(2)_1 \times U(1)_Y$, incorporating a strongly coupled inert scalar sector and discrete symmetries. The scalar sector includes two inert Higgs doublets, the SM Higgs doublet, a bidoublet from the $SU(2)_2 \times SU(2)_1$ sector, and a real scalar singlet responsible for the discrete symmetry breaking. This framework 
successfully addresses key issues such as the hierarchy problem and provides exotic composite states, including massive vector resonances. A novel feature of the model is the inclusion of  
the exotic scalars in the radiative generation of neutrino masses and universal seesaw mechanisms to yield the masses of the SM charged fermions lighter than the top quark, effectively resolving the SM fermion puzzle.  
The model successfully reproduces the mass spectra and mixing parameters of quarks and leptons within experimental uncertainties as can be seen in Sec. \ref{sec::4}.

The study of the trilinear Higgs self-coupling reveals significant deviations from the SM expectation. The interplay between the extended scalar sector introduces sizeable radiative corrections to the Higgs potential, leading to observable modifications to the trilinear self-coupling of the SM-Higgs. Our work indicates that, despite these corrections, the model remains consistent with current experimental constraints as can be seen in Figure \ref{fig:trilineargraphs}. Moreover, the parameter space allows for predictions that can be probed in upcoming collider experiments, particularly through precision measurements of the SM-Higgs self-coupling. This provides a crucial test of the underlying strong dynamics and its potential role in electroweak symmetry breaking. 

The model predicts significant contributions to the Higgs diphoton decay rate, $\Gamma_{h\rightarrow\gamma\gamma}$, 
arising from the one loop level exchange of charged scalars and heavy vector resonances. These contributions modify the decay amplitude through new quantum corrections, resulting in measurable deviations from the SM predictions. The computed Higgs diphoton signal strength $R_{\gamma\gamma}$ helps to constrain the scalar sector parameter space and aligns well with experimental values as is shown in Table \ref{tab:numerical scalars}. Additionally, this observable is stable against variations of the
effective $\beta$ parameter associated with the internal dynamics of the strong sector.

Furthermore, LFV processes, such as $\mu \to e\gamma$, $\mu \to 3e$, and $\mu-e$ conversion in a nucleai, are naturally induced at one loop level by interactions among inert scalar doublets, the bidoublet, and heavy neutrinos. Exploring the involved particle masses yields branching ratios within the reach of current and future experiments as indicated in Figure \ref{fig:lfv2}, thus providing robust tests for the model.

Our model provides a viable leptogenesis mechanism that exploits the interactions between exotic scalars and heavy neutrinos to generate a leptonic asymmetry in the early Universe. Notably, the partial dominant contribution to the CP asymmetry arises from the decay of the $\Sigma_2$ scalar, which plays a crucial role in lepton number violation through its interactions with heavy Majorana neutrinos in out-of-equilibrium decay processes.

The analysis of the baryon asymmetry evolution in the strong washout regime indicates that the baryon asymmetry parameter, $\Delta Y_B$, can attain values consistent with the Planck collaboration measurements. This result demonstrates that our model successfully accounts for the origin of the observed matter-antimatter asymmetry in the Universe via a leptogenesis process.

Contributions to the oblique $S$, $T$, and $U$ parameters from the extended scalar sector are consistent with experimental constraints. Mixing effects among bidoublet components introduce distinctive modifications, further reinforcing the viability of the model.

The model can naturally accommodate the diphoton excess at $95$ GeV reported by CMS, through the interplay of large scalar couplings and the interaction between the scalar singlet $\sigma$ and the vector-like fermions responsible for the universal seesaw mechanism. One of the exotic scalars in the spectrum can be used 
to reproduce the observed excess within current experimental bounds.

This model provides a unified framework addressing critical issues in modern particle physics, linking scalar sector extensions to neutrino masses, LFV processes, and precision measurements of the Higgs and electroweak sectors. Predicted deviations in scalar and fermionic observables, influenced by new charged scalars and vector resonances, stand out as distinctive features, offering testable benchmarks for next-generation experiments.

Future efforts should explore other observables, such as 

gravitational wave signals. The consistency of the model with current data and novel predictions underscores its relevance as a strong candidate for BSM physics.

\section*{Acknowledgments}
The authors sincerely thank M. Mora-Urrutia for her contributions during the early stages of the project and for her participation in discussions. D.SA gratefully acknowledges M. Solis Miquio for his assistance with Python. This research was supported by ANID-Chile under grants PIA/APOYO AFB230003, FONDECYT 1210131, FONDECYT 1210378, FONDECYT 1241855, and Millennium Nucleus Program grant ICN2019\_044.
\appendix

\section{The full scalar potential}
\label{app:a}

For our model, the full scalar potential can be written as

\begin{equation}
V_{\text{full}} = V + \frac{1}{\Lambda^2}V_6
\end{equation}
where $V$ to correpond Eq.~\eqref{eq:potential} and $V_6$ is the six dimensional non-renormalizable potential, given by

\begin{align}
V_6  = &  \alpha_1\left|\phi^{\dagger} \Sigma h_1\right|^2+\alpha_2\left|\phi^{\dagger} \Sigma h_2\right|^2+\alpha_3\left[\left(\phi^{\dagger} \Sigma h_1\right)^2+\left(h_1^{\dagger} \Sigma^{\dagger} \phi\right)^2+\text {H.c.}\right]
+\alpha_4\left[\left(\phi^{\dagger} \Sigma h_2\right)^2+\left(h_2^{\dagger} \Sigma^{\dagger} \phi\right)^2+\text {H.c.}\right]\nn\\
&+\alpha_5\left(\phi \phi^{\dagger}\right)^3+\alpha_6\left(\phi \phi^{\dagger}\right)^2\left(h_1 h_1^{\dagger}\right)+\alpha_7\left(\phi \phi^{\dagger}\right)^2\left(h_2 h_2^{\dagger}\right) +\alpha_8\left(\phi \phi^{\dagger}\right)^2\left(h_1 h_2{ }^{\dagger}+\text {H.c.}\right)+\alpha_9\left(\phi \phi^{\dagger}\right)^2 \operatorname{Tr}\left(\Sigma \Sigma^{\dagger}\right)\nn\\
&+\alpha_{10}\left(\phi \phi^{\dagger}\right)^2(\sigma \sigma)+\alpha_{11}\left(\phi \phi^{\dagger}\right)\left(h_1 h_1{ }^{\dagger}\right)^2 +\alpha_{12}\left(\phi \phi^{\dagger}\right)\left[\left(h_1 h_1{ }^{\dagger}\right)\left(h_2 h_2{ }^{\dagger}\right)+\left(h_1 h_2{ }^{\dagger}+\text {H.c.}\right)^2\right]\nn\\
&+\alpha_{13}\left(\phi\phi^{\dagger}\right)\left(h_1 h_1{ }^{\dagger}\right)\left(h_1 h_2{ }^{\dagger}+\text {H.c.}\right)+\alpha_{14}\left(\phi \phi^{\dagger}\right)\left(h_1 h_1^{\dagger}\right) \operatorname{Tr}\left(\Sigma \Sigma^{\dagger}\right)+\alpha_{15}\left(\phi \phi^{\dagger}\right)\left(h_1 h_1^{\dagger}\right)(\sigma \sigma)\nn\\
 &+\alpha_{16}\left(\phi \phi^{\dagger}\right)\left(h_2 h_2^{\dagger}\right)^2 +\alpha_{17}\left(\phi \phi^{\dagger}\right)\left(h_2 h_2{ }^{\dagger}\right)\left(h_1 h_2{ }^{\dagger}+\text {H.c.}\right)+\alpha_{18}\left(\phi \phi^{\dagger}\right)\left(h_2 h_2{ }^{\dagger}\right) \operatorname{Tr}\left(\Sigma \Sigma^{\dagger}\right)\nn\\
&+\alpha_{19}\left(\phi \phi^{\dagger}\right)\left(h_2 h_2{ }^{\dagger}\right)(\sigma \sigma)+\alpha_{20}\left(\phi \phi^{\dagger}\right)\left(h_1 h_2{ }^{\dagger}+\text {H.c.}\right) \operatorname{Tr}\left(\Sigma \Sigma^{\dagger}\right)+\alpha_{21}\left(\phi \phi^{\dagger}\right)\left(h_1 h_2{ }^{\dagger}+\text {H.c.}\right)(\sigma \sigma)\nn\\
&+\alpha_{22}\left(\phi \phi^{\dagger}\right) \operatorname{Tr}\left(\Sigma \Sigma^{\dagger}\right)^2 +\alpha_{23}\left(\phi \phi^{\dagger}\right) \operatorname{Tr}\left(\Sigma \Sigma^{\dagger}\right)(\sigma \sigma)+\alpha_{24}\left(\phi \phi^{\dagger}\right)(\sigma \sigma)^2+\alpha_{25}\left(h_1 h_1^{\dagger}\right)^3+\alpha_{26}\left(h_1 h_1{ }^{\dagger}\right)^2(\sigma \sigma)\nn\\
&+\alpha_{27}\left(h_1 h_1^{\dagger}\right)\left[\left(h_1 h_1^{\dagger}\right)\left(h_2 h_2^{\dagger}\right)+\left(h_1 h_2^{\dagger}+\text {H.c.}\right)^2\right]+\alpha_{28}\left(h_1 h_1^{\dagger}\right)^2\left(h_1 h_2^{\dagger}+\text {H.c.}\right)+\alpha_{29}(\sigma \sigma)^3\nn\\
&+\alpha_{30}\left[\left(h_1 h_1{ }^{\dagger}\right)\left(h_2 h_2{ }^{\dagger}\right)+\left(h_1 h_2{ }^{\dagger}+\text {H.c. }\right)^2\right]\left(h_2 h_2{ }^{\dagger}\right)+\alpha_{31}\left(h_1 h_1{ }^{\dagger}\right)\left(h_1 h_2{ }^{\dagger}+\text {H.c.}\right) \operatorname{Tr}\left(\Sigma \Sigma^{\dagger}\right)\nn\\
&+\alpha_{32}\left[\left(h_1 h_1^{\dagger}\right)\left(h_2 h_2^{\dagger}\right)+\left(h_1 h_2^{\dagger}+\text {H.c.}\right)^2\right]\left(h_1 h_2^{\dagger}+\text {H.c.}\right)+\alpha_{33}\left(h_1 h_1{ }^{\dagger}\right) \operatorname{Tr}\left(\Sigma \Sigma^{\dagger}\right)^2+\alpha_{34} \operatorname{Tr}\left(\Sigma \Sigma^{\dagger}\right)^3\nn\\
& +\alpha_{35}\left[\left(h_1 h_1^{\dagger}\right)\left(h_2 h_2^{\dagger}\right)+\left(h_1 h_2^{\dagger}+\text {H.c.}\right)^2\right] \operatorname{Tr}\left(\Sigma \Sigma^{\dagger}\right) +\alpha_{36}\left[\left(h_1 h_1^{\dagger}\right)\left(h_2 h_2^{\dagger}\right)+\left(h_1 h_2{ }^{\dagger}+\text {H.c.}\right)^2\right](\sigma \sigma)\nn\\
&+\alpha_{37}\left(h_1 h_1{ }^{\dagger}\right)\left(h_1 h_2{ }^{\dagger}+\text {H.c.}\right)(\sigma \sigma)+\alpha_{38}\left(h_1 h_1{ }^{\dagger}\right)^2 \operatorname{Tr}\left(\Sigma \Sigma^{\dagger}\right)+\alpha_{39}\left(h_2 h_2^{\dagger}\right)^2\left(h_1 h_2^{\dagger}+\text {H.c.}\right)\nn\\
&+\alpha_{40}\left(h_1 h_1{ }^{\dagger}\right) \operatorname{Tr}\left(\Sigma \Sigma^{\dagger}\right)(\sigma \sigma) +\alpha_{41}\left(h_1 h_1^{\dagger}\right)(\sigma \sigma)^2+\alpha_{42}\left(h_2 h_2{ }^{\dagger}\right)^3+\alpha_{43}\left(h_2 h_2{ }^{\dagger}\right)^2 \operatorname{Tr}\left(\Sigma \Sigma^{\dagger}\right)\nn\\
& +\alpha_{44}\left(h_2 h_2^{\dagger}\right)^2(\sigma \sigma)+\alpha_{45}\left(h_2 h_2{ }^{\dagger}\right)\left(h_1 h_2{ }^{\dagger}+\text {H.c.}\right) \operatorname{Tr}\left(\Sigma \Sigma^{\dagger}\right)+\alpha_{46}\left(h_2 h_2{ }^{\dagger}\right)\left(h_1 h_2{ }^{\dagger}+\text {H.c.}\right)(\sigma \sigma)\nn\\
&+\alpha_{47}\left(h_2 h_2^{\dagger}\right) \operatorname{Tr}\left(\Sigma \Sigma^{\dagger}\right)^2+\alpha_{48}\left(h_2 h_2{ }^{\dagger}\right) \operatorname{Tr}\left(\Sigma \Sigma^{\dagger}\right)(\sigma \sigma)+\alpha_{49}\left(h_2 h_2{ }^{\dagger}\right)(\sigma \sigma)^2+\alpha_{50} \operatorname{Tr}\left(\Sigma \Sigma^{\dagger}\right)^2(\sigma \sigma)\nn\\
&+\alpha_{51}\left(h_1 h_2{ }^{\dagger}+\text {H.c.}\right) \operatorname{Tr}\left(\Sigma \Sigma^{\dagger}\right)^2+\alpha_{52}\left(h_1 h_2{ }^{\dagger}+\text {H.c.}\right) \operatorname{Tr}\left(\Sigma \Sigma^{\dagger}\right)(\sigma \sigma) +\alpha_{53}\left(h_1 h_2^{\dagger}+\text {H.c.}\right)(\sigma \sigma)^2\nn\\
&+\alpha_{54} \operatorname{Tr}\left(\Sigma \Sigma^{\dagger}\right)(\sigma \sigma)^2
\end{align}

\section{The full expressions for gauge boson masses}
\label{app:b2}

The full expressions for the gauge boson masses given in \eq{eq:gbosonmasses} are
\begin{equation}
 \begin{aligned}
    M_A^2=&\ 0,\\
    M_Z^2=&\frac{1}{8} \left(\left(1 + c^2 \right) g_{1}^2 + c^2 g_{2}^2 + g_{y}^2 - \sqrt{ \left(1 + c^2 \right)^2 g_{1}^4 + 2 \left(-1 + c^2 \right) g_{1}^2 \left(c^2 g_{2}^2 - g_{y}^2 \right) + \left(-c^2 g_{2}^2 + g_{y}^2 \right)^2}\right) v_{\phi}^2,\\
    M_{\rho_0}^2=&\frac{1}{8} \left( \left(1 + c^2 \right) g_{1}^2 + c^2 g_{2}^2 + g_{y}^2 + \sqrt{ \left(1 + c^2 \right)^2 g_{1}^4 + 2 \left(-1 + c^2 \right) g_{1}^2 \left(c^2 g_{2}^2 - g_{y}^2 \right) + \left(-c^2 g_{2}^2 + g_{y}^2 \right)^2}\right) v_{\phi}^2,\\
    M_{W^\pm}^2=&\frac{1}{8} \left( \left(1 + c^2 \right) g_{1}^2 + c^2 g_{2}^2 - \sqrt{ \left(1 + c^2 \right)^2 g_{1}^4 + 2 c^2 \left(-1 + c^2 \right) g_{1}^2 g_{2}^2 + c^4 g_{2}^4 }\right) v_{\phi}^2,\\
    M_{\rho^\pm}^2=&\frac{1}{8} \left(\left(1 + c^2 \right) g_{1}^2 + c^2 g_{2}^2 + \sqrt{ \left(1 + c^2 \right)^2 g_{1}^4 + 2 c^2 \left(-1 + c^2 \right) g_{1}^2 g_{2}^2 + c^4 g_{2}^4 }\right) v_{\phi}^2.
\end{aligned}   
\end{equation}

\section{The effective trilinear higgs coupling}
\label{app:b}

The contribution of the Feynman diagrams in Figure \ref{fig:trilinear} using dimensional regularization is:
\begin{align}
\Gamma^{1-\text{loop}}_{h^3}(q^2) = &\ \frac{1}{16 \pi^2}C_{hHH}^3C_0(m_h^2,m_h^2,q^2,m_H^2,m_H^2,m_H^2)+\frac{1}{32\pi^2}C_{hHH}C_{hhHH}\left[B_0(q^2,m_H^2,m_H^2)+2B_0(m_h^2,m_H^2,m_H^2)\right]\nn \\
    &+\frac{1}{16 \pi^2}C_{hAA}^3C_0(m_h^2,m_h^2,q^2,m_A^2,m_A^2,m_A^2)+\frac{1}{8 \pi^2}C_{hC C}^3C_0(m_h^2,m_h^2,q^2,{m_C}^2,{m_C}^2,{m_C}^2)\\
& +\frac{1}{32\pi^2}C_{hAA}C_{hhAA}\left[B_0(q^2,m_A^2,m_A^2)+2B_0(m_h^2,m_A^2,m_A^2)\right]\nn\\
& +\frac{1}{16\pi^2}C_{hC C}C_{hhC C}\left[B_0(q^2,{m_C}^2,{m_C}^2)\right. \left.+2B_0(m_h^2,{m_C}^2,{m_C}^2)\right]\nn,
\end{align}
where $H$, $A$ and $C$ are scalar, pseudoscalar and charged-scalar particles respectively and the functions $C_0$ and $B_0$ are the Pasarino-Veltman functions \cite{tHooft:1972tcz,tHooft:1978jhc,Passarino:1978jh}.

\bibliographystyle{utphys}
\bibliography{ref.bib}

\providecommand{\href}[2]{#2}\begingroup\raggedright\begin{thebibliography}{100}

\bibitem{Planck:2018vyg}
{\bfseries Planck} Collaboration, N.~Aghanim {\em et~al.}, ``{Planck 2018
  results. VI. Cosmological parameters},''
  \href{http://dx.doi.org/10.1051/0004-6361/201833910}{{\em Astron. Astrophys.}
  {\bfseries 641} (2020) A6}, \href{http://arxiv.org/abs/1807.06209}{{\ttfamily
  arXiv:1807.06209 [astro-ph.CO]}}. [Erratum: Astron.Astrophys. 652, C4
  (2021)].

\bibitem{Fields:2019pfx}
B.~D. Fields, K.~A. Olive, T.-H. Yeh, and C.~Young, ``{Big-Bang Nucleosynthesis
  after Planck},'' \href{http://dx.doi.org/10.1088/1475-7516/2020/03/010}{{\em
  JCAP} {\bfseries 03} (2020) 010},
  \href{http://arxiv.org/abs/1912.01132}{{\ttfamily arXiv:1912.01132
  [astro-ph.CO]}}. [Erratum: JCAP 11, E02 (2020)].

\bibitem{Super-Kamiokande:1998kpq}
{\bfseries Super-Kamiokande} Collaboration, Y.~Fukuda {\em et~al.}, ``{Evidence
  for oscillation of atmospheric neutrinos},''
  \href{http://dx.doi.org/10.1103/PhysRevLett.81.1562}{{\em Phys. Rev. Lett.}
  {\bfseries 81} (1998) 1562--1567},
  \href{http://arxiv.org/abs/hep-ex/9807003}{{\ttfamily arXiv:hep-ex/9807003}}.

\bibitem{SNO:2001kpb}
{\bfseries SNO} Collaboration, Q.~R. Ahmad {\em et~al.}, ``{Measurement of the
  rate of $\nu_e+d \to p+p+e^-$ interactions produced by $^8$B solar neutrinos
  at the Sudbury Neutrino Observatory},''
  \href{http://dx.doi.org/10.1103/PhysRevLett.87.071301}{{\em Phys. Rev. Lett.}
  {\bfseries 87} (2001) 071301},
  \href{http://arxiv.org/abs/nucl-ex/0106015}{{\ttfamily
  arXiv:nucl-ex/0106015}}.

\bibitem{Weinberg:1975gm}
S.~Weinberg, ``{Implications of Dynamical Symmetry Breaking},''
  \href{http://dx.doi.org/10.1103/PhysRevD.19.1277}{{\em Phys. Rev. D}
  {\bfseries 13} (1976) 974--996}. [Addendum: Phys.Rev.D 19, 1277--1280
  (1979)].

\bibitem{Susskind:1978ms}
L.~Susskind, ``{Dynamics of Spontaneous Symmetry Breaking in the Weinberg-Salam
  Theory},'' \href{http://dx.doi.org/10.1103/PhysRevD.20.2619}{{\em Phys. Rev.
  D} {\bfseries 20} (1979) 2619--2625}.

\bibitem{Gildener:1976ai}
E.~Gildener, ``{Gauge Symmetry Hierarchies},''
  \href{http://dx.doi.org/10.1103/PhysRevD.14.1667}{{\em Phys. Rev. D}
  {\bfseries 14} (1976) 1667}.

\bibitem{Yamada:2020bqe}
M.~Yamada, ``{Gauge hierarchy problem and scalegenesis},''
  \href{http://dx.doi.org/10.22323/1.376.0077}{{\em PoS} {\bfseries CORFU2019}
  (2020) 077}, \href{http://arxiv.org/abs/2004.00142}{{\ttfamily
  arXiv:2004.00142 [hep-ph]}}.

\bibitem{Contino:2009ez}
R.~Contino, ``{New Physics at the LHC: Strong vs. weak symmetry breaking},''
  \href{http://dx.doi.org/10.1393/ncc/i2009-10427-3}{{\em Nuovo Cim.}
  {\bfseries 32} (2009) 11--18},
  \href{http://arxiv.org/abs/0908.3578}{{\ttfamily arXiv:0908.3578 [hep-ph]}}.

\bibitem{Bando:1984ej}
M.~Bando, T.~Kugo, S.~Uehara, K.~Yamawaki, and T.~Yanagida, ``{Is rho Meson a
  Dynamical Gauge Boson of Hidden Local Symmetry?},''
  \href{http://dx.doi.org/10.1103/PhysRevLett.54.1215}{{\em Phys. Rev. Lett.}
  {\bfseries 54} (1985) 1215}.

\bibitem{Bando:1984pw}
M.~Bando, T.~Kugo, and K.~Yamawaki, ``{Composite Gauge Bosons and 'Low-energy
  Theorems' of Hidden Local Symmetries},''
  \href{http://dx.doi.org/10.1143/PTP.73.1541}{{\em Prog. Theor. Phys.}
  {\bfseries 73} (1985) 1541}.

\bibitem{Bando:1985rf}
M.~Bando, T.~Kugo, and K.~Yamawaki, ``{On the Vector Mesons as Dynamical Gauge
  Bosons of Hidden Local Symmetries},''
  \href{http://dx.doi.org/10.1016/0550-3213(85)90647-9}{{\em Nucl. Phys. B}
  {\bfseries 259} (1985) 493}.

\bibitem{Bando:1987ym}
M.~Bando, T.~Fujiwara, and K.~Yamawaki, ``{Generalized Hidden Local Symmetry
  and the A1 Meson},'' \href{http://dx.doi.org/10.1143/PTP.79.1140}{{\em Prog.
  Theor. Phys.} {\bfseries 79} (1988) 1140}.

\bibitem{Bando:1987br}
M.~Bando, T.~Kugo, and K.~Yamawaki, ``{Nonlinear Realization and Hidden Local
  Symmetries},'' \href{http://dx.doi.org/10.1016/0370-1573(88)90019-1}{{\em
  Phys. Rept.} {\bfseries 164} (1988) 217--314}.

\bibitem{Chivukula:2006cg}
R.~S. Chivukula, B.~Coleppa, S.~Di~Chiara, E.~H. Simmons, H.-J. He, M.~Kurachi,
  and M.~Tanabashi, ``{A Three Site Higgsless Model},''
  \href{http://dx.doi.org/10.1103/PhysRevD.74.075011}{{\em Phys. Rev. D}
  {\bfseries 74} (2006) 075011},
  \href{http://arxiv.org/abs/hep-ph/0607124}{{\ttfamily arXiv:hep-ph/0607124}}.

\bibitem{Barbieri:2009tx}
R.~Barbieri, A.~E. Carcamo~Hernandez, G.~Corcella, R.~Torre, and
  E.~Trincherini, ``{Composite Vectors at the Large Hadron Collider},''
  \href{http://dx.doi.org/10.1007/JHEP03(2010)068}{{\em JHEP} {\bfseries 03}
  (2010) 068}, \href{http://arxiv.org/abs/0911.1942}{{\ttfamily arXiv:0911.1942
  [hep-ph]}}.

\bibitem{Foadi:2007ue}
R.~Foadi, M.~T. Frandsen, T.~A. Ryttov, and F.~Sannino, ``{Minimal Walking
  Technicolor: Set Up for Collider Physics},''
  \href{http://dx.doi.org/10.1103/PhysRevD.76.055005}{{\em Phys. Rev. D}
  {\bfseries 76} (2007) 055005},
  \href{http://arxiv.org/abs/0706.1696}{{\ttfamily arXiv:0706.1696 [hep-ph]}}.

\bibitem{Ryttov:2008xe}
T.~A. Ryttov and F.~Sannino, ``{Ultra Minimal Technicolor and its Dark Matter
  TIMP},'' \href{http://dx.doi.org/10.1103/PhysRevD.78.115010}{{\em Phys. Rev.
  D} {\bfseries 78} (2008) 115010},
  \href{http://arxiv.org/abs/0809.0713}{{\ttfamily arXiv:0809.0713 [hep-ph]}}.

\bibitem{Sannino:2009za}
F.~Sannino, ``{Conformal Dynamics for TeV Physics and Cosmology},'' {\em Acta
  Phys. Polon. B} {\bfseries 40} (2009) 3533--3743,
  \href{http://arxiv.org/abs/0911.0931}{{\ttfamily arXiv:0911.0931 [hep-ph]}}.

\bibitem{CarcamoHernandez:2010qxf}
A.~E. Carcamo~Hernandez and R.~Torre, ``{A 'Composite' scalar-vector system at
  the LHC},'' \href{http://dx.doi.org/10.1016/j.nuclphysb.2010.08.004}{{\em
  Nucl. Phys. B} {\bfseries 841} (2010) 188--204},
  \href{http://arxiv.org/abs/1005.3809}{{\ttfamily arXiv:1005.3809 [hep-ph]}}.

\bibitem{CarcamoHernandez:2010wpm}
A.~E. Carcamo~Hernandez, ``{Top quark effects in composite vector pair
  production at the LHC},''
  \href{http://dx.doi.org/10.1140/epjc/s10052-012-2154-3}{{\em Eur. Phys. J. C}
  {\bfseries 72} (2012) 2154}, \href{http://arxiv.org/abs/1008.1039}{{\ttfamily
  arXiv:1008.1039 [hep-ph]}}.

\bibitem{CarcamoHernandez:2011pfx}
A.~E. Carcamo~Hernandez, {\em {Composite Vectors and Scalars in Theories of
  Electroweak Symmetry Breaking}}.
\newblock PhD thesis, Pisa, Scuola Normale Superiore, 2011.
\newblock \href{http://arxiv.org/abs/1108.0115}{{\ttfamily arXiv:1108.0115
  [hep-ph]}}.

\bibitem{Foadi:2012bb}
R.~Foadi, M.~T. Frandsen, and F.~Sannino, ``{125 GeV Higgs boson from a not so
  light technicolor scalar},''
  \href{http://dx.doi.org/10.1103/PhysRevD.87.095001}{{\em Phys. Rev. D}
  {\bfseries 87} no.~9, (2013) 095001},
  \href{http://arxiv.org/abs/1211.1083}{{\ttfamily arXiv:1211.1083 [hep-ph]}}.

\bibitem{Hapola:2011sd}
T.~Hapola and F.~Sannino, ``{Composite Higgs to two Photons and Gluons},''
  \href{http://dx.doi.org/10.1142/S0217732311036760}{{\em Mod. Phys. Lett. A}
  {\bfseries 26} (2011) 2313--2322},
  \href{http://arxiv.org/abs/1102.2920}{{\ttfamily arXiv:1102.2920 [hep-ph]}}.

\bibitem{Belyaev:2013ida}
A.~Belyaev, M.~S. Brown, R.~Foadi, and M.~T. Frandsen, ``{The Technicolor Higgs
  in the Light of LHC Data},''
  \href{http://dx.doi.org/10.1103/PhysRevD.90.035012}{{\em Phys. Rev. D}
  {\bfseries 90} (2014) 035012},
  \href{http://arxiv.org/abs/1309.2097}{{\ttfamily arXiv:1309.2097 [hep-ph]}}.

\bibitem{CarcamoHernandez:2013ydh}
A.~E. Carcamo~Hernandez, C.~O. Dib, and A.~R. Zerwekh, ``{The Effect of
  Composite Resonances on Higgs decay into two photons},''
  \href{http://dx.doi.org/10.1140/epjc/s10052-014-2822-6}{{\em Eur. Phys. J. C}
  {\bfseries 74} (2014) 2822}, \href{http://arxiv.org/abs/1304.0286}{{\ttfamily
  arXiv:1304.0286 [hep-ph]}}.

\bibitem{CarcamoHernandez:2017pei}
A.~E. C\'arcamo~Hern\'andez, B.~D\'\i{}az~S\'aez, C.~O. Dib, and A.~Zerwekh,
  ``{Constraints on vector resonances from a strong Higgs sector},''
  \href{http://dx.doi.org/10.1103/PhysRevD.96.115027}{{\em Phys. Rev. D}
  {\bfseries 96} no.~11, (2017) 115027},
  \href{http://arxiv.org/abs/1707.05195}{{\ttfamily arXiv:1707.05195
  [hep-ph]}}.

\bibitem{Bardeen:1989ds}
W.~A. Bardeen, C.~T. Hill, and M.~Lindner, ``{Minimal Dynamical Symmetry
  Breaking of the Standard Model},''
  \href{http://dx.doi.org/10.1103/PhysRevD.41.1647}{{\em Phys. Rev. D}
  {\bfseries 41} (1990) 1647}.

\bibitem{LopezHonorez:2006gr}
L.~Lopez~Honorez, E.~Nezri, J.~F. Oliver, and M.~H.~G. Tytgat, ``{The Inert
  Doublet Model: An Archetype for Dark Matter},''
  \href{http://dx.doi.org/10.1088/1475-7516/2007/02/028}{{\em JCAP} {\bfseries
  02} (2007) 028}, \href{http://arxiv.org/abs/hep-ph/0612275}{{\ttfamily
  arXiv:hep-ph/0612275}}.

\bibitem{Ivanov:2012fp}
I.~P. Ivanov and E.~Vdovin, ``{Classification of finite reparametrization
  symmetry groups in the three-Higgs-doublet model},''
  \href{http://dx.doi.org/10.1140/epjc/s10052-013-2309-x}{{\em Eur. Phys. J. C}
  {\bfseries 73} no.~2, (2013) 2309},
  \href{http://arxiv.org/abs/1210.6553}{{\ttfamily arXiv:1210.6553 [hep-ph]}}.

\bibitem{Keus:2013hya}
V.~Keus, S.~F. King, and S.~Moretti, ``{Three-Higgs-doublet models: symmetries,
  potentials and Higgs boson masses},''
  \href{http://dx.doi.org/10.1007/JHEP01(2014)052}{{\em JHEP} {\bfseries 01}
  (2014) 052}, \href{http://arxiv.org/abs/1310.8253}{{\ttfamily arXiv:1310.8253
  [hep-ph]}}.

\bibitem{Aranda:2012bv}
A.~Aranda, C.~Bonilla, and J.~L. Diaz-Cruz, ``{Three generations of Higgses and
  the cyclic groups},''
  \href{http://dx.doi.org/10.1016/j.physletb.2012.09.011}{{\em Phys. Lett. B}
  {\bfseries 717} (2012) 248--251},
  \href{http://arxiv.org/abs/1204.5558}{{\ttfamily arXiv:1204.5558 [hep-ph]}}.

\bibitem{Aranda:2014jua}
A.~Aranda, J.~E. Barradas-Guevara, A.~Cordero-Cid, F.~de~Anda, A.~Delgado,
  O.~Felix-Beltran, and J.~Hernandez-Sanchez, ``{Flavor Physics constrains on a
  $\mathbb{Z}_5$-3HDM},'' \href{http://arxiv.org/abs/1404.7829}{{\ttfamily
  arXiv:1404.7829 [hep-ph]}}.

\bibitem{Aranda:2019vda}
A.~Aranda, D.~Hern\'andez-Otero, J.~Hern\'andez-Sanchez, V.~Keus, S.~Moretti,
  D.~Rojas-Ciofalo, and T.~Shindou, ``{Z$_3$ symmetric inert ( 2+1
  )-Higgs-doublet model},''
  \href{http://dx.doi.org/10.1103/PhysRevD.103.015023}{{\em Phys. Rev. D}
  {\bfseries 103} no.~1, (2021) 015023},
  \href{http://arxiv.org/abs/1907.12470}{{\ttfamily arXiv:1907.12470
  [hep-ph]}}.

\bibitem{Akeroyd:2016ssd}
A.~G. Akeroyd, S.~Moretti, K.~Yagyu, and E.~Yildirim, ``{Light charged Higgs
  boson scenario in 3-Higgs doublet models},''
  \href{http://dx.doi.org/10.1142/S0217751X17501457}{{\em Int. J. Mod. Phys. A}
  {\bfseries 32} no.~23n24, (2017) 1750145},
  \href{http://arxiv.org/abs/1605.05881}{{\ttfamily arXiv:1605.05881
  [hep-ph]}}.

\bibitem{Hernandez:2021iss}
A.~E.~C. Hern\'andez, S.~Kovalenko, M.~Maniatis, and I.~Schmidt, ``{Fermion
  mass hierarchy and g \ensuremath{-} 2 anomalies in an extended 3HDM Model},''
  \href{http://dx.doi.org/10.1007/JHEP10(2021)036}{{\em JHEP} {\bfseries 10}
  (2021) 036}, \href{http://arxiv.org/abs/2104.07047}{{\ttfamily
  arXiv:2104.07047 [hep-ph]}}.

\bibitem{Zerwekh:2009yu}
A.~R. Zerwekh, ``{Two Composite Higgs Doublets: Is it the Low Energy Limit of a
  Natural Strong Electroweak Symmetry Breaking Sector?},''
  \href{http://dx.doi.org/10.1142/S0217732310032627}{{\em Mod. Phys. Lett. A}
  {\bfseries 25} (2010) 423--429},
  \href{http://arxiv.org/abs/0907.4690}{{\ttfamily arXiv:0907.4690 [hep-ph]}}.

\bibitem{DiChiara:2016ybc}
S.~Di~Chiara, M.~Heikinheimo, and K.~Tuominen, ``{Vector resonances at LHC Run
  II in composite 2HDM},''
  \href{http://dx.doi.org/10.1007/JHEP03(2017)009}{{\em JHEP} {\bfseries 03}
  (2017) 009}, \href{http://arxiv.org/abs/1611.09094}{{\ttfamily
  arXiv:1611.09094 [hep-ph]}}.

\bibitem{DeCurtis:2016scv}
S.~De~Curtis, S.~Moretti, K.~Yagyu, and E.~Yildirim, ``{Perturbative unitarity
  bounds in composite two-Higgs doublet models},''
  \href{http://dx.doi.org/10.1103/PhysRevD.94.055017}{{\em Phys. Rev. D}
  {\bfseries 94} no.~5, (2016) 055017},
  \href{http://arxiv.org/abs/1602.06437}{{\ttfamily arXiv:1602.06437
  [hep-ph]}}.

\bibitem{Rojas-Abatte:2017hqm}
F.~Rojas-Abatte, M.~L. Mora, J.~Urbina, and A.~R. Zerwekh, ``{Inert
  two-Higgs-doublet model strongly coupled to a non-Abelian vector
  resonance},'' \href{http://dx.doi.org/10.1103/PhysRevD.96.095025}{{\em Phys.
  Rev. D} {\bfseries 96} no.~9, (2017) 095025},
  \href{http://arxiv.org/abs/1707.04543}{{\ttfamily arXiv:1707.04543
  [hep-ph]}}.

\bibitem{Chivukula:2011ag}
R.~S. Chivukula, E.~H. Simmons, B.~Coleppa, H.~E. Logan, and A.~Martin,
  ``{Top-Higgs and Top-Pion Phenomenology in the Top Triangle Moose Model},''
  \href{http://dx.doi.org/10.1103/PhysRevD.83.055013}{{\em Phys. Rev. D}
  {\bfseries 83} (2011) 055013},
  \href{http://arxiv.org/abs/1101.6023}{{\ttfamily arXiv:1101.6023 [hep-ph]}}.

\bibitem{SekharChivukula:2009htk}
R.~Sekhar~Chivukula, N.~D. Christensen, B.~Coleppa, and E.~H. Simmons, ``{The
  Top Triangle Moose: Combining Higgsless and Topcolor Mechanisms for Mass
  Generation},'' \href{http://dx.doi.org/10.1103/PhysRevD.80.035011}{{\em Phys.
  Rev. D} {\bfseries 80} (2009) 035011},
  \href{http://arxiv.org/abs/0906.5567}{{\ttfamily arXiv:0906.5567 [hep-ph]}}.

\bibitem{Osland:2020onj}
P.~Osland, A.~A. Pankov, and I.~A. Serenkova, ``{Updated constraints on $Z'$
  and $W'$ bosons decaying into bosonic and leptonic final states using the run
  2 ATLAS data},'' \href{http://dx.doi.org/10.1103/PhysRevD.103.053009}{{\em
  Phys. Rev. D} {\bfseries 103} no.~5, (2021) 053009},
  \href{http://arxiv.org/abs/2012.13930}{{\ttfamily arXiv:2012.13930
  [hep-ph]}}.

\bibitem{Pankov:2021vzs}
A.~A. Pankov, I.~A. Serenkova, and V.~A. Bednyakov, ``{Updated Constraints on
  Heavy Resonances Using the run 2 Data at LHC},''
  \href{http://dx.doi.org/10.18524/1810-4215.2021.34.244250}{{\em Odessa
  Astron. Pub.} {\bfseries 34} (2021) 18--22}.

\bibitem{Bhattacharyya:2015nca}
G.~Bhattacharyya and D.~Das, ``{Scalar sector of two-Higgs-doublet models: A
  minireview},'' \href{http://dx.doi.org/10.1007/s12043-016-1252-4}{{\em
  Pramana} {\bfseries 87} no.~3, (2016) 40},
  \href{http://arxiv.org/abs/1507.06424}{{\ttfamily arXiv:1507.06424
  [hep-ph]}}.

\bibitem{Workman:2022ynf}
{\bfseries Particle Data Group} Collaboration, R.~L. Workman and Others,
  ``{Review of Particle Physics},''
  \href{http://dx.doi.org/10.1093/ptep/ptac097}{{\em PTEP} {\bfseries 2022}
  (2022) 083C01}.

\bibitem{Tao:1996vb}
Z.-j. Tao, ``{Radiative seesaw mechanism at weak scale},''
  \href{http://dx.doi.org/10.1103/PhysRevD.54.5693}{{\em Phys. Rev. D}
  {\bfseries 54} (1996) 5693--5697},
  \href{http://arxiv.org/abs/hep-ph/9603309}{{\ttfamily arXiv:hep-ph/9603309}}.

\bibitem{Ma:2006km}
E.~Ma, ``{Verifiable radiative seesaw mechanism of neutrino mass and dark
  matter},'' \href{http://dx.doi.org/10.1103/PhysRevD.73.077301}{{\em Phys.
  Rev. D} {\bfseries 73} (2006) 077301},
  \href{http://arxiv.org/abs/hep-ph/0601225}{{\ttfamily arXiv:hep-ph/0601225}}.

\bibitem{Xing:2020ijf}
Z.-z. Xing, ``{Flavor structures of charged fermions and massive neutrinos},''
  \href{http://dx.doi.org/10.1016/j.physrep.2020.02.001}{{\em Phys. Rept.}
  {\bfseries 854} (2020) 1--147},
  \href{http://arxiv.org/abs/1909.09610}{{\ttfamily arXiv:1909.09610
  [hep-ph]}}.

\bibitem{deSalas:2020pgw}
P.~F. de~Salas, D.~V. Forero, S.~Gariazzo, P.~Mart\'\i{}nez-Mirav\'e, O.~Mena,
  C.~A. Ternes, M.~T\'ortola, and J.~W.~F. Valle, ``{2020 global reassessment
  of the neutrino oscillation picture},''
  \href{http://dx.doi.org/10.1007/JHEP02(2021)071}{{\em JHEP} {\bfseries 02}
  (2021) 071}, \href{http://arxiv.org/abs/2006.11237}{{\ttfamily
  arXiv:2006.11237 [hep-ph]}}.

\bibitem{DiMicco:2019ngk}
J.~Alison {\em et~al.}, ``{Higgs boson potential at colliders: Status and
  perspectives},'' \href{http://dx.doi.org/10.1016/j.revip.2020.100045}{{\em
  Rev. Phys.} {\bfseries 5} (2020) 100045},
  \href{http://arxiv.org/abs/1910.00012}{{\ttfamily arXiv:1910.00012
  [hep-ph]}}.

\bibitem{Abouabid:2024gms}
H.~Abouabid {\em et~al.}, ``{HHH whitepaper},''
  \href{http://dx.doi.org/10.1140/epjc/s10052-024-13376-3}{{\em Eur. Phys. J.
  C} {\bfseries 84} (2024) 1183},
  \href{http://arxiv.org/abs/2407.03015}{{\ttfamily arXiv:2407.03015
  [hep-ph]}}.

\bibitem{ATLAS:2022jtk}
{\bfseries ATLAS} Collaboration, G.~Aad {\em et~al.}, ``{Constraints on the
  Higgs boson self-coupling from single- and double-Higgs production with the
  ATLAS detector using pp collisions at s=13 TeV},''
  \href{http://dx.doi.org/10.1016/j.physletb.2023.137745}{{\em Phys. Lett. B}
  {\bfseries 843} (2023) 137745},
  \href{http://arxiv.org/abs/2211.01216}{{\ttfamily arXiv:2211.01216
  [hep-ex]}}.

\bibitem{Arhrib:2015hoa}
A.~Arhrib, R.~Benbrik, J.~El~Falaki, and A.~Jueid, ``{Radiative corrections to
  the Triple Higgs Coupling in the Inert Higgs Doublet Model},''
  \href{http://dx.doi.org/10.1007/JHEP12(2015)007}{{\em JHEP} {\bfseries 12}
  (2015) 007}, \href{http://arxiv.org/abs/1507.03630}{{\ttfamily
  arXiv:1507.03630 [hep-ph]}}.

\bibitem{Kanemura:2016lkz}
S.~Kanemura, M.~Kikuchi, and K.~Yagyu, ``{One-loop corrections to the Higgs
  self-couplings in the singlet extension},''
  \href{http://dx.doi.org/10.1016/j.nuclphysb.2017.02.004}{{\em Nucl. Phys. B}
  {\bfseries 917} (2017) 154--177},
  \href{http://arxiv.org/abs/1608.01582}{{\ttfamily arXiv:1608.01582
  [hep-ph]}}.

\bibitem{Moyotl:2016fdk}
A.~Moyotl, S.~Chamorro, H.~Castilla-Valdez, and M.~A. P\'erez, ``{New physics
  effects in the Higgs trilinear self-coupling through one-loop radiative
  corrections},'' \href{http://arxiv.org/abs/1610.06299}{{\ttfamily
  arXiv:1610.06299 [hep-ph]}}.

\bibitem{Kanemura:2017wtm}
S.~Kanemura, M.~Kikuchi, K.~Sakurai, and K.~Yagyu, ``{Gauge invariant one-loop
  corrections to Higgs boson couplings in non-minimal Higgs models},''
  \href{http://dx.doi.org/10.1103/PhysRevD.96.035014}{{\em Phys. Rev. D}
  {\bfseries 96} no.~3, (2017) 035014},
  \href{http://arxiv.org/abs/1705.05399}{{\ttfamily arXiv:1705.05399
  [hep-ph]}}.

\bibitem{Kanemura:2004mg}
S.~Kanemura, Y.~Okada, E.~Senaha, and C.~P. Yuan, ``{Higgs coupling constants
  as a probe of new physics},''
  \href{http://dx.doi.org/10.1103/PhysRevD.70.115002}{{\em Phys. Rev. D}
  {\bfseries 70} (2004) 115002},
  \href{http://arxiv.org/abs/hep-ph/0408364}{{\ttfamily arXiv:hep-ph/0408364}}.

\bibitem{Kanemura:2002vm}
S.~Kanemura, S.~Kiyoura, Y.~Okada, E.~Senaha, and C.~P. Yuan, ``{New physics
  effect on the Higgs selfcoupling},''
  \href{http://dx.doi.org/10.1016/S0370-2693(03)00268-5}{{\em Phys. Lett. B}
  {\bfseries 558} (2003) 157--164},
  \href{http://arxiv.org/abs/hep-ph/0211308}{{\ttfamily arXiv:hep-ph/0211308}}.

\bibitem{Bhattacharyya:2014oka}
G.~Bhattacharyya and D.~Das, ``{Nondecoupling of charged scalars in Higgs decay
  to two photons and symmetries of the scalar potential},''
  \href{http://dx.doi.org/10.1103/PhysRevD.91.015005}{{\em Phys. Rev. D}
  {\bfseries 91} (2015) 015005},
  \href{http://arxiv.org/abs/1408.6133}{{\ttfamily arXiv:1408.6133 [hep-ph]}}.

\bibitem{Logan:2014jla}
H.~E. Logan, ``{TASI 2013 lectures on Higgs physics within and beyond the
  Standard Model},'' \href{http://arxiv.org/abs/1406.1786}{{\ttfamily
  arXiv:1406.1786 [hep-ph]}}.

\bibitem{CarcamoHernandez:2023dyz}
A.~E. C\'arcamo~Hern\'andez, J.~Marchant~Gonz\'alez, D.~Salinas-Arizmendi, and
  M.~L. Mora-Urrutia, ``{Phenomenological aspects of the fermion and scalar
  sectors of a S4 flavored 3-3-1 model},''
  \href{http://dx.doi.org/10.1016/j.nuclphysb.2024.116588}{{\em Nucl. Phys. B}
  {\bfseries 1005} (2024) 116588},
  \href{http://arxiv.org/abs/2305.13441}{{\ttfamily arXiv:2305.13441
  [hep-ph]}}.

\bibitem{CarcamoHernandez:2024ycd}
A.~E. C\'arcamo~Hern\'andez, D.~Salinas-Arizmendi, J.~Vignatti, and A.~Zerwekh,
  ``{Phenomenology of an Extended $1+2$ Higgs Doublet Model with $S_3$ Family
  Symmetry},'' \href{http://arxiv.org/abs/2408.01497}{{\ttfamily
  arXiv:2408.01497 [hep-ph]}}.

\bibitem{ATLAS:2022vkf}
{\bfseries ATLAS} Collaboration, G.~Aad {\em et~al.}, ``{A detailed map of
  Higgs boson interactions by the ATLAS experiment ten years after the
  discovery},'' \href{http://dx.doi.org/10.1038/s41586-022-04893-w}{{\em
  Nature} {\bfseries 607} no.~7917, (2022) 52--59},
  \href{http://arxiv.org/abs/2207.00092}{{\ttfamily arXiv:2207.00092
  [hep-ex]}}. [Erratum: Nature 612, E24 (2022)].

\bibitem{ATLAS:2022tnm}
{\bfseries ATLAS} Collaboration, G.~Aad {\em et~al.}, ``{Measurement of the
  properties of Higgs boson production at $\sqrt{s} = 13$ TeV in the
  $H\to\gamma\gamma$ channel using $139$ fb$^{-1}$ of $pp$ collision data with
  the ATLAS experiment},''
  \href{http://dx.doi.org/10.1007/JHEP07(2023)088}{{\em JHEP} {\bfseries 07}
  (2023) 088}, \href{http://arxiv.org/abs/2207.00348}{{\ttfamily
  arXiv:2207.00348 [hep-ex]}}.

\bibitem{Saha:2022cnz}
{\bfseries CMS} Collaboration, P.~Saha, ``{Recent Measurements of~Higgs Boson
  Properties in~the~Diphoton Decay Channel with~the~CMS Detector},''
  \href{http://dx.doi.org/10.1007/978-981-19-2354-8_33}{{\em Springer Proc.
  Phys.} {\bfseries 277} (2022) 183--186}.

\bibitem{Ardu:2022sbt}
M.~Ardu and G.~Pezzullo, ``{Introduction to Charged Lepton Flavor Violation},''
  \href{http://dx.doi.org/10.3390/universe8060299}{{\em Universe} {\bfseries 8}
  no.~6, (2022) 299}, \href{http://arxiv.org/abs/2204.08220}{{\ttfamily
  arXiv:2204.08220 [hep-ph]}}.

\bibitem{Toma:2013zsa}
T.~Toma and A.~Vicente, ``{Lepton Flavor Violation in the Scotogenic Model},''
  \href{http://dx.doi.org/10.1007/JHEP01(2014)160}{{\em JHEP} {\bfseries 01}
  (2014) 160}, \href{http://arxiv.org/abs/1312.2840}{{\ttfamily arXiv:1312.2840
  [hep-ph]}}.

\bibitem{Vicente:2014wga}
A.~Vicente and C.~E. Yaguna, ``{Probing the scotogenic model with lepton flavor
  violating processes},'' \href{http://dx.doi.org/10.1007/JHEP02(2015)144}{{\em
  JHEP} {\bfseries 02} (2015) 144},
  \href{http://arxiv.org/abs/1412.2545}{{\ttfamily arXiv:1412.2545 [hep-ph]}}.

\bibitem{Abada:2022dvm}
A.~Abada, N.~Bernal, A.~E. C\'arcamo~Hern\'andez, S.~Kovalenko, T.~B. de~Melo,
  and T.~Toma, ``{Phenomenological and cosmological implications of a
  scotogenic three-loop neutrino mass model},''
  \href{http://dx.doi.org/10.1007/JHEP03(2023)035}{{\em JHEP} {\bfseries 03}
  (2023) 035}, \href{http://arxiv.org/abs/2212.06852}{{\ttfamily
  arXiv:2212.06852 [hep-ph]}}.

\bibitem{MEG:2016leq}
{\bfseries MEG} Collaboration, A.~M. Baldini {\em et~al.}, ``{Search for the
  lepton flavour violating decay $\mu ^+ \rightarrow \mathrm {e}^+ \gamma $
  with the full dataset of the MEG experiment},''
  \href{http://dx.doi.org/10.1140/epjc/s10052-016-4271-x}{{\em Eur. Phys. J. C}
  {\bfseries 76} no.~8, (2016) 434},
  \href{http://arxiv.org/abs/1605.05081}{{\ttfamily arXiv:1605.05081
  [hep-ex]}}.

\bibitem{MEGII:2025gzr}
{\bfseries MEG II} Collaboration, K.~Afanaciev {\em et~al.}, ``{New limit on
  the {\ensuremath{\mu}}+-{\ensuremath{>}}e+{\ensuremath{\gamma}} decay with
  the MEG II experiment},'' \href{http://arxiv.org/abs/2504.15711}{{\ttfamily
  arXiv:2504.15711 [hep-ex]}}.

\bibitem{Lindner:2016bgg}
M.~Lindner, M.~Platscher, and F.~S. Queiroz, ``{A Call for New Physics : The
  Muon Anomalous Magnetic Moment and Lepton Flavor Violation},''
  \href{http://dx.doi.org/10.1016/j.physrep.2017.12.001}{{\em Phys. Rept.}
  {\bfseries 731} (2018) 1--82},
  \href{http://arxiv.org/abs/1610.06587}{{\ttfamily arXiv:1610.06587
  [hep-ph]}}.

\bibitem{Chiang:1993xz}
H.~C. Chiang, E.~Oset, T.~S. Kosmas, A.~Faessler, and J.~D. Vergados,
  ``{Coherent and incoherent (mu-, e-) conversion in nuclei},''
  \href{http://dx.doi.org/10.1016/0375-9474(93)90259-Z}{{\em Nucl. Phys. A}
  {\bfseries 559} (1993) 526--542}.

\bibitem{Kitano:2002mt}
R.~Kitano, M.~Koike, and Y.~Okada, ``{Detailed calculation of lepton flavor
  violating muon electron conversion rate for various nuclei},''
  \href{http://dx.doi.org/10.1103/PhysRevD.76.059902}{{\em Phys. Rev. D}
  {\bfseries 66} (2002) 096002},
  \href{http://arxiv.org/abs/hep-ph/0203110}{{\ttfamily arXiv:hep-ph/0203110}}.
  [Erratum: Phys.Rev.D 76, 059902 (2007)].

\bibitem{Kosmas:2001mv}
T.~S. Kosmas, S.~Kovalenko, and I.~Schmidt, ``{Nuclear muon- e- conversion in
  strange quark sea},''
  \href{http://dx.doi.org/10.1016/S0370-2693(01)00657-8}{{\em Phys. Lett. B}
  {\bfseries 511} (2001) 203},
  \href{http://arxiv.org/abs/hep-ph/0102101}{{\ttfamily arXiv:hep-ph/0102101}}.

\bibitem{SINDRUM:1987nra}
{\bfseries SINDRUM} Collaboration, U.~Bellgardt {\em et~al.}, ``{Search for the
  Decay $\mu^+ \to e^+ e^+ e^-$},''
  \href{http://dx.doi.org/10.1016/0550-3213(88)90462-2}{{\em Nucl. Phys. B}
  {\bfseries 299} (1988) 1--6}.

\bibitem{Mori:2016vwi}
{\bfseries MEG} Collaboration, T.~Mori, ``{Final Results of the MEG
  Experiment},'' \href{http://dx.doi.org/10.1393/ncc/i2016-16325-7}{{\em Nuovo
  Cim. C} {\bfseries 39} no.~4, (2017) 325},
  \href{http://arxiv.org/abs/1606.08168}{{\ttfamily arXiv:1606.08168
  [hep-ex]}}.

\bibitem{Blondel:2013ia}
A.~Blondel {\em et~al.}, ``{Research Proposal for an Experiment to Search for
  the Decay $\mu \to eee$},'' \href{http://arxiv.org/abs/1301.6113}{{\ttfamily
  arXiv:1301.6113 [physics.ins-det]}}.

\bibitem{Wintz:1998rp}
P.~Wintz, ``{Results of the SINDRUM-II experiment},'' {\em Conf. Proc. C}
  {\bfseries 980420} (1998) 534--546.

\bibitem{Bernstein:2013hba}
R.~H. Bernstein and P.~S. Cooper, ``{Charged Lepton Flavor Violation: An
  Experimenter's Guide},''
  \href{http://dx.doi.org/10.1016/j.physrep.2013.07.002}{{\em Phys. Rept.}
  {\bfseries 532} (2013) 27--64},
  \href{http://arxiv.org/abs/1307.5787}{{\ttfamily arXiv:1307.5787 [hep-ex]}}.

\bibitem{Sakharov:1967dj}
A.~D. Sakharov, ``{Violation of CP Invariance, C asymmetry, and baryon
  asymmetry of the universe},''
  \href{http://dx.doi.org/10.1070/PU1991v034n05ABEH002497}{{\em Pisma Zh. Eksp.
  Teor. Fiz.} {\bfseries 5} (1967) 32--35}.

\bibitem{Datta:2021gyi}
A.~Datta, R.~Roshan, and A.~Sil, ``{Scalar triplet flavor leptogenesis with
  dark matter},'' \href{http://dx.doi.org/10.1103/PhysRevD.105.095032}{{\em
  Phys. Rev. D} {\bfseries 105} no.~9, (2022) 095032},
  \href{http://arxiv.org/abs/2110.03914}{{\ttfamily arXiv:2110.03914
  [hep-ph]}}.

\bibitem{Hambye:2005tk}
T.~Hambye, M.~Raidal, and A.~Strumia, ``{Efficiency and maximal CP-asymmetry of
  scalar triplet leptogenesis},''
  \href{http://dx.doi.org/10.1016/j.physletb.2005.11.007}{{\em Phys. Lett. B}
  {\bfseries 632} (2006) 667--674},
  \href{http://arxiv.org/abs/hep-ph/0510008}{{\ttfamily arXiv:hep-ph/0510008}}.

\bibitem{Lu:2016ucn}
W.-B. Lu and P.-H. Gu, ``{Leptogenesis, radiative neutrino masses and inert
  Higgs triplet dark matter},''
  \href{http://dx.doi.org/10.1088/1475-7516/2016/05/040}{{\em JCAP} {\bfseries
  05} (2016) 040}, \href{http://arxiv.org/abs/1603.05074}{{\ttfamily
  arXiv:1603.05074 [hep-ph]}}.

\bibitem{Pilaftsis:1997jf}
A.~Pilaftsis, ``{CP violation and baryogenesis due to heavy Majorana
  neutrinos},'' \href{http://dx.doi.org/10.1103/PhysRevD.56.5431}{{\em Phys.
  Rev. D} {\bfseries 56} (1997) 5431--5451},
  \href{http://arxiv.org/abs/hep-ph/9707235}{{\ttfamily arXiv:hep-ph/9707235}}.

\bibitem{Peskin:1990zt}
M.~E. Peskin and T.~Takeuchi, ``{A New constraint on a strongly interacting
  Higgs sector},'' \href{http://dx.doi.org/10.1103/PhysRevLett.65.964}{{\em
  Phys. Rev. Lett.} {\bfseries 65} (1990) 964--967}.

\bibitem{Altarelli:1990zd}
G.~Altarelli and R.~Barbieri, ``{Vacuum polarization effects of new physics on
  electroweak processes},''
  \href{http://dx.doi.org/10.1016/0370-2693(91)91378-9}{{\em Phys. Lett. B}
  {\bfseries 253} (1991) 161--167}.

\bibitem{Peskin:1991sw}
M.~E. Peskin and T.~Takeuchi, ``{Estimation of oblique electroweak
  corrections},'' \href{http://dx.doi.org/10.1103/PhysRevD.46.381}{{\em Phys.
  Rev. D} {\bfseries 46} (1992) 381--409}.

\bibitem{Altarelli:1991fk}
G.~Altarelli, R.~Barbieri, and S.~Jadach, ``{Toward a model independent
  analysis of electroweak data},''
  \href{http://dx.doi.org/10.1016/0550-3213(92)90376-M}{{\em Nucl. Phys. B}
  {\bfseries 369} (1992) 3--32}. [Erratum: Nucl.Phys.B 376, 444 (1992)].

\bibitem{Barbieri:2004qk}
R.~Barbieri, A.~Pomarol, R.~Rattazzi, and A.~Strumia, ``{Electroweak symmetry
  breaking after LEP-1 and LEP-2},''
  \href{http://dx.doi.org/10.1016/j.nuclphysb.2004.10.014}{{\em Nucl. Phys. B}
  {\bfseries 703} (2004) 127--146},
  \href{http://arxiv.org/abs/hep-ph/0405040}{{\ttfamily arXiv:hep-ph/0405040}}.

\bibitem{CarcamoHernandez:2015mkh}
A.~E. C\'arcamo~Hern\'andez, I.~de~Medeiros~Varzielas, and E.~Schumacher,
  ``{Fermion and scalar phenomenology of a two-Higgs-doublet model with
  $S_3$},'' \href{http://dx.doi.org/10.1103/PhysRevD.93.016003}{{\em Phys. Rev.
  D} {\bfseries 93} no.~1, (2016) 016003},
  \href{http://arxiv.org/abs/1509.02083}{{\ttfamily arXiv:1509.02083
  [hep-ph]}}.

\bibitem{CarcamoHernandez:2015smi}
A.~E. C\'arcamo~Hern\'andez, S.~Kovalenko, and I.~Schmidt, ``{Precision
  measurements constraints on the number of Higgs doublets},''
  \href{http://dx.doi.org/10.1103/PhysRevD.91.095014}{{\em Phys. Rev. D}
  {\bfseries 91} (2015) 095014},
  \href{http://arxiv.org/abs/1503.03026}{{\ttfamily arXiv:1503.03026
  [hep-ph]}}.

\bibitem{Grimus:2008nb}
W.~Grimus, L.~Lavoura, O.~M. Ogreid, and P.~Osland, ``{The Oblique parameters
  in multi-Higgs-doublet models},''
  \href{http://dx.doi.org/10.1016/j.nuclphysb.2008.04.019}{{\em Nucl. Phys. B}
  {\bfseries 801} (2008) 81--96},
  \href{http://arxiv.org/abs/0802.4353}{{\ttfamily arXiv:0802.4353 [hep-ph]}}.

\bibitem{CMS:2018cyk}
{\bfseries CMS} Collaboration, A.~M. Sirunyan {\em et~al.}, ``{Search for a
  standard model-like Higgs boson in the mass range between 70 and 110 GeV in
  the diphoton final state in proton-proton collisions at $\sqrt{s}=$ 8 and 13
  TeV},'' \href{http://dx.doi.org/10.1016/j.physletb.2019.03.064}{{\em Phys.
  Lett. B} {\bfseries 793} (2019) 320--347},
  \href{http://arxiv.org/abs/1811.08459}{{\ttfamily arXiv:1811.08459
  [hep-ex]}}.

\bibitem{CMS:2023yay}
{\bfseries CMS} Collaboration, ``{Search for a standard model-like Higgs boson
  in the mass range between 70 and 110$~\mathrm{GeV}$ in the diphoton final
  state in proton-proton collisions at $\sqrt{s}=13~\mathrm{TeV}$},''.

\bibitem{Dev:2023kzu}
P.~S.~B. Dev, R.~N. Mohapatra, and Y.~Zhang, ``{Explanation of the 95 GeV
  \ensuremath{\gamma}\ensuremath{\gamma} and bb\textasciimacron{} excesses in
  the minimal left-right symmetric model},''
  \href{http://dx.doi.org/10.1016/j.physletb.2024.138481}{{\em Phys. Lett. B}
  {\bfseries 849} (2024) 138481},
  \href{http://arxiv.org/abs/2312.17733}{{\ttfamily arXiv:2312.17733
  [hep-ph]}}.

\bibitem{Bonilla:2023wok}
C.~Bonilla, A.~E. Carcamo~Hernandez, S.~Kovalenko, H.~Lee, R.~Pasechnik, and
  I.~Schmidt, ``{Fermion mass hierarchy in an extended left-right symmetric
  model},'' \href{http://dx.doi.org/10.1007/JHEP12(2023)075}{{\em JHEP}
  {\bfseries 12} (2023) 075}, \href{http://arxiv.org/abs/2305.11967}{{\ttfamily
  arXiv:2305.11967 [hep-ph]}}.

\bibitem{BrahimAit-Ouazghour:2024img}
B.~Ait-Ouazghour, M.~Chabab, and K.~Goure, ``{Unified Interpretation of 95 GeV
  Excesses in the Two Higgs Doublet type II Seesaw Model},''
  \href{http://arxiv.org/abs/2410.11140}{{\ttfamily arXiv:2410.11140
  [hep-ph]}}.

\bibitem{Sharma:2024vhv}
P.~Sharma, A.-T. Mulaudzi, K.~Mosala, T.~Mathaha, M.~Kumar, B.~Mellado,
  A.~Crivellin, M.~Titov, M.~Ruan, and Y.~Fang, ``{Discovery Potential of
  Future Electron-Positron Colliders for a 95 GeV Scalar},''
  \href{http://arxiv.org/abs/2407.16806}{{\ttfamily arXiv:2407.16806
  [hep-ph]}}.

\bibitem{Banik:2024ugs}
S.~Banik, G.~Coloretti, A.~Crivellin, and H.~E. Haber, ``{Correlating
  A\textrightarrow{}\ensuremath{\gamma}\ensuremath{\gamma} with electric dipole
  moments in the two Higgs doublet model in light of the diphoton excesses at
  95~GeV and 152~GeV},''
  \href{http://dx.doi.org/10.1103/PhysRevD.111.075021}{{\em Phys. Rev. D}
  {\bfseries 111} no.~7, (2025) 075021},
  \href{http://arxiv.org/abs/2412.00523}{{\ttfamily arXiv:2412.00523
  [hep-ph]}}.

\bibitem{Belyaev:2023xnv}
A.~Belyaev, R.~Benbrik, M.~Boukidi, M.~Chakraborti, S.~Moretti, and S.~Semlali,
  ``{Explanation of the hints for a 95 GeV Higgs boson within a 2-Higgs Doublet
  Model},'' \href{http://dx.doi.org/10.1007/JHEP05(2024)209}{{\em JHEP}
  {\bfseries 05} (2024) 209}, \href{http://arxiv.org/abs/2306.09029}{{\ttfamily
  arXiv:2306.09029 [hep-ph]}}.

\bibitem{Biekotter:2023jld}
T.~Biek\"otter, S.~Heinemeyer, and G.~Weiglein, ``{The CMS di-photon excess at
  95 GeV in view of the LHC Run 2 results},''
  \href{http://dx.doi.org/10.1016/j.physletb.2023.138217}{{\em Phys. Lett. B}
  {\bfseries 846} (2023) 138217},
  \href{http://arxiv.org/abs/2303.12018}{{\ttfamily arXiv:2303.12018
  [hep-ph]}}.

\bibitem{Huong:2025uwx}
D.~T. Huong, A.~E. C\'arcamo~Hern\'andez, H.~T. Hung, T.~T. Hieu, and N.~A.
  P\'erez-Julve, ``{Extended IDM theory with low scale seesaw mechanisms},''
  \href{http://arxiv.org/abs/2502.19488}{{\ttfamily arXiv:2502.19488
  [hep-ph]}}.

\bibitem{Martin:2009iq}
A.~D. Martin, W.~J. Stirling, R.~S. Thorne, and G.~Watt, ``{Parton
  distributions for the LHC},''
  \href{http://dx.doi.org/10.1140/epjc/s10052-009-1072-5}{{\em Eur. Phys. J. C}
  {\bfseries 63} (2009) 189--285},
  \href{http://arxiv.org/abs/0901.0002}{{\ttfamily arXiv:0901.0002 [hep-ph]}}.

\bibitem{tHooft:1972tcz}
G.~'t~Hooft and M.~J.~G. Veltman, ``{Regularization and Renormalization of
  Gauge Fields},'' \href{http://dx.doi.org/10.1016/0550-3213(72)90279-9}{{\em
  Nucl. Phys. B} {\bfseries 44} (1972) 189--213}.

\bibitem{tHooft:1978jhc}
G.~'t~Hooft and M.~J.~G. Veltman, ``{Scalar One Loop Integrals},''
  \href{http://dx.doi.org/10.1016/0550-3213(79)90605-9}{{\em Nucl. Phys. B}
  {\bfseries 153} (1979) 365--401}.

\bibitem{Passarino:1978jh}
G.~Passarino and M.~J.~G. Veltman, ``{One Loop Corrections for e+ e-
  Annihilation Into mu+ mu- in the Weinberg Model},''
  \href{http://dx.doi.org/10.1016/0550-3213(79)90234-7}{{\em Nucl. Phys. B}
  {\bfseries 160} (1979) 151--207}.

\end{thebibliography}\endgroup

\end{document}